\documentclass{article}[12pt]
\usepackage{amsmath,amssymb}
\usepackage{epsfig}

\usepackage{graphicx}
\begin{document}

\title{Monte Carlo Methods for Estimating Interfacial Free Energies and Line Tensions}

\author{Kurt Binder$^1$\thanks{1) Institut f\"ur Physik, Johannes Gutenberg-Universit\"at Mainz, Staudinger Weg 7, D-55099 Mainz, Germany}, Benjamin Block$^1$, Subir K. Das$^2$\thanks{2) Theoretical Sciences Unit, Jawaharlal Nehru Centre for Advanced Scientific Research, Jakkur, Bangalore, 560064, India}, Peter Virnau$^1$, and David Winter$^1$}

\maketitle

\begin{abstract}
Excess contributions to the free energy due to interfaces occur for many problems encountered in the statistical physics of condensed matter when coexistence between different phases is possible (e.g.\ wetting phenomena, nucleation, crystal growth, etc.). This article reviews two methods to estimate both interfacial free energies and line tensions by Monte Carlo simulations of simple models, (e.g.\ the Ising model, a symmetrical binary Lennard-Jones fluid exhibiting a miscibility gap, and a simple Lennard-Jones fluid). One method is based on thermodynamic integration. This method is useful to study flat and inclined interfaces for Ising lattices, allowing also the estimation of line tensions of three-phase contact lines, when the interfaces meet walls (where ``surface fields'' may act). A generalization to off-lattice systems is described as well.
The second method is based on the sampling of the order parameter distribution of the system throughout the two-phase coexistence region of the model. Both the interface free energies of flat interfaces and of (spherical or cylindrical) droplets (or bubbles) can be estimated, including also systems with walls, where sphere-cap shaped wall-attached droplets occur. The curvature-dependence of the interfacial free energy is discussed, and estimates for the line tensions are compared to results from the thermodynamic integration method. Basic limitations of all these methods are critically discussed, and an outlook on other approaches is given.
\end{abstract}

\section{Introduction and Background}
The statistical mechanics of interfaces between coexisting phases dates back to van der Waals \cite{1} and Gibbs \cite{2}. Some results, like Young's equation \cite{3} for the contact angle under which the vapor-liquid interface of a sessile (macroscopic) droplet meets the supporting wall in thermal equilibrium, are known since more than 200 years. Understanding the statistical mechanics of interfacial phenomena is relevant for important problems such as nucleation \cite{4,5,6,7,8,9,10,11}, wetting and spreading of fluid films \cite{12,13,14,15,16,17,18,19,20,21,22,23,24}, metallurgy \cite{25} and crystal growth \cite{26,27} and diverse technological applications that these phenomena have (ranging from oil recovery \cite{28} over the efficient deposition of pesticides on plant leaves \cite{29} and microfluidic devices \cite{30} to the prediction of rain clouds resulting from nucleation of ice crystals in the atmosphere \cite{31}, etc.).

In view of this long history and widespread applications and broad significance of this subject it is surprising that still many basic aspects are not yet well understood. E.g., regarding interfacial profiles and widths it is still problematic to disentangle the ``intrinsic'' profile and width \cite{16} from the broadening \cite{32,33,34} caused by capillary waves \cite{35,36,37}. While experimental methods exist to measure the interfacial free energy of (macroscopically flat) fluid-fluid interfaces and liquid-vapor interfaces, the measurement of interfacial tensions involving a solid-phase is difficult \cite{38}: thus such quantities often are extracted from contact angle measurements \cite{38} assuming the validity of Young's equation \cite{3} (rather than checking it). Since accurate contact angle measurements are severely hampered by contact angle hysteresis between advancing and receding three-phase contact lines \cite{12,21,23,24} it is difficult to obtain precise experimental information on either the effects of curvature on the interfacial free energy of a (nanoscopically small) droplet or on the excess free energy due to the three-phase contact line, the line tension \cite{39,40,41}. As reviewed by Mugele et al. \cite{41}, early attempts to measure the line tensions yielded estimates that presumably are several orders of magnitude too large. And although the concept of a line tension was already introduced by Gibbs \cite{2} and since then has been thoroughly studied (see e.g. Indekeu \cite{42} and Dietrich et al. \cite{43,44}), there are still severe conceptual problems \cite{45}, and there is an ongoing debate \cite{46,47} on the physical significance of the line tension for describing the contact angle of nanodroplets \cite{48}. Likewise, there has been a longstanding debate (see Schrader et al. \cite{49} and Block et al. \cite{50} for references) on the sign and magnitude of the ``Tolman length'' $\delta$ \cite{51} introduced to describe the curvature dependent surface tension $\gamma(R)$,  $R$ being the radius of a spherical droplet
\begin{equation}\label{eq1}
\gamma (R)= \gamma (\infty)/(1+2 \delta/R).
\end{equation}
While Tolman \cite{51} originally suggested that $\delta$ is a positive constant and a length of molecular dimensions, it soon was recognized that Eq.~\ref{eq1} cannot hold for model systems that exhibit a symmetry between the coexisting phases, such as the lattice gas model which has particle-hole symmetry, or a strictly symmetrical binary (A,B) mixture \cite{52}. Since turning particles into holes (and vice versa) turns a droplet surrounded by vapor into a bubble surrounded by liquid, it thus changes the sign of $R$. The requirement that $\gamma(R)$ is invariant against such a transformation excludes Eq.~\ref{eq1} for such models, and one finds instead \cite{50} another length $\ell$ describing a $1/R^2$ dependence,
\begin{equation}\label{eq2}
\gamma (R) = \gamma (\infty) /[1+2(\ell /R)^2].
\end{equation}
On the basis of a simulation study, ten Wolde and Frenkel \cite{53} suggested that Eq.~\ref{eq2} in fact also holds for asymmetric fluids, such as the ordinary Lennard-Jones model for a vapor to liquid transition. However, a suggestion that $\delta \equiv 0$ for all fluids is at odds with many density functional calculations (see e.g. Talanquer and Oxtoby \cite{54} and Granasy \cite{55}) that actually $\delta $ is nonzero in general but strongly $R$-dependent (and $\delta (R \rightarrow \infty)$ is slightly negative for a droplet surrounded by vapor). Of course, density functional theory is not exact, e.g.\ it ignores the effects due to capillary-wave fluctuations of the interfaces and the underlying assumptions of a mean-field character \cite{56} cause the fact that the radial density profile of a droplet has a singular behavior (the radius of a metastable droplet diverges, although its free energy vanishes, when the state of the vapor reaches the spinodal density), which is spurious for systems with short range forces \cite{7,8,11}. In view of these problems it is desirable to have simulation approaches which do not suffer from these problems. In the present paper, we shall review work \cite{50} which suggests that for fluids lacking particular symmetries actually a combination of Eqs.~\ref{eq1} and \ref{eq2} is a reasonable description,
\begin{equation}\label{eq3}
\gamma (R) = \gamma(\infty) /[1+2 \delta/R+2(\ell/R)^2],
\end{equation}
where the length $\ell$ is of molecular dimensions, while $\delta$ actually is one order of magnitude smaller, and negative (for a droplet). This result \cite{50} agrees with recent simulation approaches applying rather different methods \cite{57,58} and is essentially equivalent to a relation $\delta (R) = \delta (\infty) + const/R$ proposed by the density functional theory \cite{50}.

However, also for computer simulation approaches the accurate prediction of interfacial free energies has been a challenging problem. One difficulty is that one needs to study systems with rather large linear dimensions, to avoid (or at least control) finite size effects, that are due to the constraints that boundary conditions have on long range interfacial fluctuations, such as capillary waves at interfaces between fluid phases. Even more difficult is the study of solid-liquid interfaces \cite{59,60,61,62,63,64,65,66,67,68,69,70,71,72,73,74,75,76,77}, since the periodic boundary conditions that are normally applied in the directions parallel to the interface may create a misfit resulting in an elastic distortion of the crystal structure of the solid, which would lead to severe systematic errors. Even when the finite size effects \cite{75,76,77,78,79,80,81,82,83,84,85,86,87} are under control, one must consider that interfaces are mesoscopic ``objects'' and their fluctuations are rather slow, and hence a substantial effort of statistical sampling is required. Thus, the present review can only be a progress report, and the selection of material is clearly biased by the expertise and interests of the authors.

In the next section, we focus on the estimation of interface free energies from thermodynamic integration, using the energy excess of a system with interfaces relative to a system without interfaces as observable that is sampled. This approach is particularly popular and simple for lattice models such as the Ising model \cite{88,89,90}, where standard relations for fluid systems (e.g.\ based on the anisotropy of the pressure tensor \cite{91}) are not applicable. As an example, we shall discuss the estimation of $f_{int}(\vartheta)$, the interfacial free energy of a domain wall between oppositely oriented spins in the Ising model, considering an inclination of the interface plane by an angle $\vartheta$ relative to a (100) lattice plane in the simple cubic model \cite{89}. The method can straightforwardly be extended to obtain the excess free energy of the Ising models due to walls \cite{92,93,94} (where ``surface fields'' \cite{95,96,97,98} may act). In this way, the difference in wall free energies of coexisting phases that enter the Young \cite{3} equation, can be directly estimated \cite{94,99}, and this approach can be carried over to off-lattice models of fluids as well \cite{100,101}. We shall describe such applications in Sec.~3.

Then we turn (Sec.~4) to an alternative method, where one samples the distribution of the order parameter of a system under conditions where two phases can coexist in thermal equilibrium, traversing the coexistence region using umbrella sampling techniques \cite{102,103,104} or related methods \cite{105,106,107,108,109,110,111,112,113,114}. From such studies the interfacial free energy of flat interfaces of fluid systems can be studied conveniently even when it is very small, e.g.\ in the region near the critical point \cite{106,115,116,117,118,119,120,121,122}. Recently this method has been extended to study the free energy of spherical \cite{49,50} and cylindrical droplets (and bubbles) \cite{50}).

Also an extension to wall-attached droplets has been proposed, which allows to test \cite{94,99} concepts on heterogeneous nucleation \cite{123} and to provide estimates for the line tension (Sec.~5). Finally we give in Sec.~6 a brief comparison to other methods and an outlook on open problems.

\section{Interfacial free energies of flat interfaces from thermodynamic integration}
\subsection{Comments of the preparation of systems containing interfaces and the problem of finite size effects}
As is well known, for standard Monte Carlo importance sampling \cite{110,113,114,124} the bulk free energy $F$ of a model system is not straightforwardly accessible as an ``output'' of the computation, while the internal energy $E$ is directly accessible as the thermal average of the Hamiltonian $\mathcal{H}$, $E(T)= \langle \mathcal{H}\rangle_{\textrm{NVT}}$, in the NVT ensemble of a fluid (or a lattice system such as the Ising model, for instance, where the number $N$ of spins in the lattice and the volume $V$ are strictly proportional, rather than being two separate independent variables). Using then the standard thermodynamic relation $(\beta \equiv 1/k_BT)$
\begin{equation}\label{eq4}
\bigg(\frac {\partial \beta F}{\partial \beta} \bigg)_{N,V} = E(T),
\end{equation}
free energy differences between two states at different temperatures are readily obtained from
\begin{equation}\label{eq5}
F(T)=F(T_0)+ \int \limits _{1/k_BT_0}^{1/k_BT} d \beta ' E(T'), \quad \beta '=1/k_BT'.
\end{equation}
This simple method has been occasionally used for Ising systems\cite{124,125} noting that the free energy is trivially known both for $T \rightarrow \infty$, $F(T\rightarrow \infty)=E(T)-TS(T)\rightarrow - Nk_BT \ln2$, and for $T\rightarrow 0,\quad F(T\rightarrow 0)=E(0)$. In practice, then one does not use $T_0=0$ in Eq.~\ref{eq5} but rather chooses a nonzero temperature $T_0$ for which the entropy of the system is already negligibly small, so that $F(T_0) \approx E(T_0)$ is a sufficiently accurate approximation \cite{124,125}. As a caveat, we note that for off-lattice systems the use of reference states for which the free energy is trivially known is clearly more subtle \cite{113}. A general disadvantage of Eq.~\ref{eq5} is that for an accurate numerical integration the discretization of the temperature interval from $T_0$ to $T$ needs a very large number of intermediate states $T'$, e.g., in the study of surface-induced ordering for the face-centered cubic Ising antiferromagnet a number of a few hundred temperatures was used \cite{125}, in order to locate the first-order phase transition with a relative accuracy of about $3 \cdot 10^{-5}$.

\begin{figure}\centering
\includegraphics[width=0.6\linewidth]{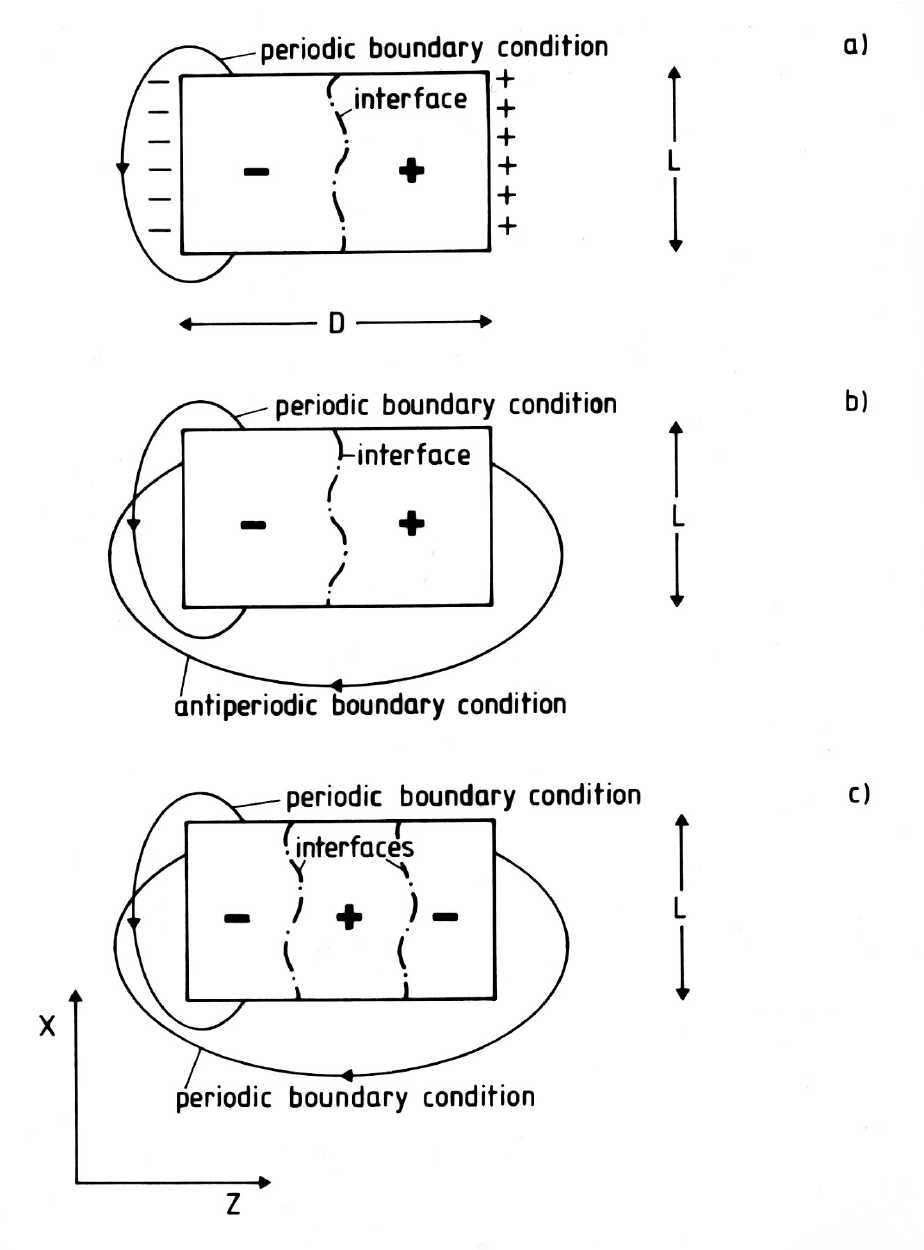}
\caption{\label{fig1} Schematic sketch of three possible simulation geometries to study interfaces in Ising systems, choosing $L \times D$ lattices (in $d=2)$ or $L \times L \times D$ lattices (in $d=3$), respectively : (a) the ``surface field'' boundary condition; (b) the antiperiodic boundary condition; (c) the fully periodic boundary condition. Note that in $x$ (and y, in $d=3$ dimensions) periodic boundary conditions are applied throughout, to ensure translational invariance in the directions parallel to the interface. The interfaces between coexisting domains at positive (+) and negative (-) magnetization are shown schematically as dash-dotted lines. Note that the state (c) is stable if the total magnetization in the system is conserved (and chosen zero or close to zero).}
\end{figure}

In order to extend this approach to obtain interfacial free energies, one needs to choose suitable boundary conditions, in order to prepare a system at phase coexistence, with one (or two) interfaces separating the coexisting phases \{e.g., domains with positive (+) and negative (-) spontaneous magnetization in an Ising ferromagnet, see Fig.\ref{fig1}\}.

The choice of a negative surface field $H_1$ in the first lattice row (or plane) and a positive surface field (of the same absolute strength) $H_D=-H_1$ in the last row (or plane), has the disadvantage that the local magnetization $m_n$ in the n'th plane or row ($n=1,2,\ldots,D)$ of the lattice will somewhat deviate from its bulk value near these free surfaces, over a distance of the order of the correlation length $\xi$ in the system \cite{95,96,97,98}. The same holds for the related ``fixed spin'' boundary condition (instead of surface fields one has spins $s_i=-1$ at all sites in the layer $n=0$ adjacent to the first layer on the left and spins $s_i=+1$ at all sites in the layer $n=D+1$ adjacent to the last layer on the right). In fact, for square (or simple cubic) lattices and nearest neighbor exchange $J$ the fixed spin boundary condition corresponds simply to a specific choice of the surface field ($H_1=-J$) for the surface field boundary condition. No disturbance at the boundaries of the lattice is obtained for the antiperiodic boundary condition Fig.~\ref{fig1}b, of course. This choice is only possible for systems with a trivial symmetry between the coexisting phases, such as the spin-reversal symmetry of the Ising ferromagnet, or the $A \leftrightarrow B$ exchange symmetry of a symmetric binary mixture (A,B). The choices of wall fields stabilizing the coexisting phases (Fig.~\ref{fig1}a) or the fully periodic boundary conditions (Fig.~\ref{fig1}c) are also possible for systems lacking such a symmetry, of course. E.g., the choice of Fig.~\ref{fig1}a has been made for the study of interfaces in an off-lattice model for a demixed compressible binary (A,B) alloy where the A-rich and B-rich phases have different lattice parameters but the same crystal structure \cite{126}. Similarly, the choice of Fig.~\ref{fig1}c has been made for the study of interfaces in a model for colloid-polymer mixtures \cite{87}.

In order to extend the application of Eqs.~\ref{eq4}, \ref{eq5} to interfacial free energies, one needs to single out the interfacial contribution (which is proportional to $L$ or $L^2$, in $d=2$ or $d=3$ dimensions) from the bulk free energy (which is proportional to $LD$ or $L^2D$, respectively). This is done by considering suitable differences between the energy of a system containing an interface (Fig.~\ref{fig1}a,b) or two interfaces (Fig.~\ref{fig1}c) and a corresponding system without any interfaces but the same linear dimensions. E.g., we simulate a system with boundary fields $H_1=H_D$, denoting its energy as $E_{++}(T)$, while we denote the energy of the system of Fig.~\ref{fig1}a as $E_{-+}(T)$. Likewise, we denote the energy of the system in Fig.~\ref{fig1}b, as $E_{AP}(T)$ while a corresponding system with periodic boundary conditions throughout (and uniform magnetization, which may be either the positive or the negative spontaneous magnetization) has energy $E_P(T)$. Then the interfacial free energy $F_{\textrm{int}}(T)$ is obtained from
\begin{equation}\label{eq6}
F_{\textrm{int}}(T) = F_{\textrm{int}}(T_0) + \int \limits ^{1/k_BT}_{1/k_BT_0}d \beta ' [E_{-+}(T) - E_{++}(T)],
\end{equation}
if the geometry of Fig.~\ref{fig1}a is used, or
\begin{equation}\label{eq7}
F_{\textrm{int}} = F_{\textrm{int}}(T_0) + \int \limits _{1/k_BT_0}^{1/k_BT} d \beta ' [E_{AP}(T)-E_P(T)].
\end{equation}
For an Ising ferromagnet we have for $T_0 \rightarrow 0$ and an interface perpendicular to a lattice direction of the square $(d=2)$ or simple cubic $(d=3)$ lattice,
\begin{equation}\label{eq8}
E_{-+}(T_0)-E_{++}(T_0)=E_{AP}(T_0)-E_P(T_0)= F_{\textrm{int}}  (T_0)=2JL^{d-1},
\end{equation}
due to the ``broken bonds'' (connecting oppositely oriented spins) across the interface.

Assuming that the two interfaces in Fig.~\ref{fig1}c are not interacting, a similar reasoning (considering the energy difference between the system of Fig.~\ref{fig1}c and a corresponding system with also periodic boundary conditions but in a pure phase) would simply yield $2F_{\textrm{int}}(T)$.

We emphasize that this straightforward approach relies on two basic ingredients:
\begin{itemize}
\item[(i)] there is a symmetry between the two coexisting phases (denoted by ``+'' and ``-'' in Fig.~\ref{fig1}), so when we consider differences such as $E_{-+}(T) - E_{++}(T)$, the bulk contribution to the free energy cancels out exactly.
    \item[(ii)] We can reach a reference state (at a temperature $T_0$) where entropic contributions are fully negligible, without passing through an intervening first-order phase transition when we carry out the integration from $1/k_BT_0$ to $1/k_BT$.
\end{itemize}

These prerequisites are not fulfilled in many cases of interest, of course, such as for off-lattice models. E.g., for a binary symmetric Lennard-Jones fluid (as studied by Das et al. \cite{50,100,101}) Eqs.~\ref{eq6}, \ref{eq7} still would be applicable, but no reference state, where the interfacial free energy is known, exists. In fact, at low temperatures the system does not stay fluid but undergoes a (first order) transition to the crystal phase. Note also, that even for the crystal (due to the use of classical statistical mechanics) care is needed in the discussion of its entropy at low temperatures, and often one connects via thermodynamic integration to an Einstein crystal \cite{127}, whose statistical mechanics is trivial to evaluate. Furthermore, most basic models (e.g., the Lennard-Jones model of a simple fluid, or the Asakura-Oosawa model \cite{128} for a colloid polymer mixture, etc.) lack a symmetry between coexisting phases. Then no analog of antiperiodic boundary conditions exists, while one can still use boundary fields that energetically prefer one of the coexisting phases (e.g., this is done in the study of interface localization transition using the Asakura-Oosawa model for colloid-polymer mixtures \cite{129}).

\begin{figure}\centering
\includegraphics[width=0.6\linewidth]{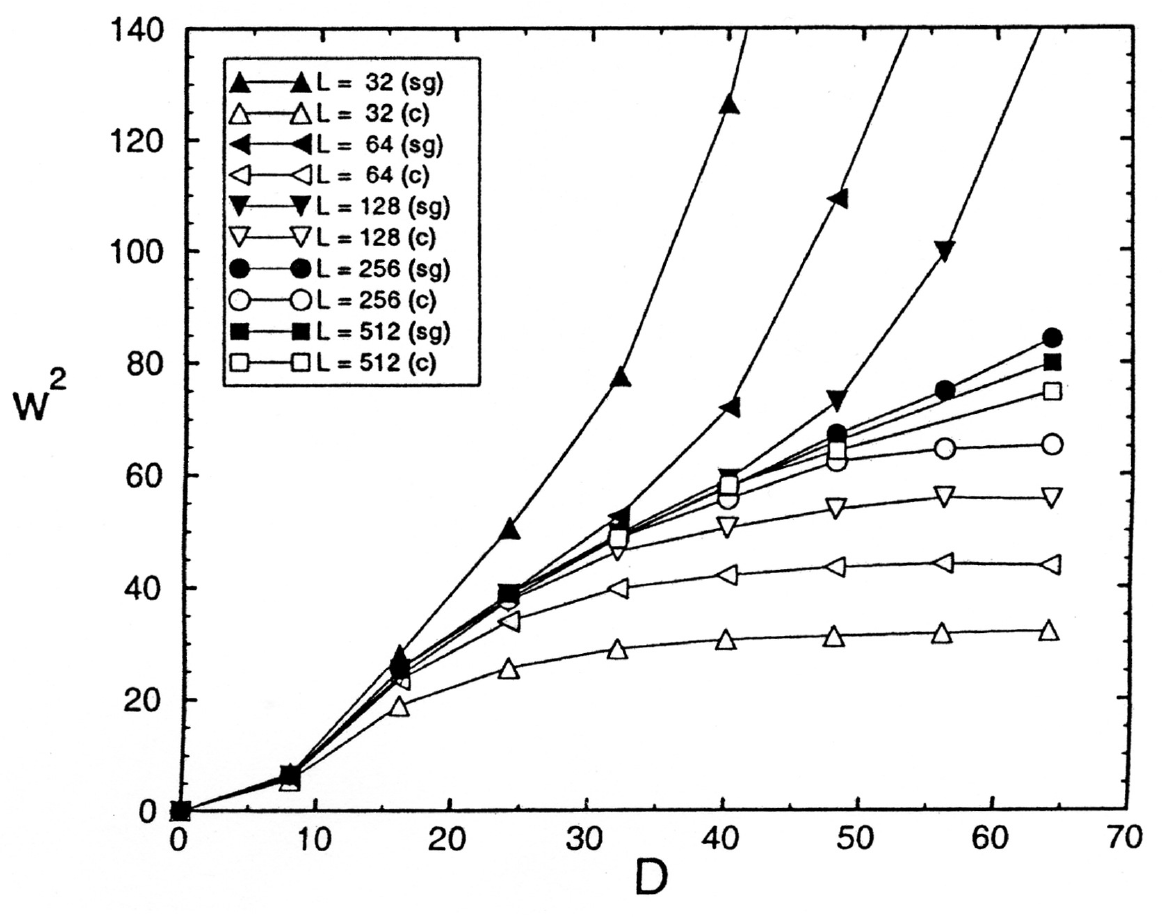}
\caption{\label{fig2} Plot of the mean square interfacial width $w^2$ of an interface between coexisting A-rich and B-rich phases of a symmetrical polymer mixture vs.\ the perpendicular linear dimension $D$ for several choices of the parallel linear dimension $L$ (L = 32,64,128,256 and 512 lattice spacings, respectively), using the geometry of Fig.~\ref{fig1}a. Monte Carlo simulation results for the bond fluctuation model of two types of polymer chains with chain lengths $N_A=N_B=N=32$ are shown, at a volume fraction of $\phi =0.5$ occupied lattice sites, and a temperature $T=0.48 T_c$. Two choices of statistical ensembles are included: the canonical ensemble, including identity exchanges in the Monte Carlo move (open symbols), marked by (c) and the semi-grand-canonical ensemble (filled symbols), marked by (sg). Statistical errors are typically of the order of the size of thy symbols, and curves are drawn as guides to the eye only. From Werner et al. \cite{82}.}
\end{figure}

While the simulation geometries shown schematically in Fig.~\ref{fig1} allow the estimation of interfacial free energies via thermodynamic integration only in exceptional cases, they are widely used to study other interfacial properties, e.g.\ the interfacial profile and its width \cite{33,49,68,69,70,71,72,73,74,75,76,77,78,79,80,81,82,83,84,85,86,87,90,126,130}. However, it is not always recognized that these properties do depend strongly on the chosen simulation geometry, in particular when the interface is rough (interfaces between coexisting fluid phases always are rough, and the same holds for interfaces in the $d=2$ Ising model; in contrast, the interface in the $d=3$ Ising model is rough only for temperatures exceeding the roughening transition temperature $T_R$ \cite{131}). In the geometry of Fig.~\ref{fig1}a, the average position of the interface is at $z=D/2$, but the interface becomes delocalized in the limit $D\rightarrow \infty$, similar to the case of Figs.~\ref{fig1}b,c, where the z-coordinate(s) of the interface(s) are not fixed. This translational degree of freedom of the interface as a whole along the z-axis means that there is a logarithmic contribution $-k_BT \ln D$ to $F_{\textrm{int}}(T)$ when one uses the geometry of Fig.~\ref{fig1}b, and hence there is a logarithmic correction to the interfacial tension, for $L \rightarrow \infty$, $D\rightarrow \infty$,
\begin{equation}\label{eq9}
f_{\textrm{int}}(T,L,D)= F_{\textrm{int}}(T)/(k_BTL) = f_{\textrm{int}} (T) - \ln D/L, \quad d = 2\; ,
\end{equation}
\begin{equation}\label{eq10}
f_{\textrm{int}}(T,L,D)= F_{\textrm{int}}(T)/(k_BTL^2) = f_{\textrm{int}} (T) - \ln D/L^2 ,  \quad d = 3\;,
\end{equation}
where higher order terms (such as const/$L$ or const$/L^2$, respectively) have been ignored. However, for quantities such as the mean square interfacial widths $w^2$ of the interface the size effects are much more dramatic: Fig.~\ref{fig2} gives an example \cite{82}, where in the geometry of Fig.~\ref{fig1}a the order parameter profile $m(z)$ of an interface between coexisting A-rich and B-rich phases of a model polymer mixture was studied \cite{82}. Here $m(z)$ is defined by
\begin{equation}\label{eq11}
m(z)=[\rho_A(z)-\rho_B(z)]/[\rho_A(z)+ \rho_B(z)] \;,
\end{equation}
$\rho_A(z), \rho_B(z)$ being the average densities of A(B) monomers. It turned out that $m(z)$ can be described very well by a tanh profile
\begin{equation}\label{eq12}
m(z)=m_b \tanh [(z-D/2)/w],
\end{equation}
where $m_b$ is close to unity; although Eq.~\ref{eq12} is a standard mean field result for ``intrinsic'' interfacial profiles, where the "intrinsic width" ($w_0$) is a well-defined molecular length, Fig.~\ref{fig2} shows that the width $w$ found in the simulation has nothing to do with $w_0$: there is a dependence of $w$ on both L and D, and on the type of the statistical ensemble used. In the semi-grand-canonical ensemble, the z-coordinate of the interface position may fluctuate around its average value $z_0=D/2$, while in the canonical ensemble this fluctuation is suppressed. Taking (in the semi-grand-canonical ensemble) the limit $D \rightarrow \infty$ at fixed $L$, one hence expects that $w^2 \propto D^2$, the interface as a whole may essentially fluctuate freely, apart from the immediate vicinity of the confining boundaries at $z=0$ and $z=D$. If one takes the opposite limit, $L \rightarrow \infty$ at fixed $D$, a phenomenological theory \cite{81,82} rather yields $w^2\propto D$, while in the canonical ensemble $w^2$ at fixed $L$ reaches finite plateaus when $D\rightarrow \infty$, $w^2 \propto \ln L$. Both laws $w^2 \propto D, \; w^2 \propto \ln L$ can be traced back to capillary wave-type fluctuations \cite{35,36,37}. In fact, if one combines capillary wave theory and the ``intrinsic profile'' via a convolution approximation \cite{132} one obtains $(D\rightarrow \infty)$
\begin{equation}\label{eq13}
w^2=w_0^2 + \frac {k_BT \ln L}{4 f_{\textrm{int}}(T)} + \frac {k_BT \ln (q_{\textrm{max}}/2\pi)}{4 f_{\textrm{int}}(T)},
\end{equation}
where $q_{\textrm{max}}$ is a short wavelength cutoff for the capillary wave spectrum. Eq.~\ref{eq13} shows that from a plot of $w^2$ vs. $\ln L$ the interfacial free energy $f_{\textrm{int}}(T)$ can be estimated. However, it is not possible to also estimate $w_0$, since in general $q_{\textrm{max}}$ is not known (or perhaps not even precisely defined). But it has been shown for some models \cite{75,76,84,86,87} that the use of Eq.~\ref{eq13} yields estimates for the interfacial tension that are compatible with estimates from other methods.

\begin{figure}\centering
\includegraphics[width=0.6\linewidth]{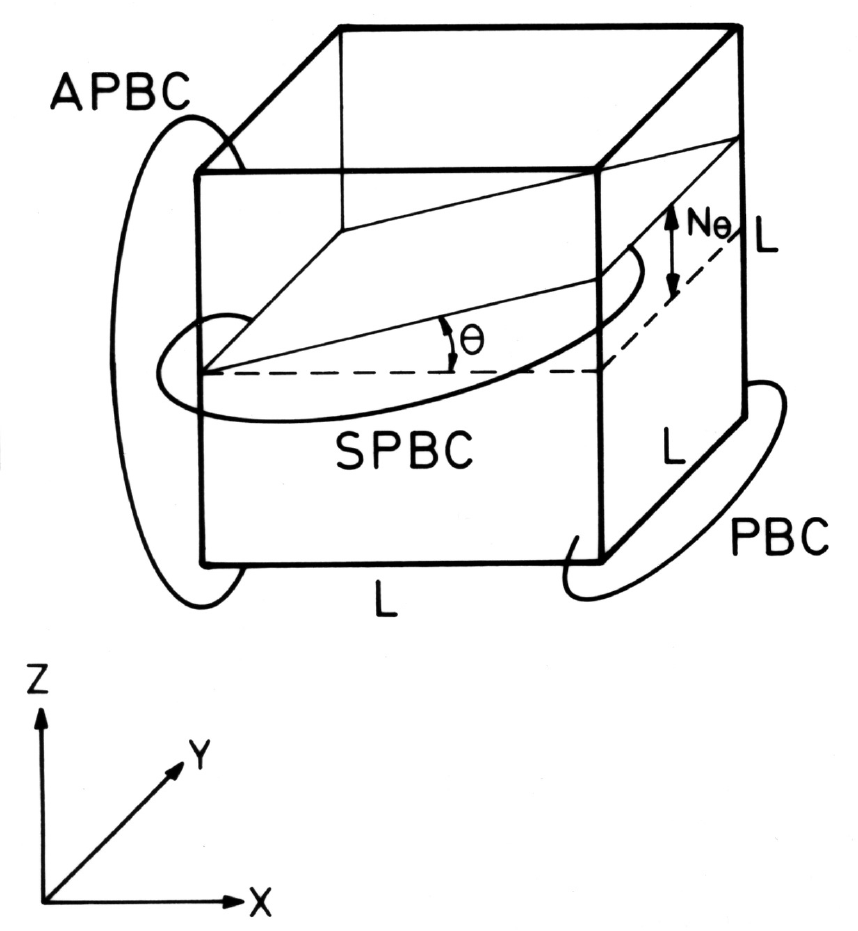}
\caption{\label{fig3} Boundary conditions used for the study of an $L \times L \times L$ Ising system, imposing a tilted interface. Antiperiodic boundary conditions (APBC) used in the z-direction, an ordinary periodic boundary condition (PBC) is used in the y-direction while a screw periodic boundary condition (SPBC) is used in the x-direction, involving a shift in the z-direction by $N_{\theta}$ lattice spacings.}
\end{figure}

\subsection{Tilted interfaces and the concept of interfacial stiffness}

Returning to the Ising model, we now discuss the fact that in lattice systems the interfacial free energy does not only depend on temperature, but also on the orientation of the interface relative to the lattice directions. Fig.~\ref{fig3} shows a generalization of the antiperiodic boundary condition (Fig.~\ref{fig1}b) to impose a tilted interface. This is done by combining the antiperiodic boundary condition (APBC) along the z-direction with a screw-periodic boundary condition (SPBC) in x-direction. I.e., the standard periodic boundary condition is not applied at planes $z=const$, but along the z-axis there occurs a shift by $N_\theta$ lattice units, causing a tilt angle $\theta$ defined via $\theta = arctan (N_\theta/L)$, while in y-direction still a standard periodic boundary condition is used. This combination of boundary conditions enforces (at temperatures where the Ising system is in the ordered phase) an interface in the system which is inclined (relative to the xy-plane) by an angle $\theta$, with the y-axis as the rotation axis.

\begin{figure}\centering
\includegraphics[width=0.6\linewidth]{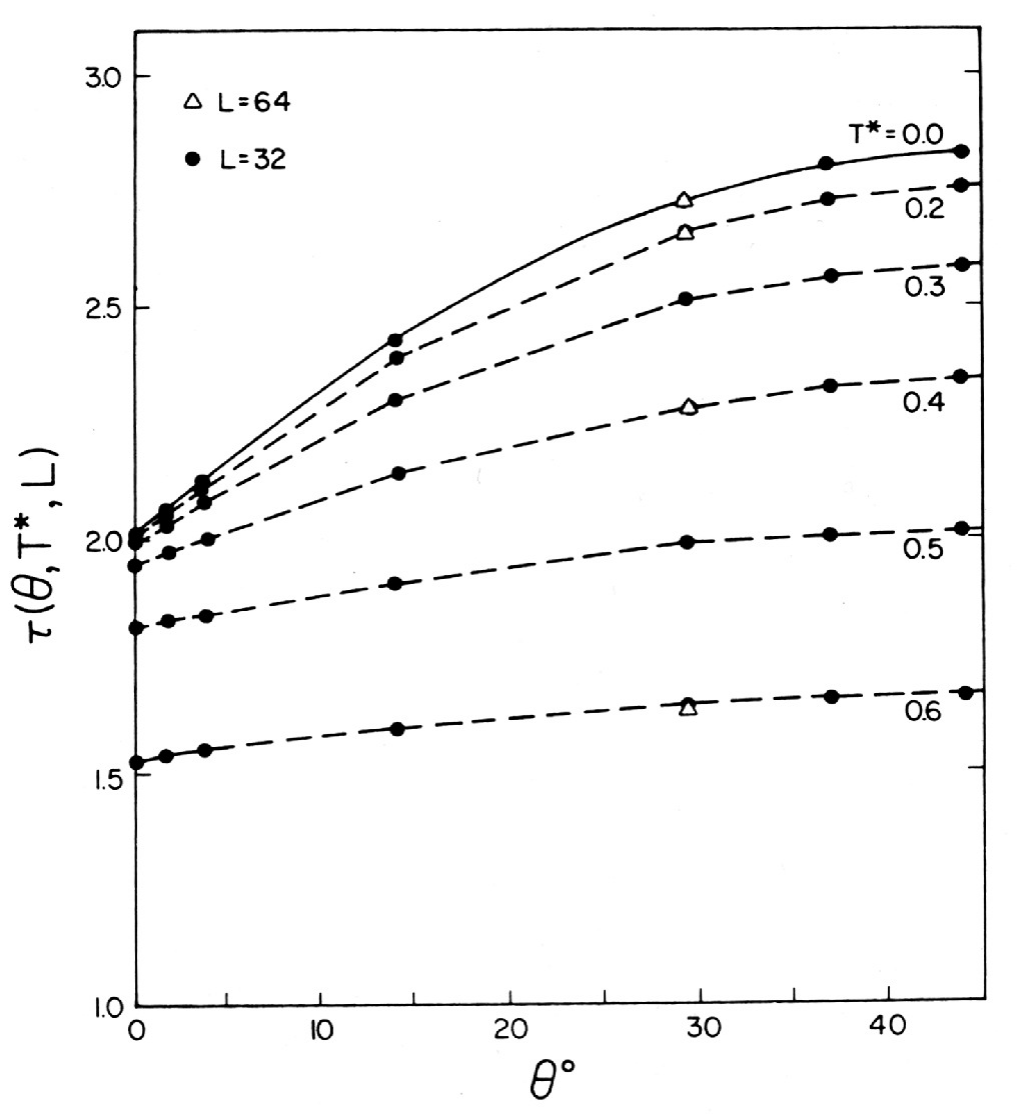}
\caption{\label{fig4} Anisotropic interfacial free energy $\tau (\theta,T^*,L)$ of the three-dimensional simple cubic Ising ferromagnet plotted versus the angle $\theta$, for several values of the reduced temperature $T^*=T/T_c$, as indicated; for $\theta = 30$° two choices ($L=64$ and $L=32$) of the linear dimension of the $L\times L\times L$ lattice are included, to show that at the chosen temperatures finite size effects in fact are negligible. Note that $\tau$ is measured in units of $J$. From Mon et al.~\cite{89}.}
\end{figure}

Of course, the most general case would involve a SPBC in the y-direction as well, involving both a shift $N_{\theta_x}$ for the SPBC in x-direction and a shift $N_{\theta_y}$ for the SPBC in y-direction. In fact, such a complete study of the angle-dependent interfacial free energy $F_{int}(T,\theta_x,\theta_y)$ is mandatory for an understanding of the equilibrium shape of a (macroscopic) domain of down-spins surrounded by a ``sea'' of up-spins, in terms of the Wulff \cite{133} construction. However, we are not aware of any attempts towards such a complete treatment, and also the work of Mon et al. \cite{89} only deals with the estimation of $\tau(\theta,T^*,L) =F_{int}(T,\theta_x=\theta, \theta_y=0)/L^2$ at low temperatures $(T^*=T/T_c$ with \cite{134} $J/k_BT_c=0.221655)$, $0\leq T^*\leq 0.6$ (Fig.~\ref{fig4}). However, for $T>T_R$ (with \cite{89} $T_R^* \approx 0.54 \pm 0.02$) the angular dependence of $\tau(\theta,T^*,L)$ is very small \cite{89}, while for $T\rightarrow 0$ it is quite appreciable (Fig.~\ref{fig4}).

We use here the notation $\tau(\theta,T^*,L)$ to alert the reader towards the fact that one needs to watch out carefully on finite size effects (although one does not detect any in Fig.~\ref{fig4}, finite size effects are appreciable near $T_R^*$ at small $\theta$). In fact, the behavior of $\tau(\theta,T^*,L)$ near $T_R^*$ is rather subtle, since, for $\theta \rightarrow 0$
\begin{equation}\label{eq14}
\tau (\theta,T^*,L \rightarrow \infty) = \tau (0,T^*) + \tau '(0,T^*) |\theta | +  \tau '' (0,T^*) \frac {\theta ^2}{2}, \quad T<T^*_R,
\end{equation}
\begin{equation}\label{eq15}
\tau (\theta,T^*,L \rightarrow \infty) = \tau (0,T^*) + \tau '' (0,T^*) (\theta^2/2), \;T>T_R^*\;.
\end{equation}
For $T<T_R^*$, the interfacial tension has a cusp-shaped singularity at $\theta \rightarrow 0$, while for $T >T_R^*$ it is analytic. The derivative $\tau '(0,T^*)$ has the physical meaning of a step free energy
\begin{equation}\label{eq16}
f_{\textrm{step}}(T) = (\partial \tau(\theta ,T^*,L \rightarrow \infty)/\partial \theta) _{T^*,\theta =0},
\end{equation}
and the roughening transition is characterized by the vanishing of the step free energy (just as the transition in the bulk is characterized by the vanishing of the standard interfacial tension between the coexisting phases which become indistinguishable at $T_c$ \cite{32}). The roughening transition is also characterized by a diverging correlation length $\xi_R(T)$ of ``height fluctuations'' (i.e., ``excursions'' of the local interface height $z(x,y)$ relative to its average $z=z_0$) \cite{131},
\begin{equation}\label{eq17}
\xi_R(T) \propto 1/f_{\textrm{step}}(T) \propto \exp [c/(T_R-T)^{1/2}],
\end{equation}
where $c$ is a constant.

\begin{figure}\centering
\includegraphics[width=0.6\linewidth]{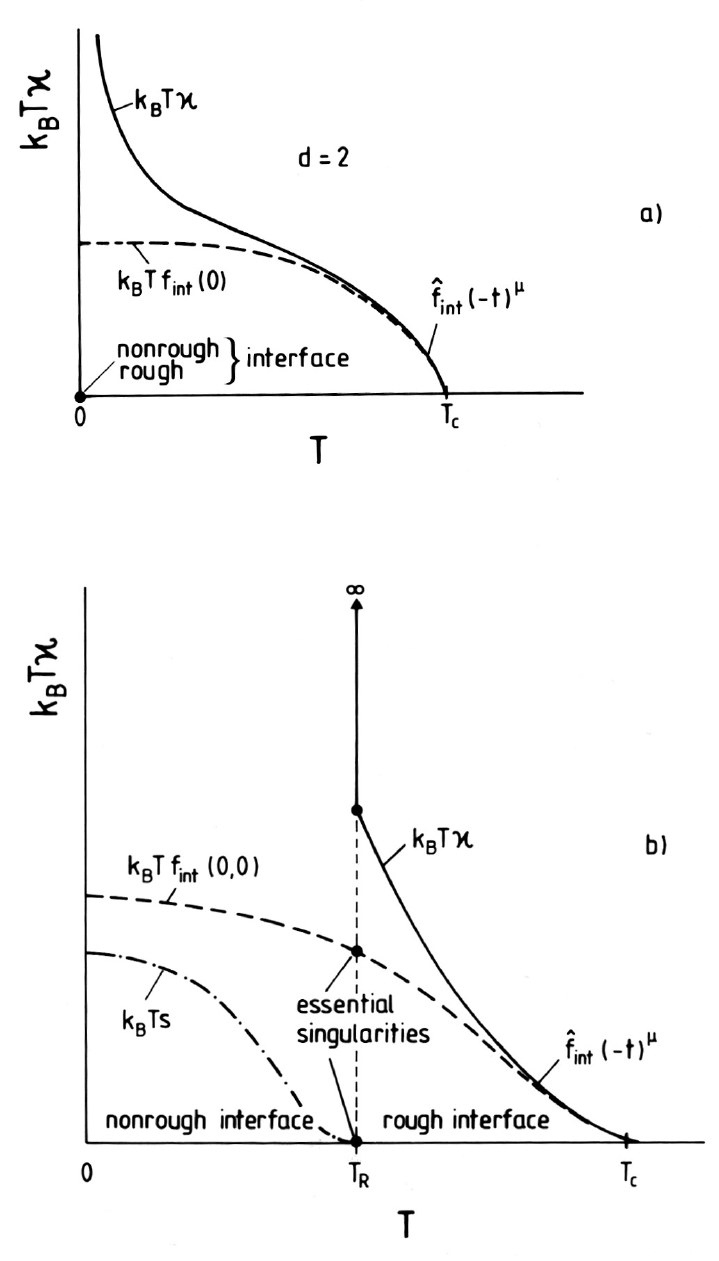}
\caption{\label{fig5} Schematic temperature variation of the interfacial stiffness $k_BT\kappa $ and interfacial free energy, for an interface oriented perpendicularly to a lattice direction of a square (a) and simple cubic (b) lattice, respectively. While for $d=2$ the interface is rough for all nonzero temperatures, in $d=3$ it is rough only for temperatures $T$ exceeding the roughening transition temperature $T_R$. For $T < T_R$ there exists a nonzero free energy of surface steps (denoted as $k_BTs$ in this figure) which vanishes at $T_R$ with an essential singularity. While $\kappa$ is infinite throughout the non-rough phase $\kappa$ reaches a universal value as $T \rightarrow T^*_R$. Note that $\kappa$ and $f_{\textrm{int}}$ to leading order become identical for $T \rightarrow T_c^-$, both in $d=2$ and in $d=3$. Note that $T_R=0$ in $d=2$, and here $k_BTf_{\textrm{int}}(T,\theta =0)/L^{d-1}$ throughout.}
\end{figure}

In a finite system the singular behavior for $T \rightarrow T_R$ is rounded off as soon as $\xi_R(T)$ becomes of the order of $L$. We now define for $T >T_R$ the interfacial stiffness as \cite{131,132,135}
\begin{equation}\label{eq177}
\kappa (T^*) = [\tau (0,T)+ \tau^{''}(0,T)]/k_BT.
\end{equation}
In systems with anisotropic but rough interfaces it is $\kappa (T)$ rather than $f_{\textrm{int}}/k_BT$ that describes the broadening of interfacial profiles by capillary waves \{Eq.~\ref{eq13}\}. Fig.~\ref{fig5} compares the temperature dependence of $f_{\textrm{int}}(T)$ and of $\kappa (T)$ for both d=2 and d=3 dimensions. It has already been mentioned that (one-dimensional) interfaces are rough at all nonzero temperatures, although at $T=0$ the interface (along a lattice direction of the square lattice) clearly is non-rough. Thus $T_R=0$ can be considered as the roughening transition temperature, and in fact $k_BT \kappa (T \rightarrow 0)$ diverges (Fig.~\ref{fig5}a), while $k_BT f_{\textrm{int}}(\theta =0, T \rightarrow 0=2)$, as expected from Eq.~\ref{eq8}. For interfaces in standard fluid systems, of course, the isotropy of space dictates that the interface free energy is completely independent of interface orientation, and no distinction between the interfacial stiffness (which enters as a factor setting the scale for the capillary wave Hamiltonian) and the interfacial free energy exists.

\subsection{Surface excess free energies due to external walls}

The thermodynamic integration approach of Sec.~2.1 can be straightforwardly carried over to compute surface excess free energies due to external boundaries of a system, such as the walls of a container holding a fluid, or the free surface of an Ising ferromagnet. The Ising case actually is the most instructive and simple example, and hence shall be considered first.

Keeping in mind the interpretation of the Ising model as a lattice gas system $(S_i=-1$ and $S_i = +1$ representing the cases of empty and occupied lattice sites i, respectively), it is natural to consider thin Ising films with free surfaces for which in the first layer ($n=1$) a surface field $H_1$ and in the last layer ($n=D$) a surface field $H_D$ act. These surface fields translate in the lattice gas interpretation, into short-range potentials attracting the particles of the fluid to the walls (which represent the missing layer at $n=0$ and $n=D+1$, respectively).

The generic Hamiltonian that one uses hence is \cite{92,93,94,95,96,97,98,99}
\begin{eqnarray}\label{eq18}
\mathcal{H}= -J \sum \limits _{\langle i,j\rangle} S_iS_j -J_s \sum \limits_{\langle i,j\rangle\; \textrm{both from the}  \atop_ \textrm{surfaces} n = 1 \; \textrm{or} \; n=D} S_iS_j - H\sum \limits_i S_i \nonumber \\ - H_1 \sum \limits _{i \in n=1} S_i - H_D \sum \limits _{i \in n=D} S_j\quad , \quad \quad S_i = \pm 1.
\end{eqnarray}
Here we consider the nearest neighbor Ising ferromagnet on the simple cubic lattice, choosing a $L \times L \times D$ geometry with two free surface layers of size $L \times L$ at layers $n=1$ and $n=D$ (taking again the lattice spacing as our unit of length). Periodic boundary conditions are applied only in x and y directions, while there occur no interactions with spins in layers $n=0$ and $n=D+1$. It then is useful to allow in the layers $n=1$ and $n=D$ for exchange constants $J_s$ different from the bulk exchange J. It is well known that the model with $J_s=J$ (and $T >T_R$ \cite{92}) exhibits at some critical field $H_{1c}(T)$ a critical wetting transition \cite{136,137}. However, choosing $J_s>J$ one finds a regime \{above a line $J_{sc}(T)/J$\} where the order of the wetting transition is 1st order rather than 2nd order \cite{136,137}. Since in real systems 1st order wetting transitions are rather common \cite{15,17,18,19,24} and critical wetting is the rare exception \cite{19}, it is of interest to have a model at one's disposal where by variation of a parameter $(J_s/J)$ one can control the order of the wetting transition. All these wetting transitions occur for bulk field $H=0$, of course.

In the limit of very thick films $(D \rightarrow \infty)$, where correlations of spins near one wall with spins near the other wall can be neglected, the free energy per spin $f(T,H,H_1,H_D,D)$ of the model can then be decomposed as \cite{95,96,97,98}
\begin{eqnarray}\label{eq19}
f(T,H,H_1,H_D,D)= f_b(T,H)+ \frac 1 D f_s(T,H,H_1)+\frac 1 D f_s(T,H,H_D).
\end{eqnarray}
Here, $f_b(T,H)$ is the free energy per spin of a bulk Ising system, which depends on neither $H_1$ nor $H_D$, of course. The surface excess free energy of the left wall (where $H_1$ acts) is $f_s(T,H,H_1)$, while  $f_s(T,H,H_D)$ refers to the surface excess free energy of the right wall, where $H_D$ acts. As is well known, wetting transitions show up as singularities of the respective surface excess free energies \cite{15,17,22}.

We assume now that $H_1 <0$ and consider the limit $H \rightarrow 0^+$, so we have (at temperatures below the bulk critical temperature $T_{cb}$) a positive spontaneous magnetization $m_{\textrm{coex}}(T) >0$,
\begin{equation}\label{eq20}
m_b(T,H)=-(\partial f_b(T,H)/\partial H)_T\;, m_{\textrm{coex}}(T)=m_b(T,H \rightarrow 0^+).
\end{equation}
We denote the excess free energy of the left wall that belongs to states with positive magnetization in the bulk as $f_s^{(+)}(T,H,H_1)$. In the regime of incomplete wetting, $f_s^{(+)}(T,H,H_1)$ then is the excess free energy of a surface where the region of positive magnetization in the film extends even close to the left wall (although the field $H_1$ would energetically favor a negative magnetization). In the wet phase, however, we have a (macroscopically thick) domain of the negative magnetization adjacent to the left wall, separated by an interface (as studied in Sec.~2.1) from the domain with positive magnetization (which is the majority phase in the considered limit $H \rightarrow 0^+$). Consequently, the surface excess free energy of a wet surface is
\begin{equation}\label{eq21}
f_s^{\textrm{wet}}(T,0,H_1)=f_s^{(-)}(T,0,H_1) + f_{\textrm{int}}(T).
\end{equation}
Here $f_s^{(-)} (T,0,H_1)$ is the excess free energy of a surface where both the bulk magnetization and the surface field $H_1$ are negative, reached for the limit $H \rightarrow 0^-$, of course. The wetting transition of this surface occurs when the free energies of the wet and the non-wet surfaces are equal,
\begin{equation}\label{eq22}
f_s^{(+)} (T,0,H_1)=f_s^{(-)} (T,0,H_1)+f_{\textrm{int}}(T)\quad.
\end{equation}
Note that Eq.~\ref{eq21} explicitly incorporates the fact that the interface separating the negative domain (near the wall) from the positive domain is located at such a large distance from the wall, that the free energy contributions of the wall \{$f_s^{(-)}(T,0,H_1)$\} and the interface \{$f_{\textrm{int}}(T)$\} can be simply added, without considering any interaction forces between the interface and the wall. When we consider the Ising-lattice gas analogy, we readily see that Eq.~\ref{eq22} just is the standard relation between wall-liquid $\gamma_{w \ell}(T)$ and wall-vapor $\gamma_{wv}(T)$ free energies at liquid-vapor phase coexistence for a wet wall,
\begin{equation}\label{eq23}
\gamma_{wv}(T)= \gamma_{w\ell}(T) + \gamma (T),
\end{equation}
where the vapor-liquid interfacial tension is now denoted as $\gamma (T)$ to make contact with the notation of Sec.~1.

For the case of incomplete wetting, the generalization of Eqs.~\ref{eq22}, \ref{eq23} simply is the well-known Young \cite{3} equation for the contact angle $\theta$,
\begin{equation}\label{eq24}
\gamma_{wv}(T)-\gamma_{w \ell}= \gamma (T) \cos \theta \quad ,
\end{equation}
or
\begin{equation}\label{eq25}
f_s^{(+)}(T,0,H_1)-f_s^{(-)} (T,0,H_1)=f_{\textrm{int}}(T) \cos \theta ,
\end{equation}
respectively. The wetting transition is approached from the region of incomplete wetting when $\theta \rightarrow 0$ \cite{12,13,14,15,16,17,18,19,29,21,22,23,24}.

For Ising ferromagnets, spin reversal symmetry implies a symmetry relation for the surface excess free energies,
\begin{equation}\label{eq26}
f_s^{(-)} (T,0,H_1)=f_s^{(+)}(T,0,H_1).
\end{equation}
An interesting special case is found for $H_1=0$, namely
\begin{equation}\label{eq27}
f_s^{(-)} (T,0,0) - f_s^{(+)}(T,0,0)=0.
\end{equation}

Combining this result with Eq.~\ref{eq25} shows that $H_1=0$ corresponds to $\cos \theta=0$, i.e. $\theta = \pi/2$. As a result, we conclude that for the estimation of the contact angle, for which only the difference $f_s^{(+)}(T,0,H_1)-f_s^{(-)} (T,0,H_1)$ matters, \{Eq.~\ref{eq25}, \ref{eq27}\} can serve as a convenient initial state for a thermodynamic integration.

In practice one can slightly simplify the problem even further, treating a thin film with antisymmetric surface fields, $H_D=-H_1$, to conclude
\begin{equation}\label{eq28}
f_s^{(-)}(T,0,-H_1)=f_s^{(+)}(T,0,H_D) ,
\end{equation}
since the surfaces at $n=1$ and $n=D$ can be treated on an equal footing. Now one makes use of the relations
\begin{equation}\label{eq29}
m_1=(\partial f_s(T,H,H_1)/\partial H_1)_T , \quad m_D=-(\partial f_s(T,H,H_D)/\partial H_D)_T,
\end{equation}
for the desired thermodynamic integrations, since both layer magnetization $m_1$ and $m_D$ can be straightforwardly sampled. Thus we conclude, using Eqs.~\ref{eq26}-\ref{eq29},
\begin{equation}\label{eq30}
f_s^{(+)}(T,0,H_1)-f_s^{(-)} (T,0,H_1)= \int \limits _0^{m_1} [m_D(T,0,H_1')-m_1(T,0,H_1')]d H_1'.
\end{equation}

\begin{figure}\centering
\includegraphics[width=0.6\linewidth]{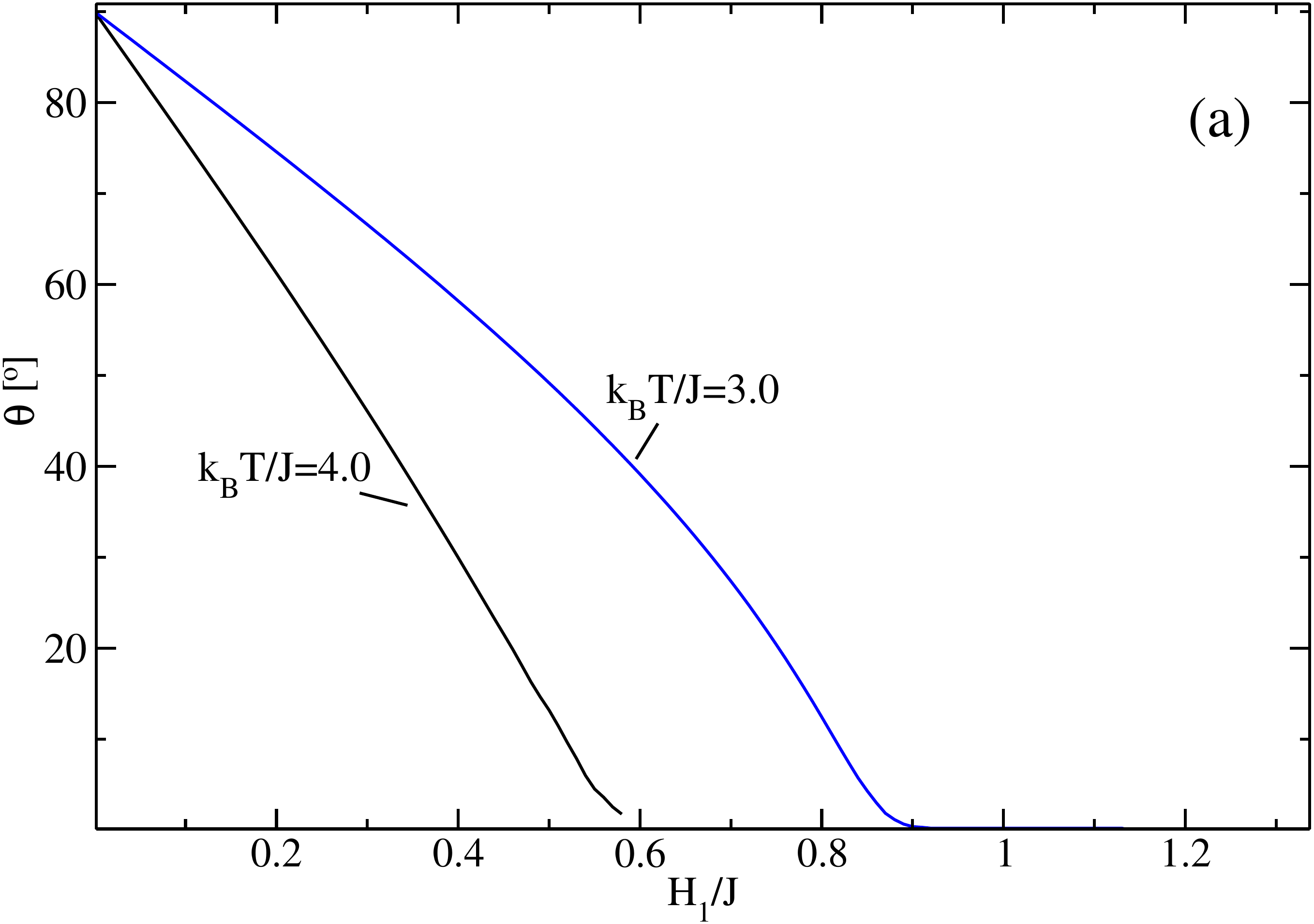}\vspace{0.5cm}

\includegraphics[width=0.6\linewidth]{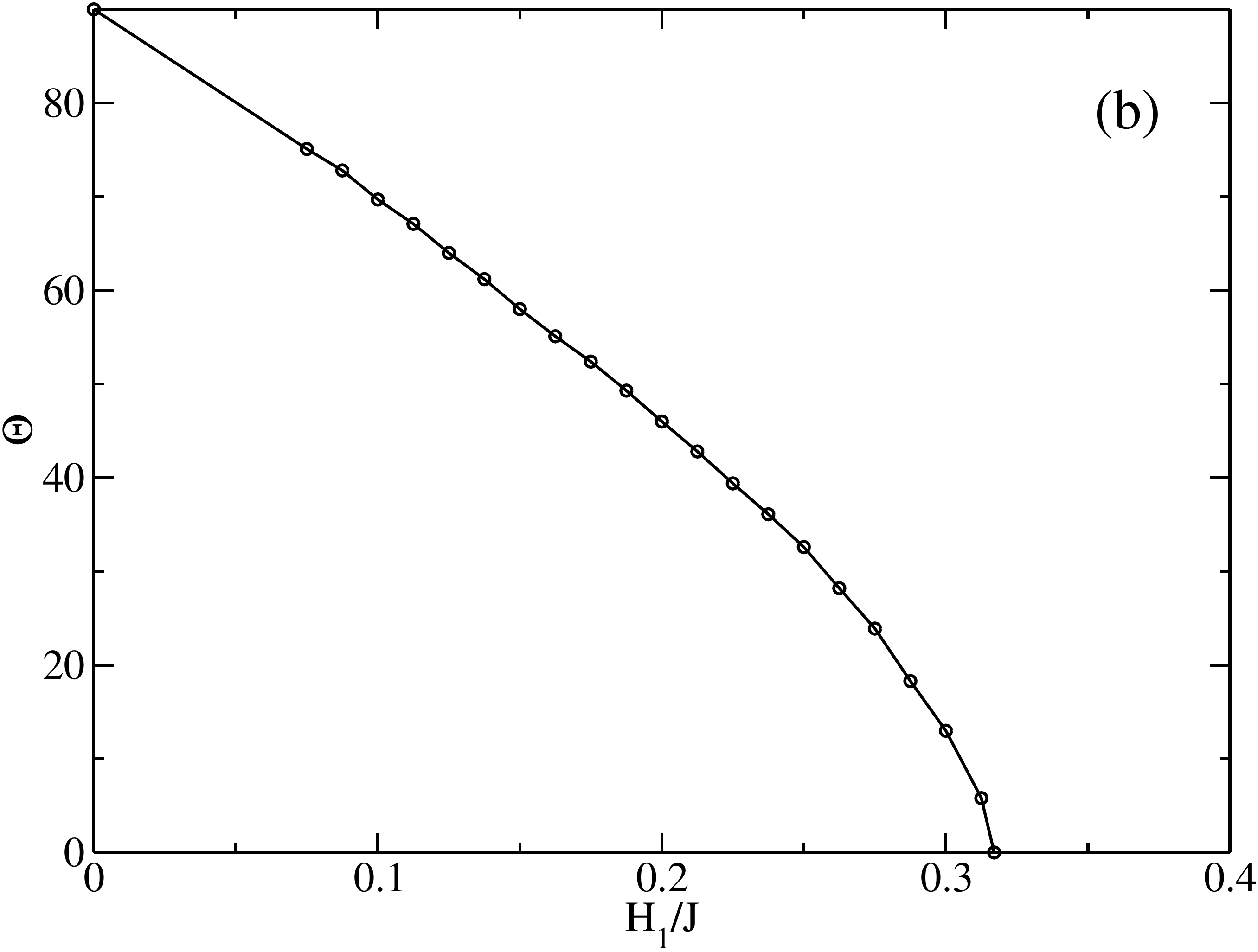}
\caption{\label{fig6} Contact angle $\theta$ plotted vs.~$H_1/J$, for the case $J_s/J=1$ (a) and $J_s/J=1.4$ (b). In case (a), both temperatures $k_BT/ J=3.0$ (Using a $L \times L \times D$ system with $L=D=60)$ and $k_BT/J=4.0 (L=D=100)$ are included, while case (b) used $L=D=64$ and $k_BT/J=4.0$. Note that at critical wetting strong fluctuations and finite size effects render the data very close to $H_{1c}/J$ unreliable.}
\end{figure}

One performs for Eq.~\ref{eq30} a calculation where one studies a thin film with a positive magnetization in the bulk and varies the surface fields $H_1',H_D'=-H_1'$ from $H_1'=0$ to $H_1'=H_1$ in small steps. Fig.~\ref{fig6} shows, as an example, the variation of the contact angle $\theta$ as a function of $H_1/J$ for $k_BT/J=4.0$ and $J_s/J=1.0$ (a) and $J_s/J=1.4$ (b). From previous work applying other methods \cite{136,137} it is known that for $J_s/J=1.0$ critical wetting at this temperature occurs for $H_{1c}/J\approx 0.55 \pm 0.015$ (while for $k_BT/J=3.0$ critical wetting occurs for $H_{1c}/J\approx 0.84 \pm 0.02$) and for $J_s/J=1.4$ a first order wetting transition occurs for $H_{1c}/J\approx  0.3$ (due to hysteresis \cite{136}, only a very rough estimation was possible). The thermodynamic integration method, using small steps $\Delta H_1/J=0.0125$ \cite{93}, allows for a much better accuracy, yielded $H_{1c}/J = 0.318 \pm 0.005$. Near critical wetting, one predicts that \cite{15,19,22,24}
\begin{equation}\label{eq31}
\Delta f_s \equiv [f_s^{(-)} (T,0,H_1)-f_s^{(+)}(T,0,H_1)]/f_{\textrm{int}}(+) - 1 \propto |H_{1c}(T)-H_1|^{2 \nu''} \;,
\end{equation}
where $\nu''=1$ according to mean field theory, while for the Ising model $\nu''\approx 4$ is expected. Since in this region $\cos \theta \approx 1-\theta^2/2$, one concludes further
\begin{equation}\label{32}
\theta \propto |H_{1c}(T)-H_1|^{\nu''} , H_1 \rightarrow H_{1c}(T)
\end{equation}
as one approaches the 2nd order wetting transition from the incomplete wetting regime. From the experience with other methods to study critical wetting with Monte Carlo methods \cite{136,137} one hence expects that not very close to $H_{1c}(T)$ the variation of $\theta$ with $H_1$ is linear, while strong curvature sets in for $H_1$ very close to $H_{1c}(T)$. The data of Fig.~\ref{fig6}a are compatible with such an interpretation, but do not allow a precise characterization of the singular behavior of $\theta$ near $H_{1c}(T)$ due to the limited accuracy of the data used in Fig.~\ref{fig6}a \cite{138}. In the case of of first order wetting, we have instead of Eq.~\ref{eq31}
\begin{equation}\label{eq33}
\Delta f_s \propto H_{1c}(T)-H_1, \quad \theta \propto \sqrt{|H_{1c}(T)-H_1|}
\end{equation}
and this relation indeed is borne out nicely by the data. Note that $f_{\textrm{int}}(T)$ in Eqs.~\ref{eq25}, \ref{eq31} was taken from the work of Hasenbusch and Pinn \cite{90,139}, see also \cite{93,94}.

\section{Estimation of Contact Angles for Off-Lattice Models of Fluids}

For off-lattice models of binary (A,B) liquid mixtures, it is also possible to choose interactions such that the symmetry against interchange of A and B is strictly preserved. This symmetry is equivalent to the spin reversal symmetry of an Ising model. The method for obtaining the contact angle of Ising models, that was described in the preceding subsection, can then easily be carried over to such symmetric models in continuous space. For simplicity, we here summarize the approach of a recent ``case study'' \cite{100,101} that did address a specific model, the binary symmetric Lennard-Jones mixture confined in a thin film with ``antisymmetric'' walls, at a temperature far below the critical temperature of unmixing for this model.

In this model, one considers $N$ point particles at positions  $\vec{r}_i, i=1, \ldots N,$ in a box of linear dimensions $L \times L \times D$, with periodic boundary conditions in x- and y-directions, and with impenetrable walls of area $L\times L$ each located at $z=0$ and $z=D$. The particles interact with pairwise potentials $u_{\alpha \beta} (r_{ij})$, where $\alpha \beta$ refers to the type of pair (AA, AB, or BB, respectively), and $r_{ij}$ is the absolute value of the distance between the particles, $r_{ij}=|\vec{r}_i-\vec{r}_j|$. Starting from the full Lennard-Jones (LJ) potential $\phi_{LJ}^{\alpha \beta}(r) = 4 \epsilon_{\alpha \beta}[(\sigma_{\alpha \beta}/r)^{12} -(\sigma _{\alpha \beta}/r)^6]$, $u_{\alpha \beta} (r_{ij})$ is chosen identical to previous work addressing the bulk phase diagram \cite{140},
\begin{equation}\label{eq34}
u_{\alpha \beta} (r_{ij} \leq r_c)=\phi_{LJ}^{\alpha \beta}(r_{ij})-\phi _{LJ}^{\alpha \beta}(r_c)-(r_{ij}-r_c) \frac {d\phi_{LJ}^{\alpha \beta}}{dr_{ij}} \bigg |_{r_{ij}=r_c} ,
\end{equation}
while $u(r_{ij} \geq r_c)=0$. This choice, Eq.~\ref{eq34}, ensures that both the potential and the force are continuous at the cutoff-distance $r_{ij}=r_c$. The potential parameters are chosen such that the mixture is fully symmetric,
\begin{equation}\label{eq35}
\sigma_{AA} = \sigma _{BB}=\sigma_{AB}=\sigma , \quad \epsilon_{AA} = \epsilon_{BB}= 2 \epsilon_{AB}=\epsilon \quad ,
\end{equation}
and we chose $r_c=2.5 \sigma$ taking units in this section such that $\sigma = 1, \; \epsilon =1, \; k_B =1$. Working at a reduced density $\rho^*= \rho \sigma ^3=N\sigma ^3/V=1$, the critical temperature of unmixing is \cite{140} $T_c=1.4230 \pm 0.0005$. Considering then a temperature $T=1$, far below $T_c$ neither the vapor-liquid transition nor the liquid-solid transition of the model create problems, the system is a dense (almost incompressible) fluid. At this temperature, phase separation into coexisting A-rich and B-rich phases is essentially complete (the concentration $x_A=N_A/N$ with $N=N_A + N_B$ takes a value $x_{A(2)}^{\textrm{coex}} =0.97$, and $x_{A(1)}^{\textrm{coex}}=1-x_{A(2)}^{\textrm{coex}} = 0.03$ by symmetry.) Phase coexistence occurs at chemical potential difference $\Delta \mu = 0$ between the particles.

The ``antisymmetric'' wall potentials are defined as follows: on the A-particles acts a purely repulsive wall at both walls, and in addition an attractive potential (of strength $\epsilon_a$) only at the left wall $(z=0)$. On the B-particles, the same repulsive potentials acts on both walls, and in addition there is an attractive potential (of strength $\epsilon_a$) only at the right wall (z=D). Thus
\begin{equation}\label{eq36}
u_A(z)= \frac {2 \pi \rho^*} {3} \bigg \{\epsilon _r [(\frac {\sigma}{z+\delta})^9 +(\frac{\sigma}{D + \delta -z})^9 ] - \epsilon_a(\frac {\sigma}{z+\delta})^3 \bigg\},
\end{equation}
\begin{equation}\label{eq37}
u_B(z)= \frac {2 \pi \rho^*} {3} \bigg \{\epsilon _r [(\frac {\sigma}{z+\delta})^9 +(\frac{\sigma}{D + \delta -z})^9 ] - \epsilon_a(\frac {\sigma}{D + \delta - z})^3 \bigg\},
\end{equation}
where an offset $\delta = \sigma/2$ is used, and $\epsilon_r = \epsilon/15$.

One then simulates the thin film in the semi-grand-canonical ensemble under phase coexistence conditions $(\Delta \mu=0)$ and varies $\epsilon_a$. The difference in wall free energies $\gamma_{wA}-\gamma_{wB}$ in the regime of incomplete wetting (considering, with no loss of generality, the A-rich phase in the bulk) is then found by a thermodynamic integration method, inspired by the treatment of Sec. 2.3. We write the free energy of the thin film in terms of the partition function $Z$ as follows $(\beta = 1/k_BT)$
\begin{equation}\label{eq38}
F=-k_BT \ln Z = -k_BT\ln \int d\vec{X} \exp \bigg \{- \beta \mathcal{H}_b(\vec{X}) - \beta \mathcal{H}^r_w(\vec{X}) - \beta \epsilon_a \mathcal {H} _w^r (\vec{X})\bigg \},
\end{equation}
where $\vec{X}$ stands for the coordinates of all the particles, $\mathcal{H}_b(\vec{X})$ describes all the particle interactions, $\vec{H}_w^r(\vec{X})$ the energy of all the particles due to the repulsive part of Eqs.~\ref{eq36}, \ref{eq37}, and $\epsilon_a\mathcal{H}_w^a(\vec{X})$ the energy of all the particles due to the attractive part of Eqs.~\ref{eq36}, \ref{eq37}. Prefactors in $Z$ that are unimportant in the present context have been not spelled out. Note that $\mathcal{H}_w^a(\vec{X})$ can be written as
\begin{equation}\label{eq39}
\mathcal{H}^a (\vec{X})=L^2(\frac {2 \pi \rho^*}{3}) [\int \limits _0^D \rho _A(z,\vec{X}) (\frac {\sigma}{z+\delta})^3 dz + \int \limits _0^D \rho _B(z,\vec{X}) (\frac {\sigma}{D+\delta -z})^3 dz]
\end{equation}
Here $\rho_A(z,\vec{X})$ is the density of A-particles in configuration $\vec{X}$ in the (infinitesimally thin) slice $L^2 dz$ from $z$ to $z+dz$, and $\rho_B(z,\vec{X})$ is defined similarly.
Note that due to the repulsive wall potentials, $\rho_A(z,\vec{X})$ and $\rho_B(z,\vec{X})$ are essentially zero outside the interval $0 \leq z \leq D$, over which the integrations in Eq.~\ref{eq38} are performed.

Since for $D\rightarrow \infty$, the free energy $F$ can be decomposed as
\begin{equation}\label{eq41}
(\partial f_s^{\textrm{left}}/\partial {\epsilon_a}) _{T,\Delta \mu=0} = \frac {2 \pi\rho^*}{3} \int \limits ^D_0 dz (\frac{\sigma}{z + \delta})^3 \langle \rho_A(z)\rangle_{\epsilon_a},
\end{equation}
and
\begin{equation}\label{eq42}
(\partial f_s^{\textrm{right}}/\partial {\epsilon_a}) _{T,\Delta \mu=0} = \frac {2 \pi\rho^*}{3} \int \limits ^D_0 dz (\frac{\sigma}{D+ \delta - z})^3 \langle \rho_B(z)\rangle_{\epsilon_a},
\end{equation}
Here the notation $\langle \ldots\rangle_{\epsilon_a}$ emphasizes that the averages are sampled at a nonzero value of $\epsilon_a$ in Eqs.~\ref{eq36}, \ref{eq37}. Eqs.~\ref{eq41}, \ref{eq42} are the generalizations of Eq.~\ref{eq29} for the Ising (lattice gas) model - if we chose in the Ising model a ``surface field'' $H_1f(n)$ which does not only act on the spins in the first layer $n=1$, but on all layers, we would have to replace $m_1$ in Eq.~\ref{eq29} by $\sum  \limits _n m_nf(n)$, of course.

Thermodynamic integration of Eqs.~\ref{eq41}, \ref{eq42} with respect to $\epsilon_a$ then yields
\begin{equation}\label{eq43}
f_s^{\textrm{left}} (\epsilon_a)= f_s^{\textrm{left}} (\epsilon_a=0)+ \frac {2 \pi \rho^{*} }{3} \int \limits _0^{\epsilon_a} d \epsilon_a' \int \limits _0^Ddz (\frac {\sigma}{z+\delta})^3 \langle \rho_A(z)\rangle _{\epsilon_a'} ,
\end{equation}
\begin{equation}\label{eq44}
f_s^{\textrm{right}} (\epsilon_a)= f_s^{\textrm{right}} (\epsilon_a=0)+ \frac {2 \pi \rho^{*}}{3} \int \limits _0^{\epsilon_a} d \epsilon_a' \int \limits _0^Ddz (\frac {\sigma}{\delta + \rho -z})^3 \langle \rho_B(z)\rangle _{\epsilon_a'}.
\end{equation}

Remember that so far we were considering solely wall free energies of the A-rich phase. But due to the symmetry of our model the wall free energies of A-rich and B-rich phases are related; similar to Eq.~\ref{eq26} we have
\begin{equation}\label{eq45}
f_s^{\textrm{left}}(\epsilon_a) \bigg |_{B-rich} = f_x^{\textrm{right}}(\epsilon_a) \bigg |_{A-rich}.
\end{equation}
Thus the desired difference can be written

\begin{eqnarray}\label{eq46}
&&\gamma_{wA}-\gamma_{wB} = f_s^{\textrm{left}} (\epsilon_a)|_{\textrm{A-rich}} - f_s^{\textrm{left}} (\epsilon_a) |_{\textrm{B-rich}} = \nonumber \\
&&f_s^{\textrm{left}}(\epsilon_a)|_{\textrm{A-rich}} - f_s^{\textrm{right}}(\epsilon_a)|_{A-rich} = \nonumber \\
&&= (\frac{2\pi\rho ^*}{3}) \int _0^{\epsilon_a} d \epsilon _a' \int \limits _0^D dz [\langle \rho _A(z)\rangle
_{\epsilon_a'} \nonumber \\
&&(\frac{\sigma}{z+\delta})^3 - \langle \rho_B(z)\rangle _{\epsilon_a'} (\frac {\sigma}{D+\delta -z})^3]\;.
\end{eqnarray}
Both statistical averages $\langle\rho_A(z)\rangle _{\epsilon_a'}$, $\langle \rho_B(z)\rangle _{\epsilon_a'}$ are sampled in the same (A-rich) phase only. Of course, $\gamma_{wA}-\gamma_{wB}=0$ for $\epsilon_a=0$, implying once more a contact angle of 90$^\circ$ for this ``neutral wall'' situation. Subtle conceptual problems \cite{45} concerning the proper identification of individual wall free energies $\gamma_{wA}, \gamma_{wB} $ due to the freedom to define exactly where in our off-lattice model the walls are located $(z=0$ and $z=D$ or $z=-\delta$ and $z=D+\delta $) will however leave this difference unaffected.

\begin{figure}\centering
\includegraphics[width=0.6\linewidth]{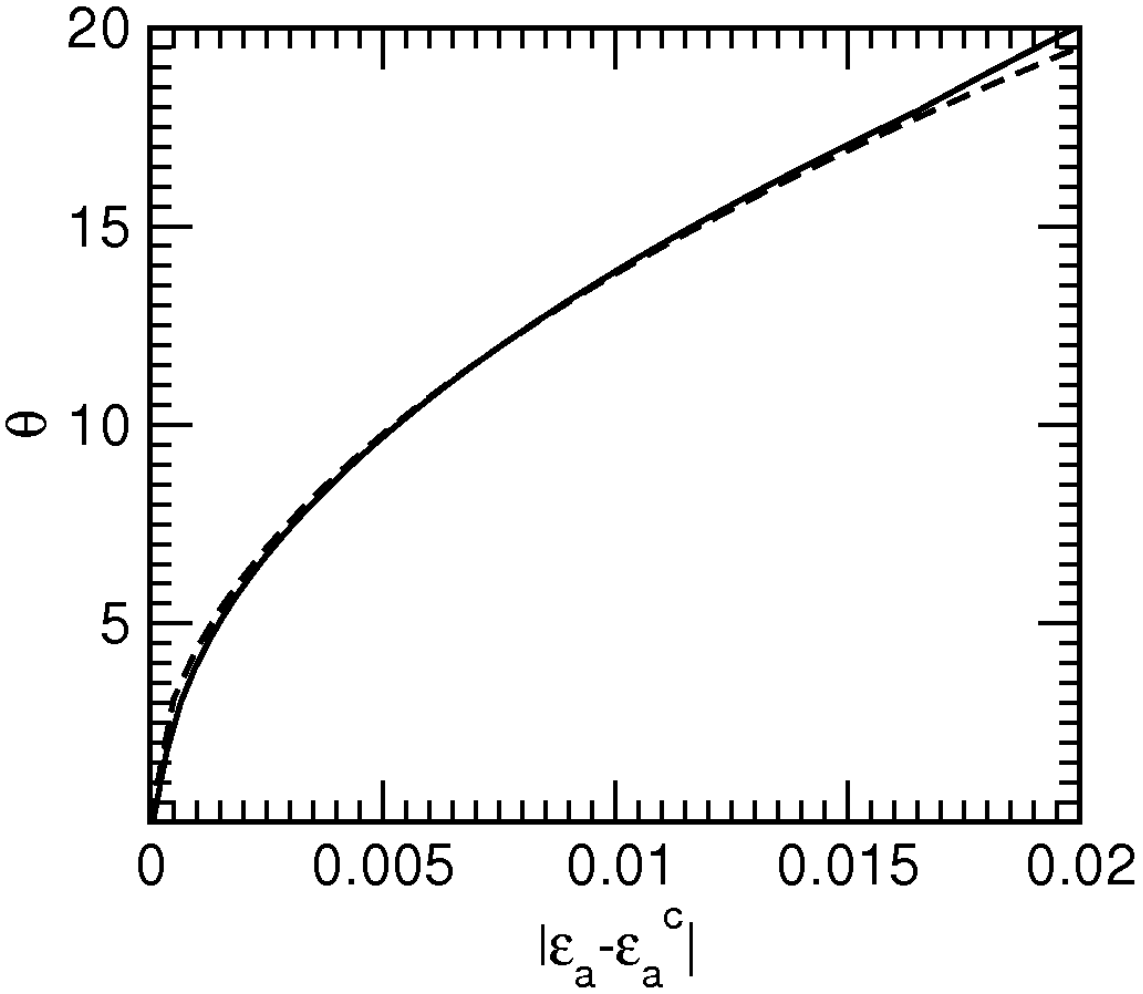}
\caption{\label{fig7} Contact angle $\theta$ of a symmetrical binary (A,B) Lennard-Jones mixture \{Eqs.~\ref{eq34}, \ref{eq35}\}, at $T=1.0$, exposed to wall potentials Eqs.~\ref{eq36}, \ref{eq37} plotted vs. the strength $\epsilon_a$ of the attractive wall potential. A fist-order wetting transition is predicted to occur at $\epsilon_{a}^{c} \approx 0.24$. The data are compatible with a relation $\theta \propto \sqrt{\epsilon_{a}^{c}-\epsilon_a},$ analogous to Eq.~\ref{eq33}. (broken curve)}
\end{figure}

If the interfacial tension $\gamma_{AB}$ between the coexisting A-rich and B-rich phases is known independently (see Sec.~4), Young's equation $\cos \theta=[\gamma_{wA}-\gamma_{wB}]/\gamma(T)$ can be used to predict the dependence of the contact angle $\theta$ on $\epsilon _a$ (Fig.~7).
In this model it is also possible to extract estimates of $\theta$ directly from the observation of slab configurations \cite{100,101} equilibrating thin films (for the same choice of bulk and wall potentials as described above) in the canonical ensemble, using $x_A=x_B=0.5$, and starting with a slab configuration with perpendicular walls at $x_1=L/4$ and $x_2= 3L/4 $ as an initial condition (the B-rich phase is then located in between $x_1 $ and $x_2$). Although the necessary equilibration of the interfaces is a time-consuming procedure, accurate estimates for $\theta$ at various values of $\epsilon_a<\epsilon_{a,c}$ can be obtained \cite{100,101}, however choices where $\theta $ is small cannot be studied (due to the need to make both $L$ and $D$ rather large). Somewhat surprisingly, it was found that the ``macroscopic'' contact angle (as obtained from the method as described above, which avoids the study of phase coexistence in the thin film geometry, focusing on estimating suitable observables exclusively in the single-phase region of the system) agrees with the contact angles found from direct observations of phase coexistence on the nanoscale very well \cite{100,101}.

Unfortunately, an extension of the above method to systems lacking this particular symmetry does not seem to be straightforward: surface potentials typically in a thin film will produce a shift of the parameters where phase coexistence occur (by a mechanism analogous to ``capillary condensation'') \cite{22}. E. g. for an asymmetric binary mixture the bulk phase coexistence occurs at a nontrivial value $\Delta \mu_{\textrm{coex}}(T)$ and in the thin film (even with a purely repulsive potential at both walls) will be shifted to a value $\Delta\mu_{\textrm{coex}}(T,D)$. Thus, this case also corresponds to a nontrivial (a priori unknown) value of the contact angle at both walls, implying that the interface separating both coexisting phases and running from the wall at $z=0$ to the wall at $z=D$ is curved. Such curved interfaces occur also in symmetric mixtures with symmetric rather than ``antisymmetric'' wall potentials (see e. g. \cite{141} for examples), or when one considers ``liquid bridges'' at vapor-liquid phase coexistence in capillaries (e.g. \cite{142}), and are hard to analyze.

An elegant way to avoid this problem, allowing the direct observation of contact angles of flat interfaces between coexisting phases that lack particular symmetries, has been proposed by Dimitrov et al.\cite{143}. One chooses a thin film geometry where phase coexistence between liquid and vapor at coexistence pressure of the bulk is enforced by coupling the system to an external liquid reservoir held at this pressure. The wall at $z=0$ has two different parts: for $0 \leq x <L/2$, it has a wall potential that is strongly attractive to the fluid particles (leading to complete wetting of the liquid), while for $L/2 <x<L$ it is repulsive (leading to drying), and a periodic boundary condition is used only along the y-direction. The liquid reservoir is at $x<0$, while at $x<L$ there is another (repulsive) wall. The other wall at $z=D$ is homogeneous for $0<x<L$, and there the wall potential is chosen such that incomplete wetting conditions occur. In this way, one can stabilize a flat interface, pinned at $z=0$ along the line $x=L/2$, and ``hitting'' the wall at $z=D$ under a nontrivial contact angle (which is the observable of interest) that depends on the potential that acts on this wall. Only near $z=0$ does one expect some curvature of the interface, essentially one creates a flat interface inclined under the contact angle, which can be varied over an extended range \cite{143}.

\section{The order parameter distribution and what we can learn from it about phase coexistence}
In this section we consider systems that exhibit a critical temperature $T_c$ such that for temperatures $T<T_c$ the system can coexist in two phases distinguished by different values of a scalar order parameter: e.g., in an anisotropic magnet (as described by the Ising model) the order parameter is the magnetization per spin $m$, and domains with magnetization $\pm m_{\textrm{coex}}$ can coexist ($m_{\textrm{coex}}$ being the absolute value of the spontaneous magnetization); in an (incompressible) binary (A,B) mixture the order parameter is the relative concentration of one species, say A, defined as $x_A=N_A/(N_A+N_B)$ where $N_A, N_B$ are the particle numbers of species A and B. We then assume that for $T<T_c$ a miscibility gap occurs, described by the coexistence of two phases with different concentrations, $x_{A(1)}^{\textrm{coex}}$ and $x_{A(2)}^{\textrm{coex}}$. We consider here explicitly again only the case of a strictly symmetric Lennard-Jones mixture (details about this model were already given in the previous section) for which $x_{A(2)}^{\textrm{coex}}=1-x_{A(1)}^{\textrm{coex}}$. Fig.~8 shows, as an example, the phase diagram of the model that we study (it has a critical temperature at $T_c=1.4230 \pm 0.0005$, while we study phase coexistence at $T=1.0$ where $x_{A(1)}^{\textrm{coex}}=0.030$, so one has coexistence between almost pure A and pure B). Finally, we deal with the simple Lennard-Jones fluid \cite{49,50} where the order parameter is the particle density $\rho=N/V$ (V being the volume of the simulation box, which we take as a cube of linear dimension $L$ with periodic boundary conditions), and for $T<T_c$, vapor (at density $\rho_v$)and liquid (at density $\rho_\ell$) can coexist. The model that is studied is defined by the Lennard-Jones (LJ) potential $u(r)$ depending on the distance $r$ between the particles,
\begin{equation}\label{eq466}
u(r)=4 \epsilon \{(\sigma /r)^{12}-(\sigma/r)^6\} + C, \quad r \leq r_c,
\end{equation}
\begin{equation}\label{eq47}
u(r)=0, \quad r >r_c,
\end{equation}
choosing units such that the depth of the potential well
($\epsilon$) and its range $(\sigma)$ both are unity, $\epsilon=1,
\sigma=1$, the cutoff $r_c=2.2^{1/6} \sigma$, and the constant C
is chosen as $C=127/16384$ so $u(r=r_c)=0$. For this model
$T_c=0.999 $ \cite{144} (also choosing Boltzmann's constant
unity).

\begin{figure}\centering
\includegraphics[width=0.6\linewidth]{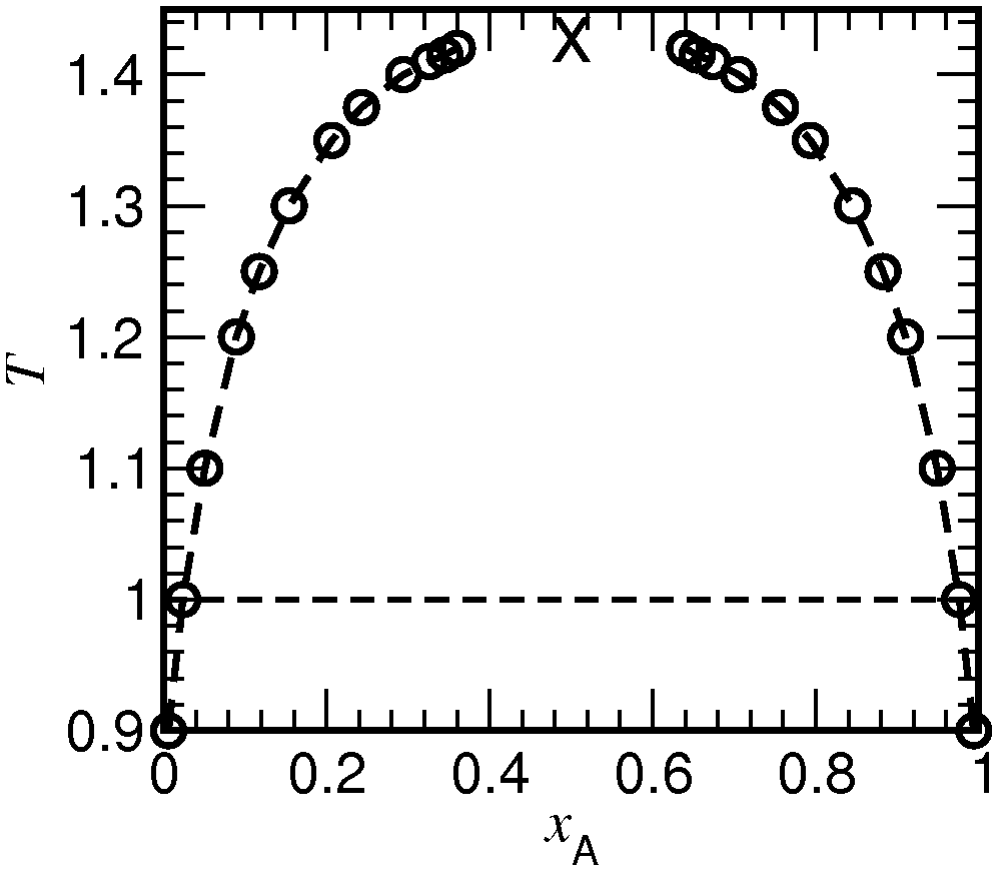}
\caption{\label{fig8} Phase diagram of the symmetric binary (A,B) Lennard-Jones mixture with $\rho^*=1$ in the plane of variables $T$ and relative concentration $x_A=N_A/N$ of the A particles, for fixed total particle number $N=6400$. The cross shows the critical point, taken from \cite{140}. The horizontal broken line indicates that phase coexistence is studied for $T=1.0$. From block et al.\cite{50}.}
\end{figure}

While for the Ising model spin reversal symmetry ensures that
phase coexistence occurs at magnetic field $H=0$, and in the
symmetric binary mixture the symmetry against interchange of
particle labels A and B ensures that phase coexistence occurs at
chemical potential difference $\Delta \mu = \mu_A-\mu_B=0$, no
such symmetry exists for the one-component Lennard-Jones fluid,
and phase coexistence occurs at a nontrivial value
$\mu_{\textrm{coex}}(T)$ of the chemical potential $\mu$ of the
particles. Carrying out simulations in the grand canonical
$(\mu VT)$ ensemble and recording the probability distribution of the density $P_L(\rho,T)$ one finds
$\mu_{\textrm{coex}}(T)$ from the equal weight rule \{in the
region of $\mu$ where $P_L(\rho,T)$ exhibits two peaks one near
$\rho_v(T)$ and the other near $\rho_\ell(T)$\}. \cite{145,146}.

\begin{figure}\centering
\includegraphics[width=0.6\linewidth]{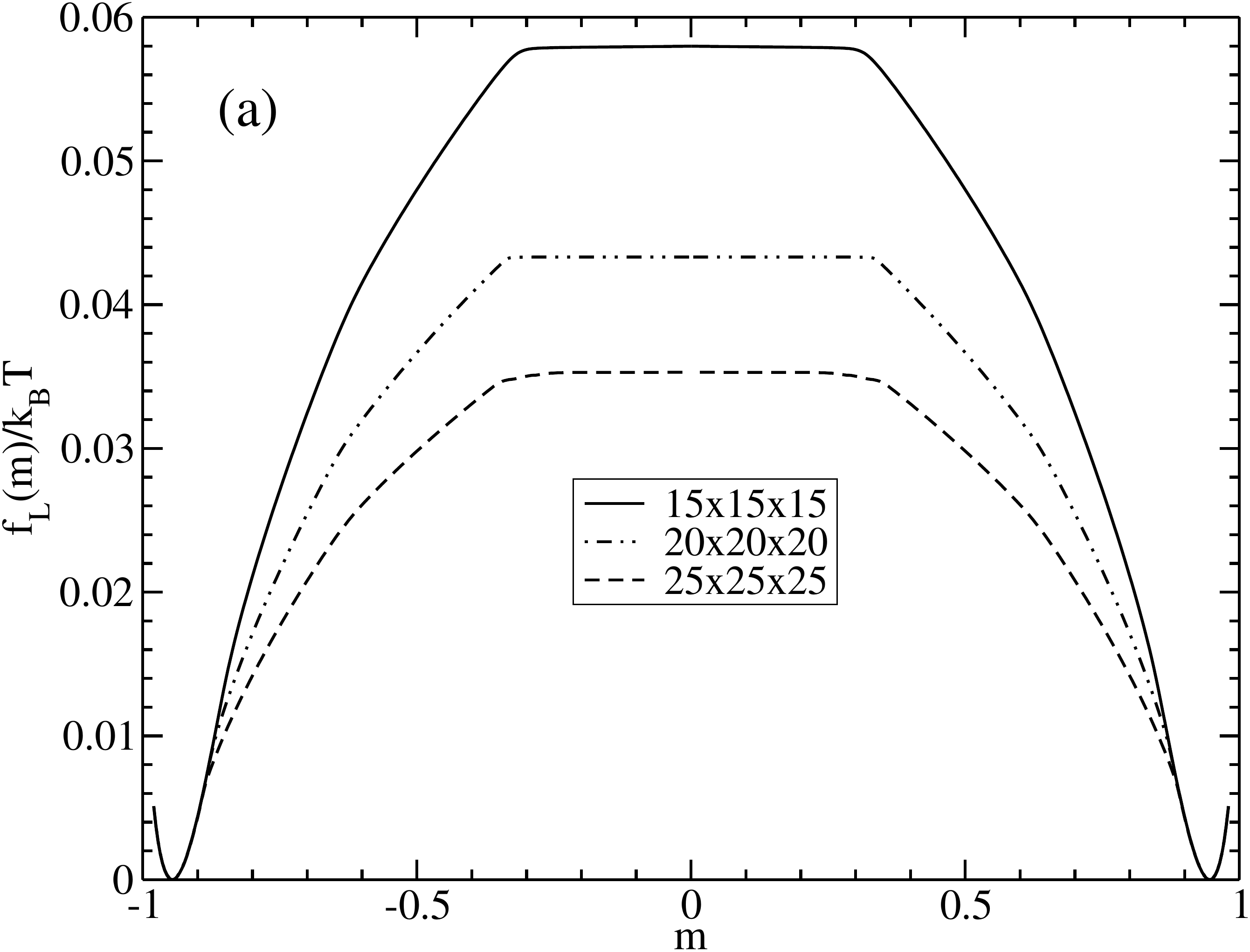}
\includegraphics[width=0.65\linewidth]{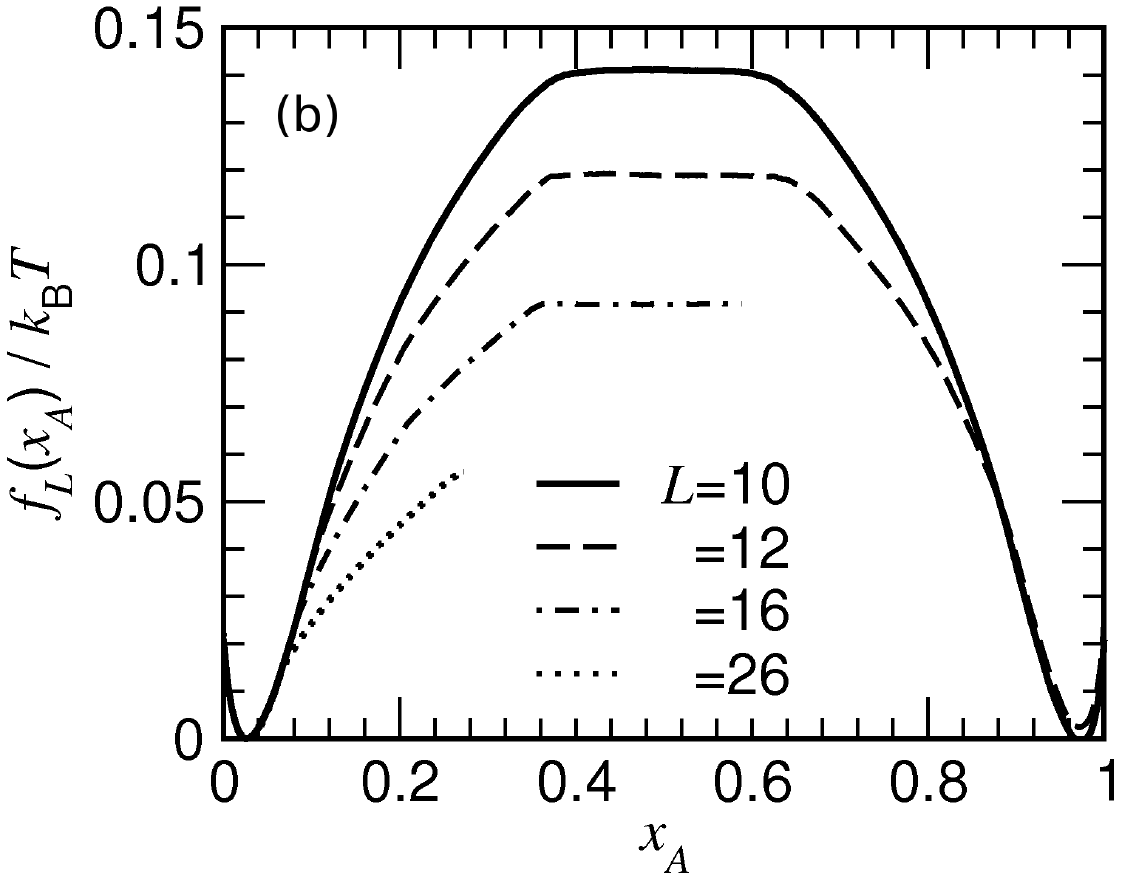}
\includegraphics[width=0.6\linewidth]{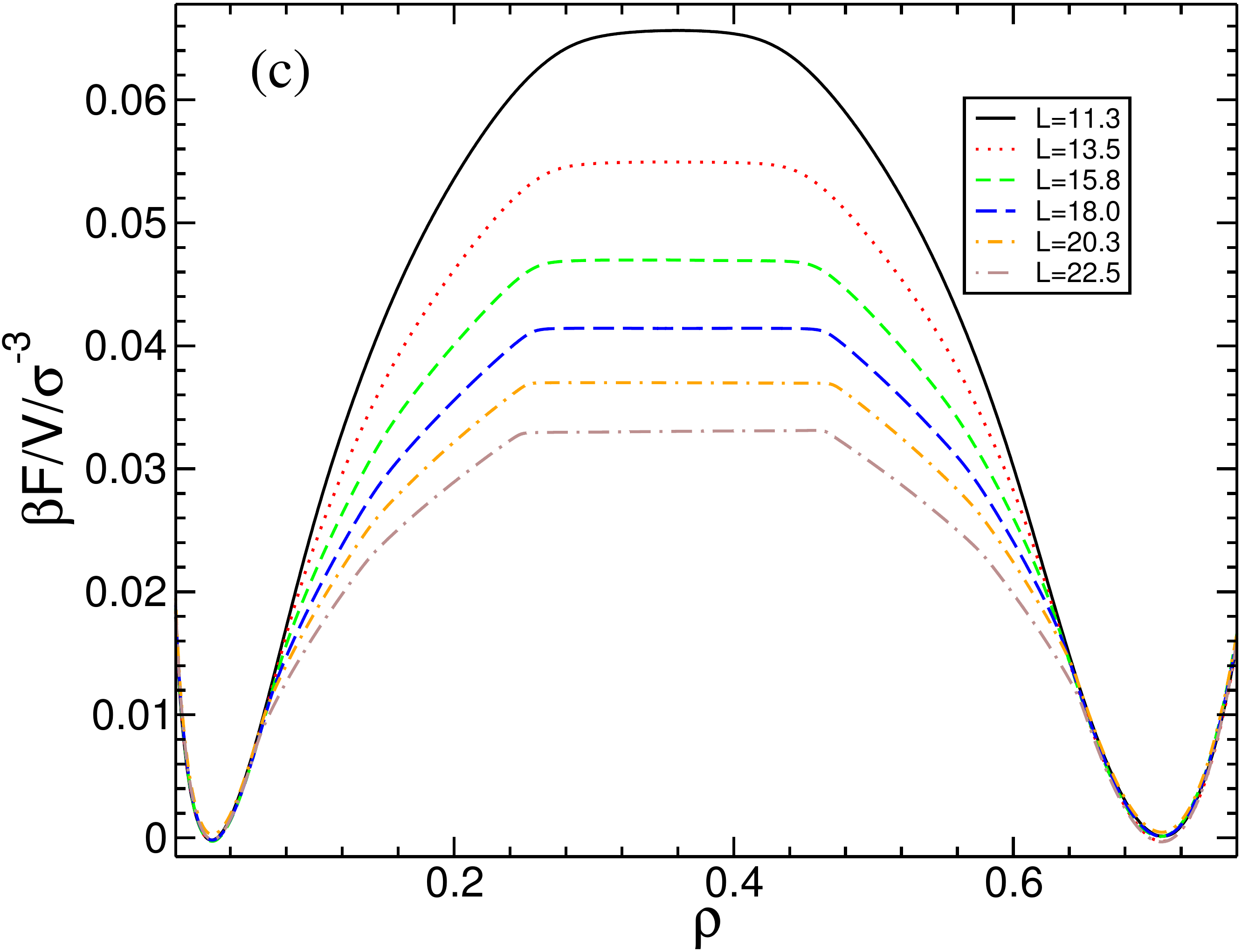}
\caption{\label{fig9} (a) Effective free energy per spin $f_L(m,T)$ of $L\times L \times L$ Ising lattices plotted vs. $m$ at $k_BT/J=3.0$ for three choices of $L$, as indicated. (b) Effective free energy $f_L (x_A,T)$ of a symmetric binary Lennard-Jones mixture in a $L \times L \times L$ cube with periodic boundary conditions for several choices of $L$ and $T=1.0$. From Block at al. \cite{50}. (c) Effective free energy $f_L(\rho,T)$ of a Lennard-Jones fluid \{Eqs.~\ref{eq47}\} at $T=0.78 T_c$, for six choices of $L$ as indicated. }
\end{figure}

Having obtained the distribution functions $P_{LHT}(m)$ for the
Ising model (at $H=0), P_{LN\Delta \mu T}(x_A)$ for the symmetric
binary mixture (at $\Delta \mu =0$), and $P_{L\mu T}(\rho)$ for
the LJ fluid (at $\mu=\mu_{\textrm{coex}}(T)$), we associate with
these distributions effective free energy functions, defined as
follows (the dimensionality $d$ is $3$ throughout in the
following; and for simplicity, we use the symbol $f$ throughout
for these densities of thermodynamic potential, irrespective of
the system)
\begin{equation}\label{eq48}
f_L(m,T)=-(k_BT/L^d) \ln [P_{LHT}(m)/P_{LHT}(m_{\textrm{coex}})],
\end{equation}
\begin{equation}\label{eq49}
f_L(x_A,T)=-(k_BT/L^d) \ln [P_{LN \Delta \mu T}(x_A)/P_{LN \Delta \mu T}(x^{\textrm{coex}}_A)]\; ,
\end{equation}
and
\begin{equation}\label{eq50}
f_L(\rho ,T)=-(k_BT/L^d) \ln [P_{L\mu T}(\rho)/P_{L\mu T}(\rho_v)]+ \mu \rho\; .
\end{equation}
Typical examples for these effective free energy functions are
shown in Fig.~\ref{fig9}. One can see a characteristic double-well
shape, with a ``hump'' in between that is more or less flat in the
center, and depends clearly on the linear dimension of the system.

The double well-shape of these effective free energies may be
considered as somewhat reminiscent of the double well shaped mean
field free energy density $f^{MF}(m,T)$ obtained when one treats
the Ising model in molecular field approximation \{and related
mean field free energy densities $f^{MF}(x_A,T), f^{MF}(\rho,T)$
for the other systems\}. We emphasize, however, that such an
analogy would be completely misleading: while these mean field
free energy densities do not depend on $L$  and describe inside
the two-phase coexistence region, homogeneous states that are
interpreted as being metastable (up to the inflection point, the
so-called spinodal) and unstable (inside of the spinodal), we deal
here with the statistical mechanics of finite systems in
inhomogeneous states in full thermal equilibrium. The whole
structure that is seen in these curves inside of the two-phase
coexistence region all comes from interfacial effects \cite{8}.

\begin{figure}\centering
\includegraphics[width=0.3\linewidth]{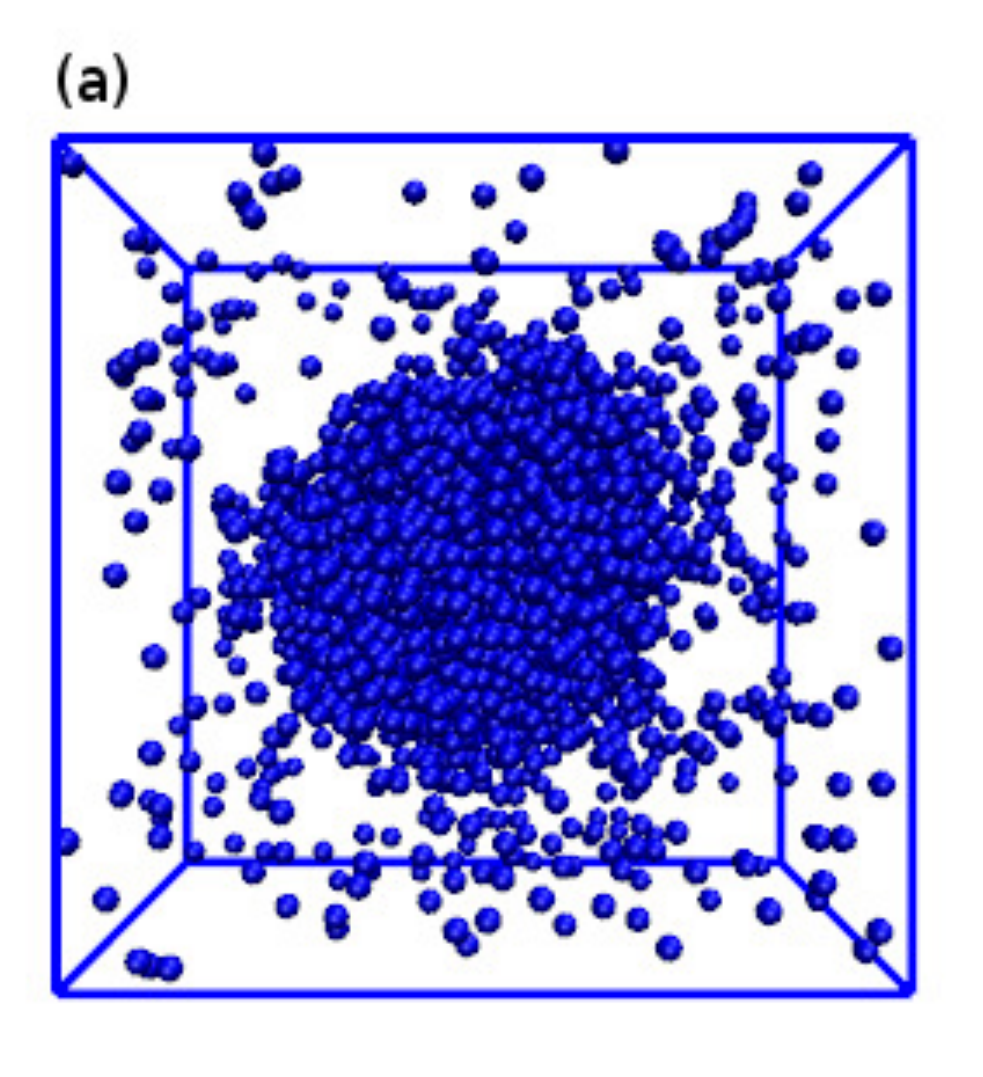}

\includegraphics[width=0.3\linewidth]{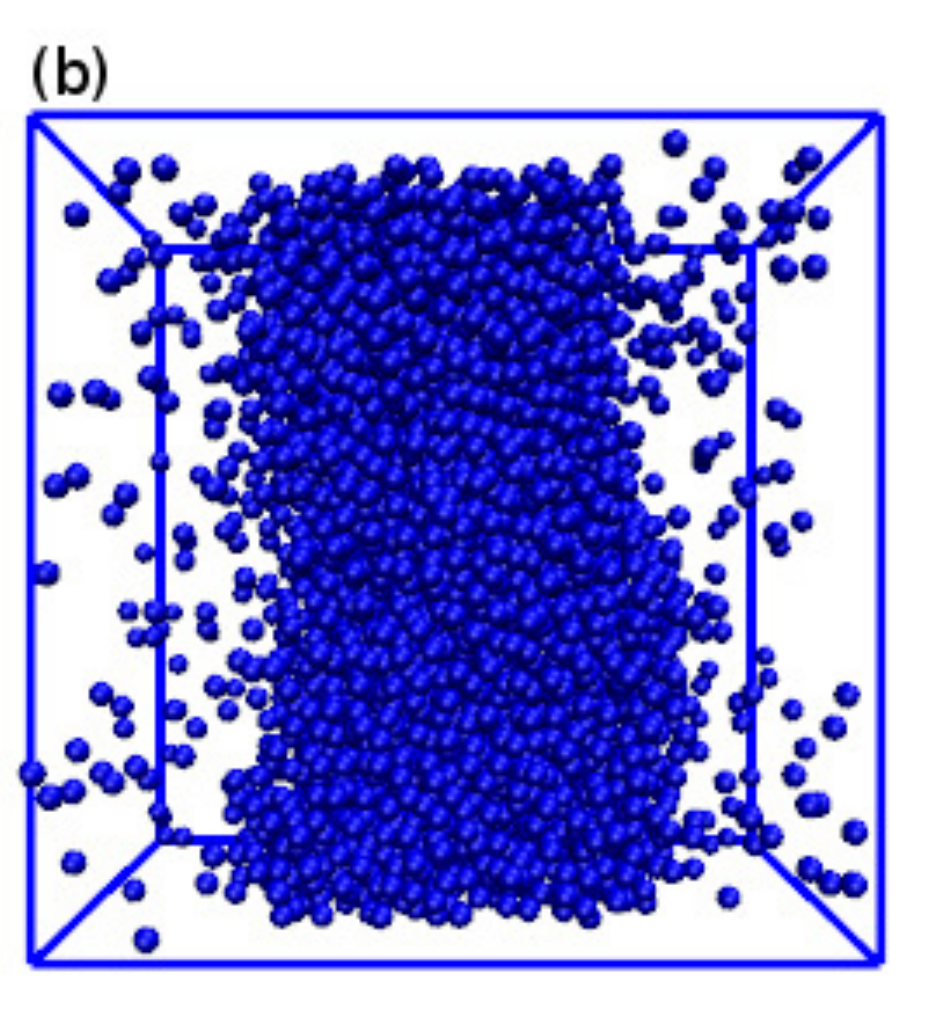}
\includegraphics[width=0.3\linewidth]{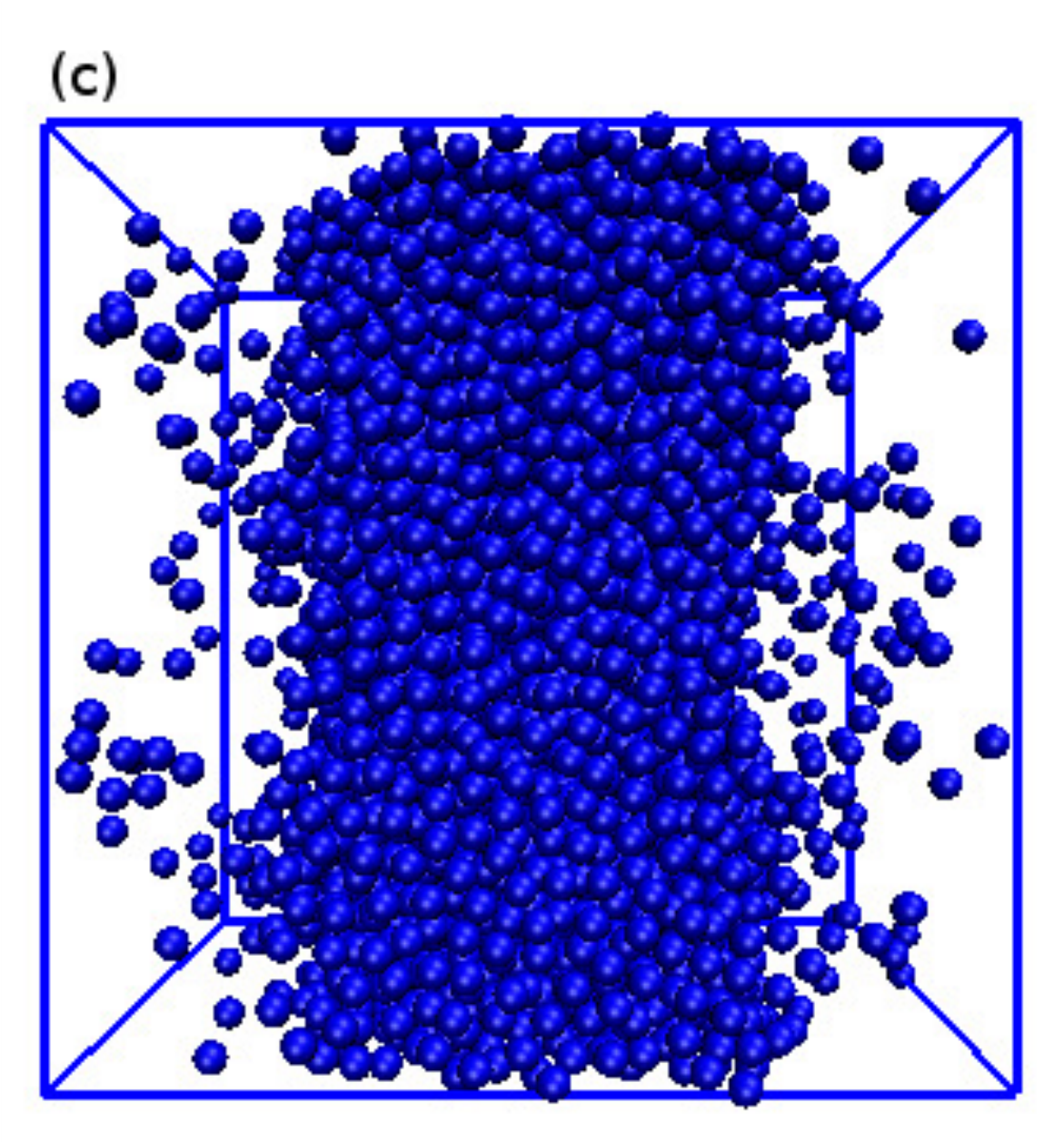}
\caption{\label{fig10} Configuration snapshots of a symmetric binary Lennard-Jones mixture at $T=1.0$ for $L \times L \times L$ systems with periodic boundary conditions with $L=24$ and three different choices of the concentration: $x_A=0.15$ (a), 0.27 (b) and 0.5(c). Only A particles are shown by dark dots, B particles are not shown for clarity. In case (a), one can recognize an (almost) spherical droplet surrounded by supersaturated ``vapor'' of A particles; case (b) shows an almost cylindrical droplet, connected into itself via the periodic boundary condition; case (c) shows a slab configuration (the A-rich phase is separated from the phase poor in A by almost planar interfaces).}
\end{figure}

This statement is easily verified when one examines snapshot
pictures of the typical configurations contributing to the average
in various regions of this effective free energy density. As an
example, Fig.~\ref{fig10} presents configuration snapshots
\cite{100} of the symmetric binary $(A,B)$ mixture for
$x_A=0.15$, 0.27, and 0.50, respectively. In the snapshot pictures
the positions of the A-particles are indicated by dots, the
positions of the B-particles are not shown at all, for the sake of
clarity. One clearly recognizes that both droplets of
approximately spherical and cylindrical shape occur, as well as
slab-like configurations bounded by two (approximately planar)
interfaces. The latter two configurations are stabilized by the
periodic boundary conditions. Note that for an $L \times L \times
L$ box the cylinder axis can be oriented along any of the x,y,z
coordinate axes, and the same symmetry holds for the orientation
vector that is perpendicular to the interfaces in the slab state.
Of course, these spherical, cylindrical and slab-like domains
exhibit these regular shapes on average only: the sampling that is
done does not at all constrain the fluctuations of these domains,
at every value of $x_A$ all configurations that are compatible
with the chosen concentration contribute according to their
statistical weight.

By choosing rectangular boxes one can constrain the orientations
of the cylinders or slabs, respectively: choosing a volume $L
\times L \times D$ with $D<L$ the cylinder axis will be
predominantly along the z-direction (since a cylinder of radius R
and height D then costs a surface free energy proportional to a
surface area $2R\pi D$ instead of $2R\pi L$). Likewise, choosing
a volume $L \times L \times D$ with $D>L$ the domain walls in the
slab state will predominantly be oriented perpendicular to the
z-axis, since the surface free energy cost (proportional to $L^2$)
then is less than for the other two orientations (for which it is
proportional to $L D > L^2$).

We now comment on the flatness of the hump in Fig.~\ref{fig9}b in
the region when the slab configuration prevails. If we assume that
the two interfaces that separate the A-rich phase from the B-rich
background do not interact with each other at all, the free energy
cost to form them is simply $2 \gamma_{AB}(L)L^2$, where
$\gamma_{AB}(L)$ is the interfacial free energy density (since the
periodic boundary conditions constrain interfacial fluctuations,
e. g. the spectrum of capillary waves is discretized, we allow
some weak dependence of $\gamma_{AB} (L)$ on $L$, requiring
however that $\gamma_{AB}(\infty)$ exists).

The statement that the droplet shapes are (on average) either
spherical or cylindrical have tacitly implied that the interfacial
free energy does not depend on the direction of the vector
perpendicular to the interface. This is the case for off-lattice
fluid-fluid interfaces (such as the interfaces of the model binary
fluid mixture studied in Fig.~\ref{fig10}) or liquid-vapor
interfaces. However, for the Ising model, due to the underlying
lattice structure, such an isotropy of interfacial properties
holds only very close to the critical point. In fact, as discussed already in Sec.~2.2, for temperatures below the roughening transition temperature $T_R$ of the (100) interface the droplet surface contains strictly flat regions. While at $T=0$ the shape that minimizes the surface free energy at given interface orientation in the Ising model simply is a cube, at nonzero $T<T_R$ the determination of the minimum free energy droplet shape is a nontrivial problem \cite{133}, since the interface contains both planar parts (``facets'') and curved regions, and since facets appear only for $T<T_R$, $T_R$ can also be interpreted as faceting transition temperature. Similar complications due to the anisotropy of the interfacial free energy are also important for crystal-fluid interfaces. However, we shall not further consider the problem of anisotropic interfacial free energies in this paper.

Another important problem concerns the transitions between the various regimes in Figs.~\ref{fig9} and \ref{fig10}. The transitions between slabs and cylinders, cylinders and spheres, and states with spherical droplets and without it are not sharp (for finite $L$) but rounded, when $x_A$ is varied. This observation is consistent with the fact, of course, that ``sharp'' phase transitions (i.e., associated with a singular variation of the free energy of the system) can occur in the thermodynamic limit only, $L \rightarrow \infty$. In this limit, the transition (at $x_A^t)$ from the homogeneous one-phase state to a state containing a droplet moves towards the coexistence curve, $x_A^t \rightarrow x_{A,1}^{\textrm{coex}}$, while the droplet to cylinder transition occurs at a volume fraction $\Psi$ of the A-rich phase given by
\begin{equation}\label{eq51}
\Psi ^{\textrm{drop-cyl}} = (4 \pi) / 81 \approx 0.155, \quad \Psi =(x_A-x_{A,1}^{\textrm{coex}})/(1-2x_{A,1}^{\textrm{coex}})
\end{equation}
and the cylinder to slab transition occurs at
\begin{equation}\label{eq52}
x_A^{\textrm{cyl -slab}} = x_{A,1}^{\textrm{coex}} + \Psi^{\textrm{cyl-slab}} (1-2x_{A,1}^{\textrm{coex}}), \quad \Psi^{\textrm{cyl-slab}}= 1/\pi \approx 0.318 \; .
\end{equation}

\begin{figure}\centering
\includegraphics[width=0.6\linewidth]{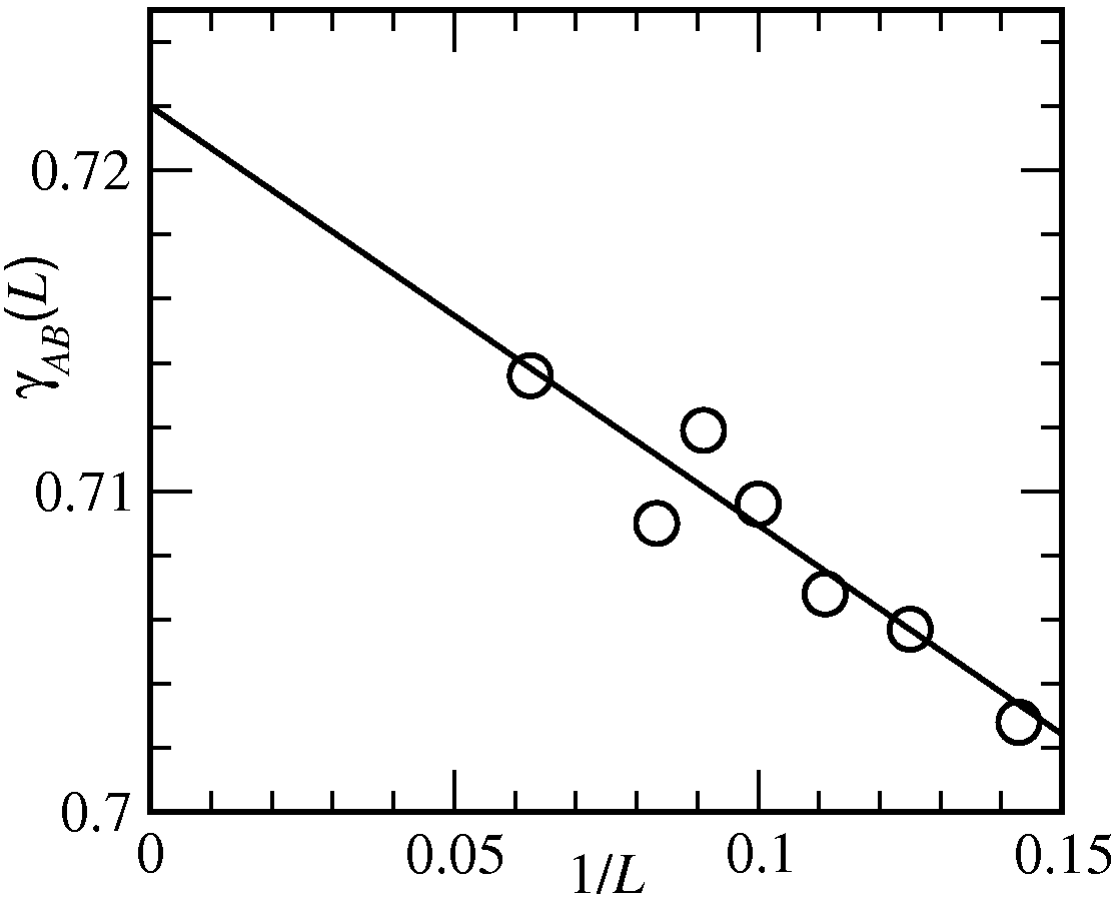}
\caption{\label{fig11} Extrapolation of the interface tension $\gamma_{AB}(L)$ of flat interfaces between A-rich and B-rich phases of Lennard-Jones mixtures, coexisting in $L \times L \times L$ systems at $k_BT/\epsilon= 1.0$ in a slab geometry (as discussed in Figs.~\ref{fig9}(b), \ref{fig10}(c)). The shown straight line fit yields $\gamma_{AB}(\infty) = 0.722 \pm 0.002$. From Block et al. \cite{50}.}
\end{figure}

Due to the symmetry of the system with respect to the interchange of A and B, analogous transitions occur at $1-\Psi_A^{\textrm{drop-cyl}}$ and $ 1-\Psi^{\textrm{cyl-slab}}$. For finite $L$, all these transitions are rounded, and hence when one analyzes the interfacial free energies of spherical or cylindrical interfaces (or even flat interfaces encountered in the slab states), it is important that one stays off from the regions where these rounded transition occur. Of course, in the strict sense this means that always only the approach towards the thermodynamic limit is a well-defined problem, both for curved and for flat interfaces. In fact, for a spherical droplet there is always a nonzero (albeit for large $L$ in practice negligibly small) probability that the droplet makes a transition to cylindrical shape or that it evaporates completely. This caveat is not just a matter of principle, but actually prevents the study of extremely small droplets, that contain just only a few particles only, also in practice - such small droplets are not well distinguished from other fluctuations that occur in the homogeneous phase.

Fig.~\ref{fig11} shows, as an example, the estimation of the interfacial tension $\gamma_{AB}= \gamma_{AB}(L\rightarrow \infty)$ for the binary LJ mixture, using an extrapolation of $\gamma_{AB}(L)$ versus $1/L$. It is seen, that in practice the dependence of $\gamma_{AB}(L)$ on $L$ can be rather weak, and a very good accuracy can in fact be reached. However, it should be noted that the precise nature of finite-size corrections to $\gamma_{AB}$ is not fully understood. When this method of estimating interfacial tensions from $f_L(m,T)$ was introduced almost 30 years ago for the Ising model in $d=2$ and $d=3$ dimensions \cite{115}, an alternative formula including a logarithmic correction was also suggested
\begin{equation}\label{eq53}
\gamma(L) = \gamma(\infty) + \frac{a}{L} + \frac{b \ln L}{L} ,
\end{equation}
with two undetermined constants a,b. However, lacking theoretical guidance on the values of these constants, a very large range of $L$ would be required to allow for an unambiguous estimation of the three constants $\gamma(\infty)$, a and b from an empirical fit to simulation data on $\gamma(L)$.
\begin{figure}\centering
\includegraphics[width=0.6\linewidth]{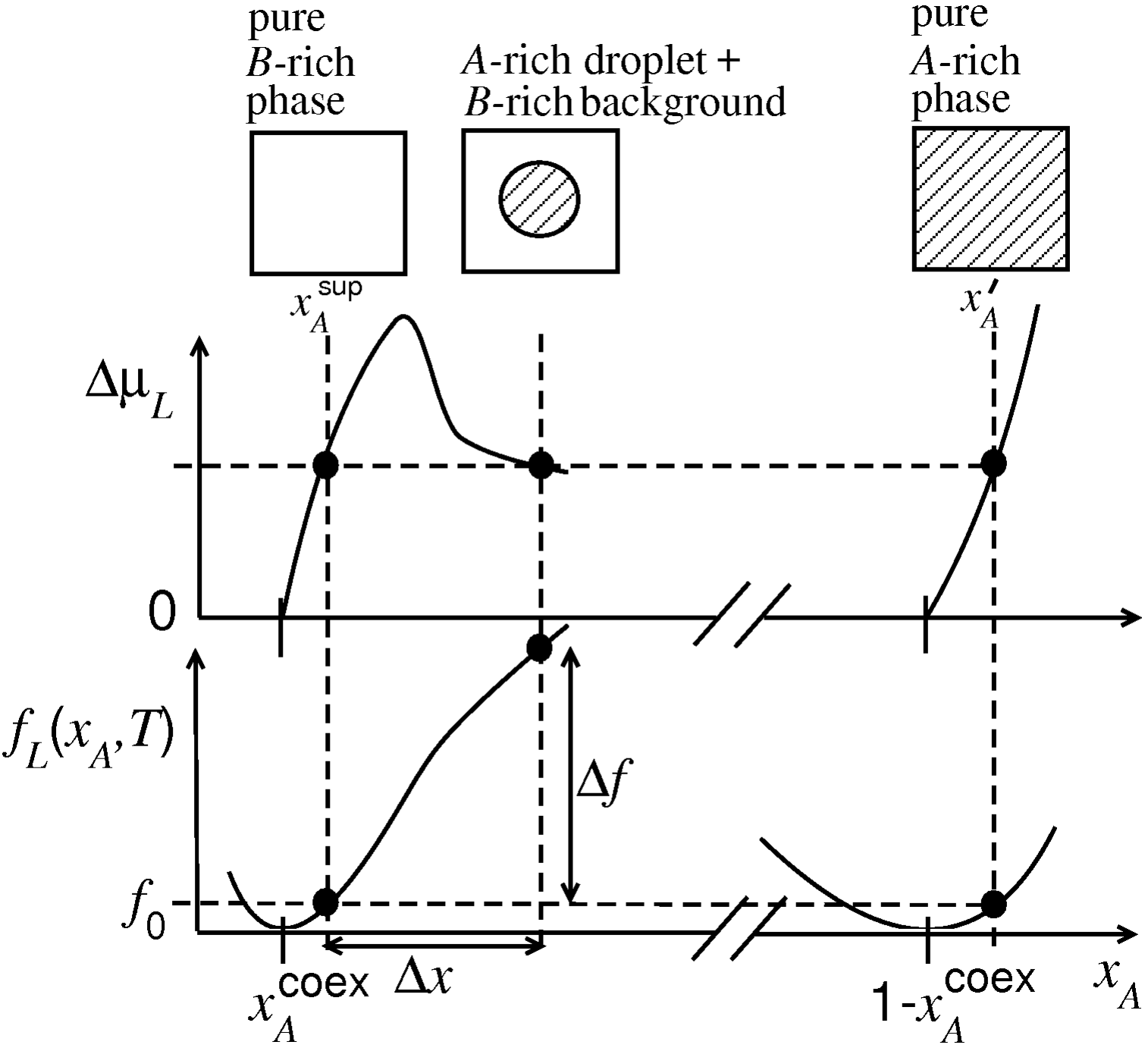}
\caption{\label{fig12} Schematic explanation of the generalized lever rule and its use to obtain droplet free energies for a symmetrical fluid binary (A,B) mixture. At the same chemical potential difference $\Delta \mu_L$ (horizontal broken line) three states can be identified: a pure B-rich phase at concentration $x_A^{\textrm{sup}} > x_A^{\textrm{coex}}$, a pure A-rich phase at concentration $x_A'>x_{A,2}^{\textrm{coex}}=1-x_A^{\textrm{coex}}$, and a state containing an A-droplet of radius $R$, surrounded by B-rich background (which must also have the concentration $x_A^{\textrm{sup}}$), at average concentration $x_A$. Defining $R$ by the condition that the concentration in the droplet equals $x_A'$, observation of $\Delta x$ yields information on $R$, while observation of $\Delta f$ yields information on the surface free energy of the droplet.}
\end{figure}

We note that despite these difficulties, there is ample evidence both in $d=2$ (where one can compare to Onsager's exact solution for $\gamma(\infty)$ \cite{143}) and in $d=3$ that the method illustrated in Fig.~\ref{fig11} is practically useful \cite{49,50,79,106,107,116,117,118,119,120,121,122}. In fact, for accurate estimations of $\gamma_{v \ell}$ of the vapor-liquid interface tension this method has become the method of choice \cite{49,11,117,118,119,120,121,122}.

Thus, while it early has been recognized that $f_L(m,T)$  \{as well as $f_L(x_A,T)$ or $f_L(\rho,T)$, respectively\} contains valuable information on the interface free energy of planar interface in the respective models, only recently it has been exploited that one can obtain also the interface free energy for the case of curved interfaces. Fig.~\ref{fig12} explains the basic idea for the case of the fluid binary mixture. One first introduces the variable conjugate to the order parameter $(m,x_A, \; \textrm{or} \; \rho$, respectively) as a derivative of the free energy with respect to the order parameter:
\begin{equation}\label{eq54}
H_L(m,T)=[\partial f_L(m,T)/\partial m]_T ,
\end{equation}
\begin{equation}\label{eq55}
\frac {1}{k_BT} \Delta \mu_L(x_A,T) = [\partial f_L(x_A,T)/\partial x_A]_T  ,
\end{equation}
and
\begin{equation}\label{eq56}
\frac {1}{k_BT} \mu _L (\rho , T)= [\partial f_L (\rho,T)/\partial \rho]_T .
\end{equation}

\begin{figure}\centering
\includegraphics[width=0.6\linewidth]{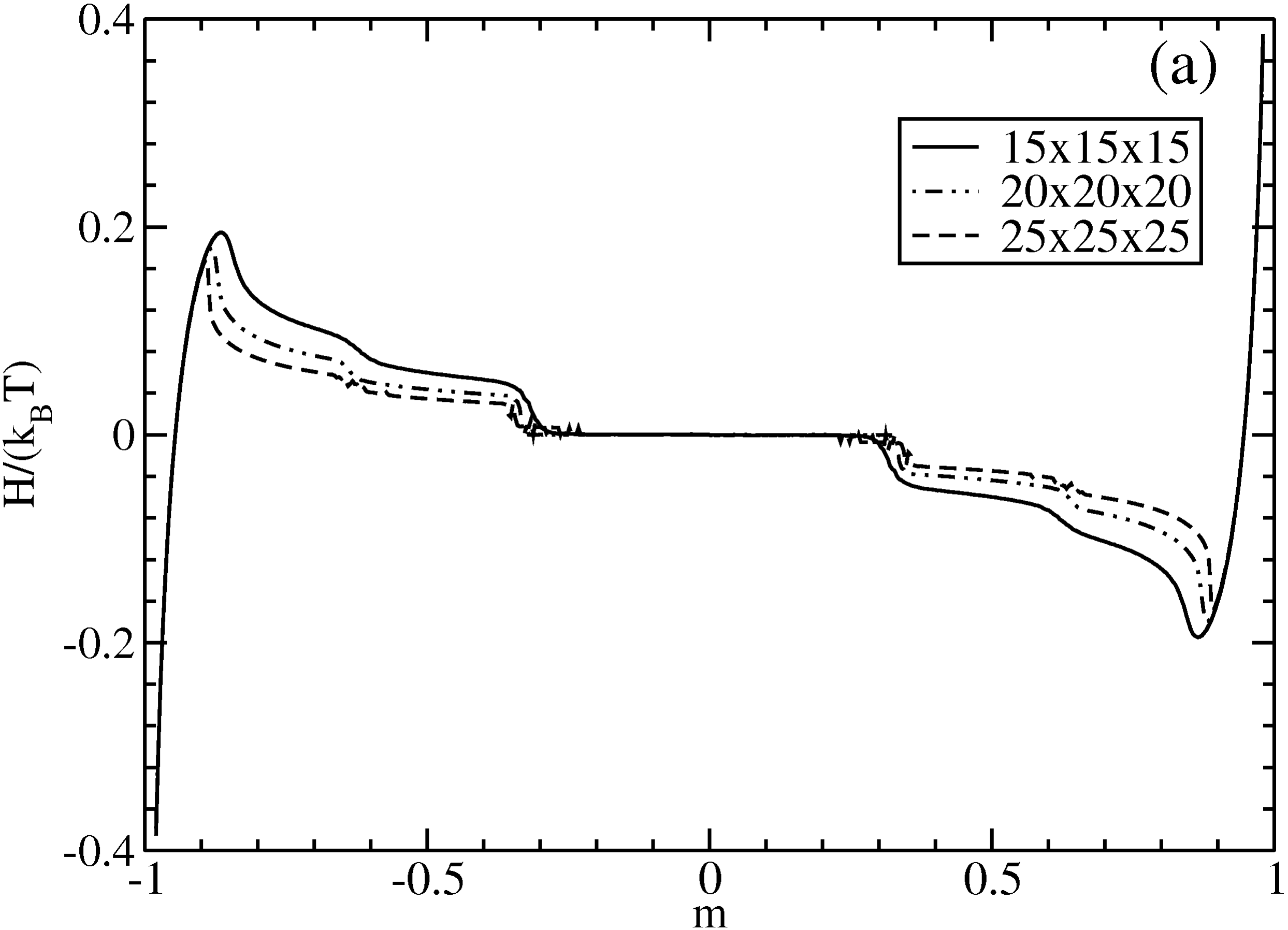}
\includegraphics[width=0.65\linewidth]{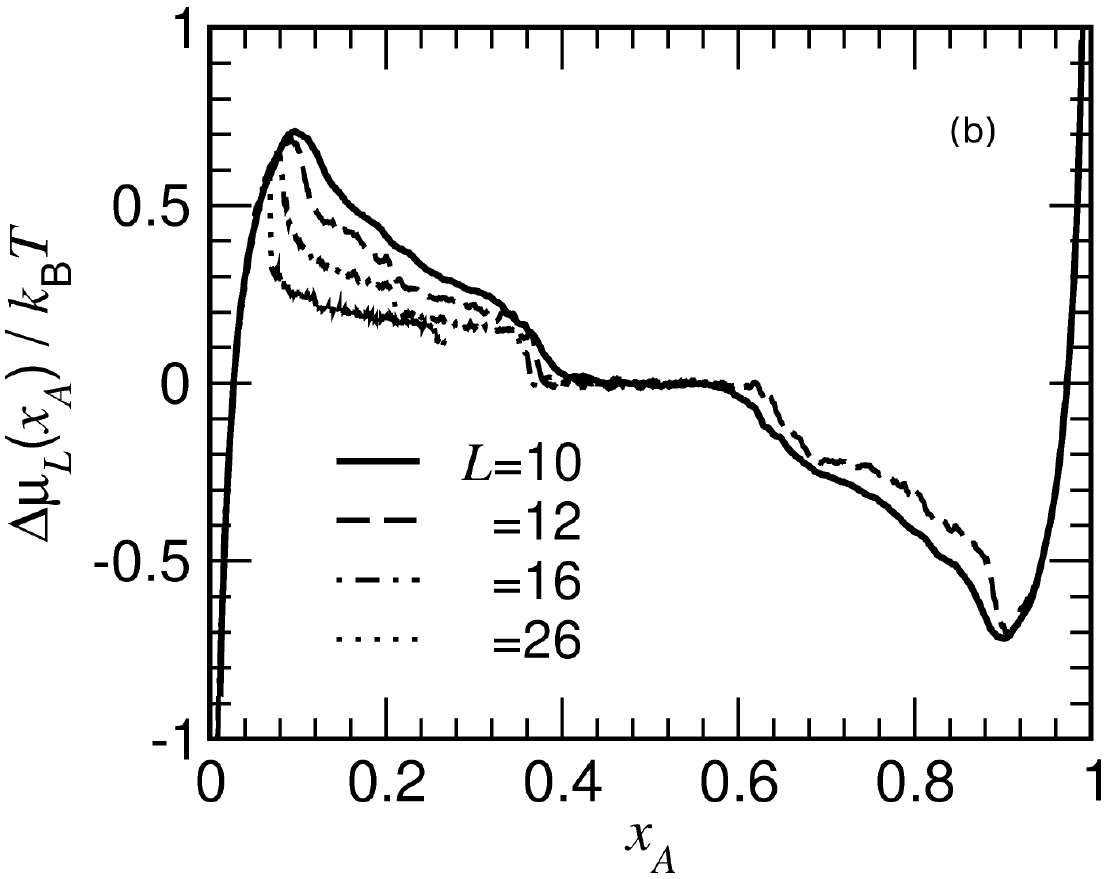}
\includegraphics[width=0.6\linewidth]{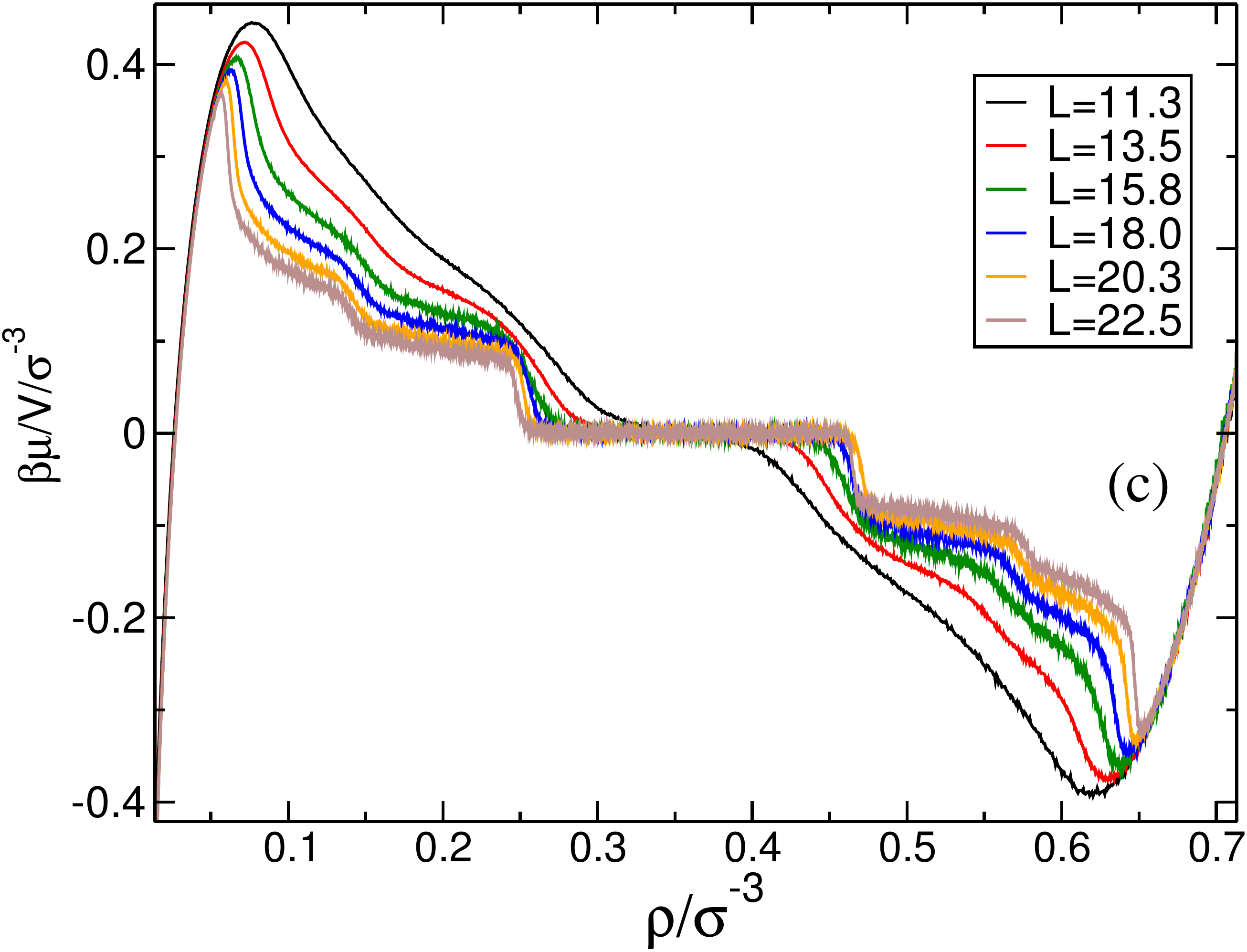}
\caption{\label{fig13} (a) Ising model isotherms at $k_BT/J=3.0$ plotting $H_L(m,T)/k_BT$ versus magnetization $m$ for several choices of $L$ as indicated. From Winter et al. \cite{94}. (b) Isotherms of the binary symmetric Lennard-Jones mixture at $T=1$, $\rho =1$, plotting $\Delta \mu _L(x_A,T)/k_BT$ vs. concentration $x_A$. Data refer to a single run at each size $L$, to illustrate the typical noise level (200 Monte Carlo steps per particle have been used for each window of the successive umbrella sampling). For the final analysis, five such runs were averaged together. From Block et al. \cite{50}. (c) Isotherms of the simple Lennard Jones fluid \{Eqs.~\ref{eq47}\} at $T=0.78 T_c$, plotting $\mu_L(\rho,T)/k_BT$ versus the density $\rho$, for six linear dimensions $L$, as indicated. Note that lengths are given in units of $\sigma (=1)$. From Block \cite{149}.}
\end{figure}

In the thermodynamic limit, $H_L(m,T)$ converges to an L-independent function $H(m,T)$ for $m <-m_{\textrm{coex}}$; this is just the description of the magnetic equation of state of the Ising magnet (or the equation of state of the binary mixture or fluid) in the bulk one-phase region, of course. Analogous statements refer to $\Delta \mu _L(x_A,T)$ and $\mu _L(\rho,T)$. Fig.~\ref{fig13} shows specific examples for these functions, defined in Eqs.~\ref{eq54}-\ref{eq56}, as obtained from simulations \cite{49,50,94,149}. The data show that the maxima and minima of these curves (which correspond to the droplet (bubble) evaporation/condensation transition \cite{150,151,152}, as noted above) indeed become sharper as $L$ increases, and move towards the coexistence curve. Arguments have been presented \cite{150,151} that in $d=3$ dimensions both the distance from coexistence and the height of these extrema scale as $L^{-3/4}$ and some numerical evidence for this relation can be found in Schrader et al. \cite{148}. For $d=2$, an analogous relation (scaling with $L^{-2/3}$) has been established rigorously \cite{152}. Thus, when the limit $L \rightarrow \infty$ has been taken, no trace of this transition (nor of the transitions from spherical to cylindrical domains (Eq.~\ref{eq51}) or from cylindrical domains to slabs (Eq.~\ref{eq52}), respectively) is left in the isotherms, while for finite $L$ all these transitions are subject to finite size rounding, as is obvious from Fig.~\ref{fig13}. However, for large $L$ the relative extent of the rounding is small (see Ref. ~\cite{151} for a phenomenological discussion), and hence for large $L$ one can identify parts of the rounding shown in Figs.~\ref{fig9} and \ref{fig13} where the droplet-vapor coexistence is not affected by these transitions, and hence the analysis outlined in Fig.~\ref{fig12} becomes applicable. Similarly, for volume fractions $\Psi$ of the minority phase in the region $\Psi^{\textrm{drop-cyl}}<\Psi <\Psi^{\textrm{cyl-slab}}$, one can use an analysis analogous to Fig.~\ref{fig12} to extract the surface free energy of cylindrical interfaces \cite{50} as long as one does not use volume fractions $\Psi$ close to the boundaries of this interval.

We here indicate only for the case of the binary mixture, how the surface free energy of droplets is extracted, inspired by the construction indicated in Fig.~\ref{fig12}. It is implied that the volumes of the two phases $V_I$ (droplet) and $V_{II}$ (surroundings) and their particle numbers $N_{A,I}, N_{A,II}, N_{BI}, N_{BII}$ are additive
\begin{equation}\label{eq57}
V = V_I+V_{II}, N_A=N_{AI}+N_{AII}, \quad N_B=N_{BI}+N_{BII}.
\end{equation}
Thus the interface in Fig.~\ref{fig12} is defined infinitesimally thin (so there is no excess volume associated with the interface) and its location is placed such that there occurs no excess of the particle number of either component. Note that we have assumed a strictly symmetric and practically incompressible mixture, so that a description in terms of a single order parameter density $x_A$ suffices. Unfortunately, the generalization of our procedures to compressible mixtures where two order parameter densities $\rho$ and $x_A$ need to be considered is not straightforward.

\begin{figure}\centering
\includegraphics[width=0.6\linewidth]{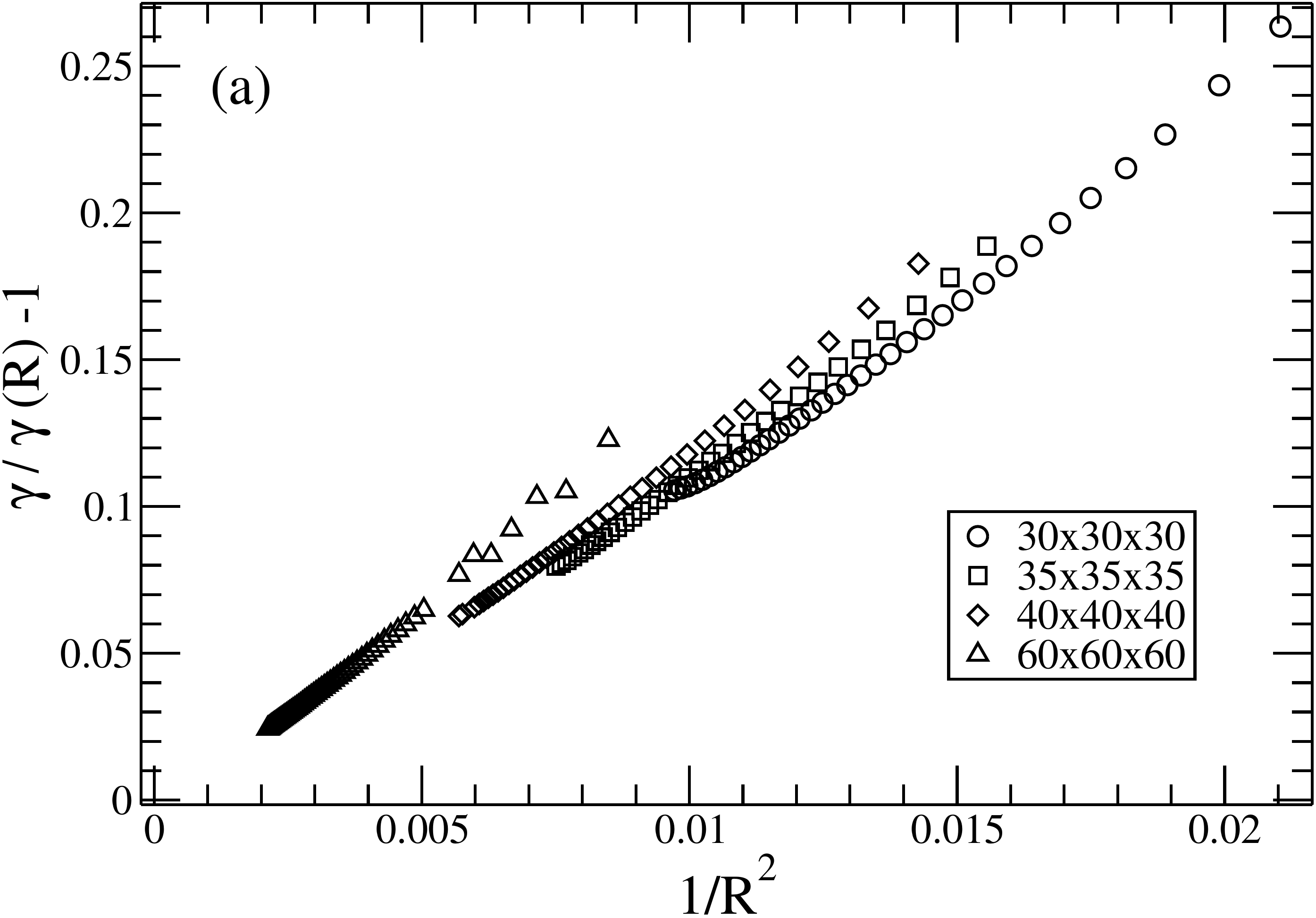}
\includegraphics[width=0.7\linewidth]{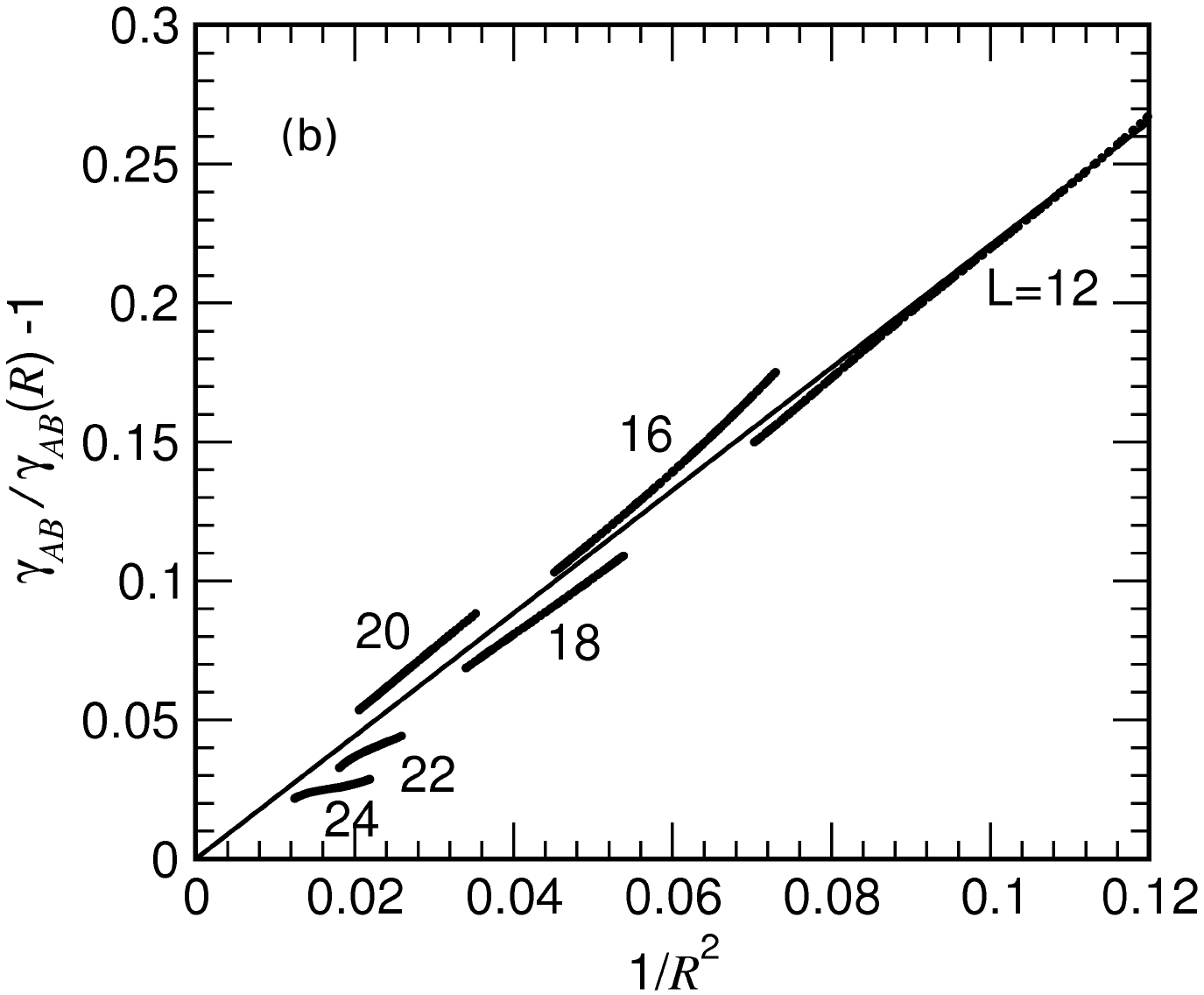}
\caption{\label{fig14} (a) Plot of $\gamma (\infty)/\gamma (R)-1$ versus $1/R^2$, for the Ising model at $k_BT/J=4.0$.  Here $\gamma_{AB}(\infty)$ was taken from the work of Hasenbusch and Pinn \cite{30}. Different symbols show different choices of the linear dimension $L$, as indicated. (b) Plot of $\gamma_{AB}(\infty) /\gamma_{AB}(R)-1$ versus $1/R^2$, for the symmetric binary Lennard-Jones mixture at $T=1$, $\rho =1$. Here, $\gamma_{AB}$ was taken from Fig.~\ref{fig11}. Different choices of $L$ are used for different parts of the curve. From Block et al. \cite{50}.}
\end{figure}

Fig.~\ref{fig12} then implies, when we assume that the average shape of the droplet is a sphere of radius $R$, that
\begin{equation}\label{eq58}
\Delta x = x_A^{\textrm{sup}}-x_A=(4 \pi R^3/3L^3)(1-2x_A^{\textrm{coex}})\;;
\end{equation}
remember that $x_A^{\textrm{sup}}$ is the concentration of the surroundings (supersaturated in A particles) of the droplet. Since the free energy density of this phase and the bulk term of the free energy density of the droplet are equal, $\Delta f $ is entirely due to the surface free energy of the droplet, and hence,
\begin{equation}\label{eq59}
\Delta f = 4 \pi R^2 \gamma _{AB}(R)/L^3\;,
\end{equation}
where we have denoted the interfacial tension of a droplet, which has the (equimolar) radius $R$, as $\gamma_{AB}(R)$, for the case of the symmetric binary mixture. Of course, Eqs.~\ref{eq58}, \ref{eq59} can be straightforwardly transcribed to the Ising model (if the approximation that the interface tension is independent of interface orientation is accurate). In the case of the vapor-liquid transition, which lacks any symmetry between the coexisting phases, of course, droplets of radius $R$ and bubbles of radius $R$ may give rise to different surface tensions $\gamma_{VL}(R)$ in both cases. As is well-known, this difference is related to the existence of a nonzero Tolman length \cite{51}, as discussed already in the introduction.

Using this method outlined in Fig.~\ref{fig12}, the curvature-dependent interfacial tension has been obtained both for the Ising model \cite{49,94}, the binary symmetric Lennard-Jones mixture \cite{50} and the simple Lennard-Jones fluid \cite{50,148}. Fig.~\ref{fig14} presents plots of $\gamma(\infty)/\gamma(R)-1$ (Ising model) and $\gamma_{AB}(\infty)/\gamma_{AB}(R)-1$ (binary mixture) versus $1/R^2$; the linear variation demonstrates that these data are compatible with Eq.~\ref{eq2}. Of course, due to the need to use, for each choice of $L$, only those data in Fig.~\ref{fig13} which are not affected by the droplet evaporation/condensation transition nor by the sphere to cylinder transition of the droplet, each choice of $L$ can yield only a small part of the desired curves $\gamma(R)$ and $\gamma_{AB}(R)$. Ideally, these small parts should superimpose such that a unique master curve results. Actually, there clearly occurs some scatter rather than such an ideal perfect superposition, but nevertheless the desired master curve is reasonably well defined, yielding
\begin{equation}\label{eq60}
\gamma(\infty)/\gamma(R)-1 \approx 10/R^2 , \quad \textrm{Ising \; model},
\end{equation}
\begin{equation}\label{eq61}
\gamma_{AB}(\infty)/\gamma_{AB}(R)-1 \approx 2.2/R^2 , \quad \textrm{binary \; mixture}.
\end{equation}

\begin{figure}\centering
\includegraphics[width=0.6\linewidth]{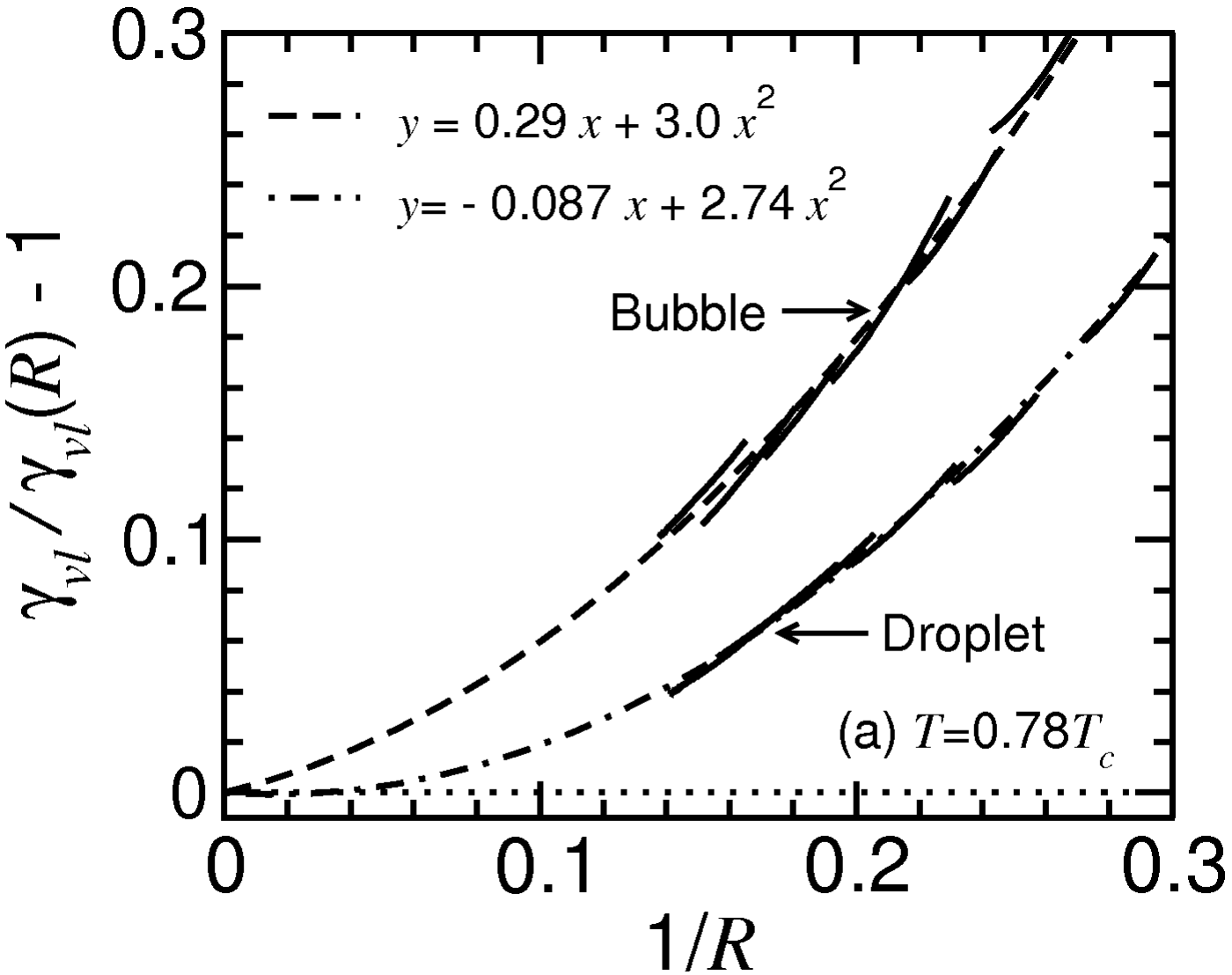}

\includegraphics[width=0.6\linewidth]{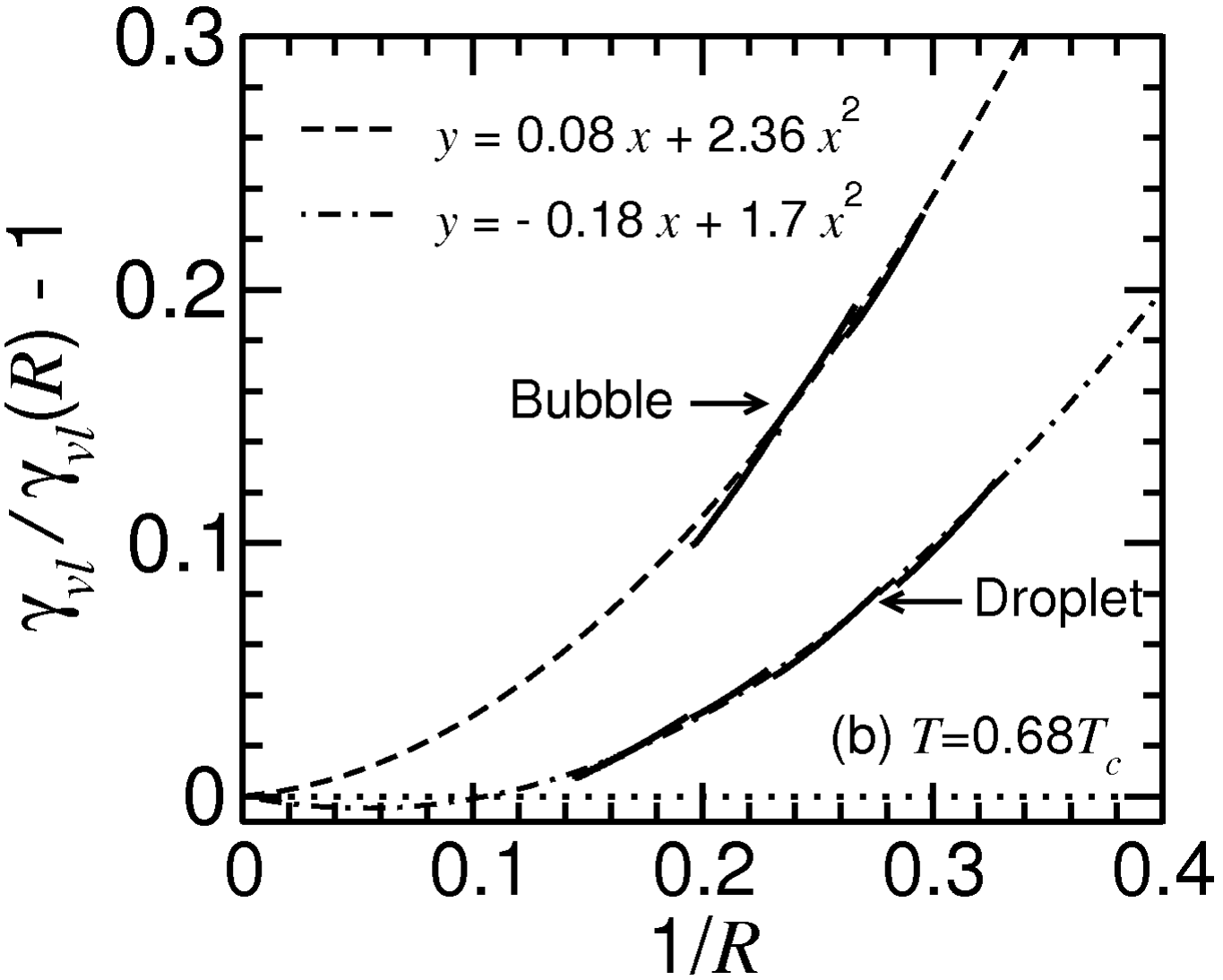}
\caption{\label{fig15} Plots of $\gamma_{v \ell}(\infty)/\gamma_{v\ell}(R)-1$ vs $1/R$ for spherical droplets and bubbles for the LJ fluid at (a) $T=0.78 T_c$, and (b) $T=0.68 T_c$. Fits to functional forms $y=ax + bx^2$, with adjustable constants a,b as quoted in the figure, are included. From Block et al. \cite{50}.}
\end{figure}

\begin{figure}\centering
\includegraphics[width=1.0\linewidth]{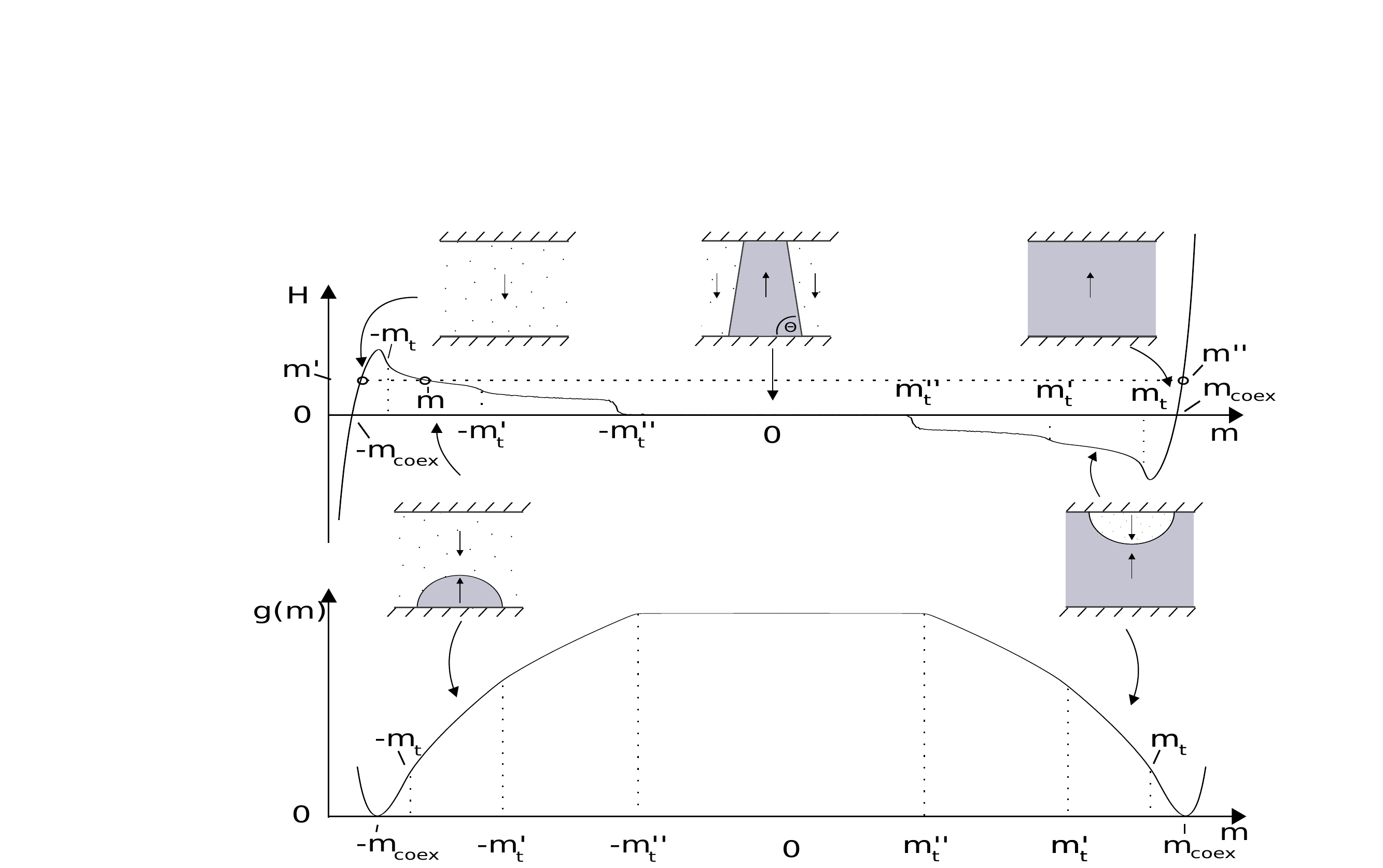}
\caption{\label{fig16} Schematic sketch of an isotherm $H_{L,D}$ versus $m$ in the Ising model in the $L \times L \times D$ geometry where in the first layer ($n=1$) a positive surface field $H_1$ acts, while in the last layer $(n=D)$ a negative surface field $H_D=-H_1$ acts (upper part). The strength of these surface fields is assumed to be in the regime of incomplete wetting \{$H_1<H_{1w}(T)$\}, and the temperature is below criticality $(T<T_c)$. The lower part shows the associated effective free energy $f_{L,D}(m,H_1,T)$ of the system. For both $H_{L,D}(m,H_1,T)$ and $f_{L,D}(m,H_1,T)$ one can distinguish different regimes: for $m <-m_t(m >m_t)$ pure phases of predominantly negative (positive) magnetization occur; for $-m_t<m<-m_{t'}$ ($m_t'<m<m_t)$ wall attached sphere-cap shaped droplets of positive (negative) magnetization occur, while for $-m_t'<m<-m_t'$ $(m_t''<m<m_t')$ the shape of the wall-attached droplets has changed from sphere cap to cylinder cap (not shown). In the region $-m_t''<m<m_t''$, the walls are planar, and thus $H_L(m,H_1,T)\equiv 0$ in this regime. The phase coexistence in the bulk occurs for $m = \pm m_{\textrm{coex}}$. The dotted horizontal line in the upper part indicates that the (pure) state with magnetization $m'$ at the ascending part can coexist not only with a pure state at the other ascending part $(m')$ but also with a mixed-phase state (m) at the descending part of the curve $H_{L,D}(m,H_1,T)$. Note that the transition at the values $\pm m_t, \pm m_t'$, and $\pm m_t''$ (highlighted by broken vertical lines) are not sharp but exhibit finite-size rounding.}
\end{figure}

\begin{figure}\centering
\includegraphics[width=0.6\linewidth]{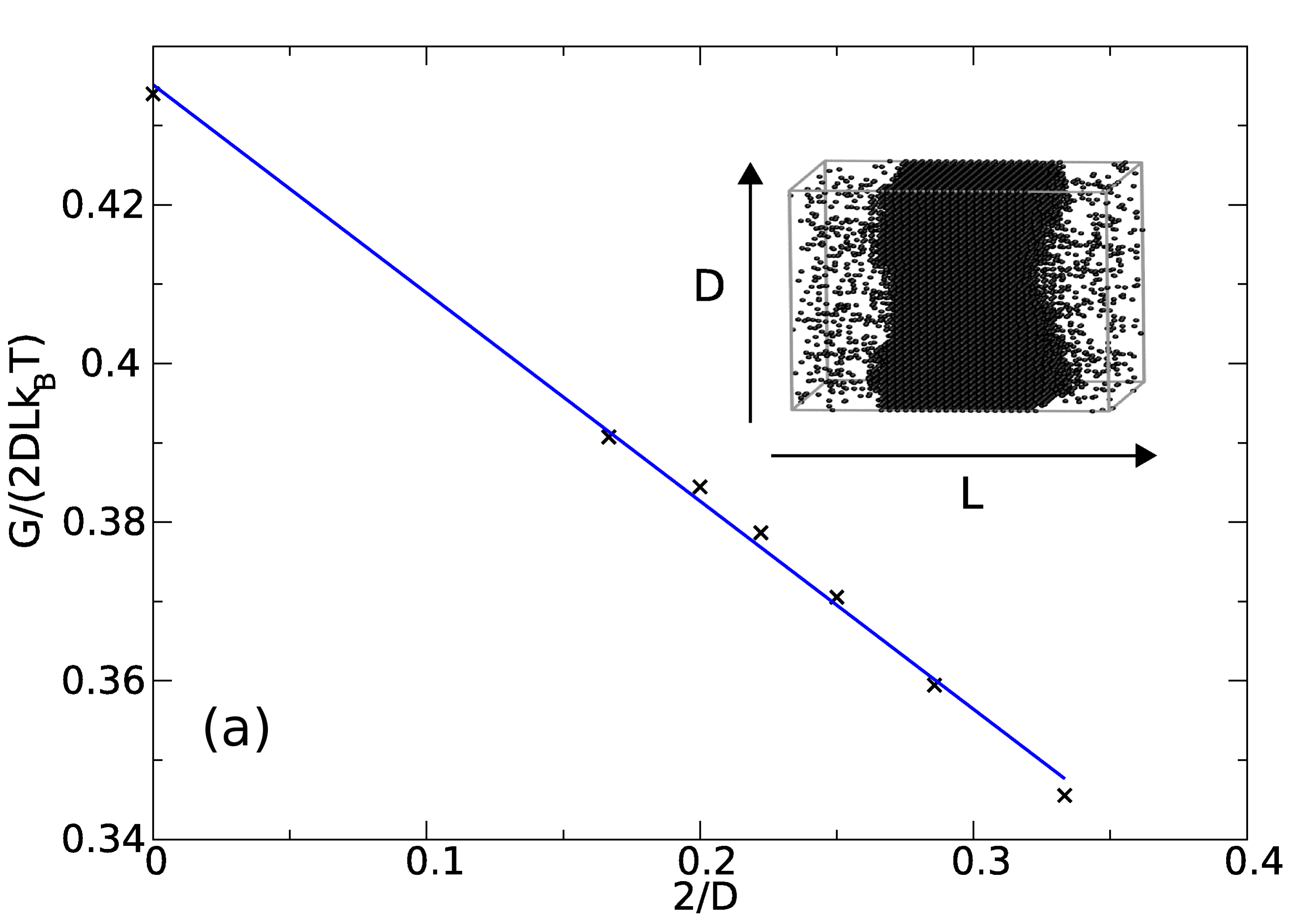}
\includegraphics[width=0.8\linewidth]{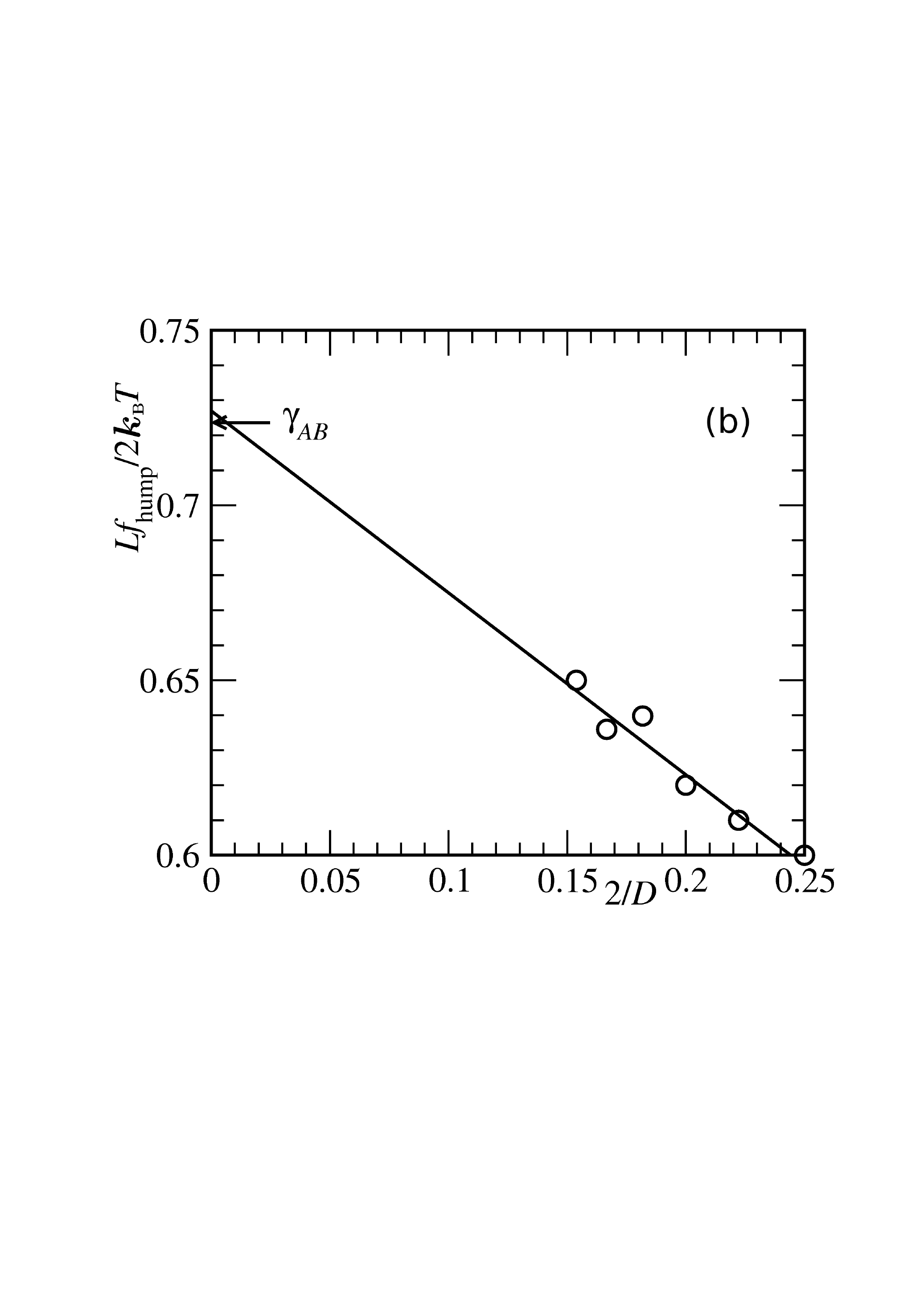}
\caption{\label{fig17} (a) Plot of free energy hump $f_L(m \approx 0, H_1=0,T)L/(2k_BT)$ for the Ising model at $k_BT/J=3.0$ (a) and of $f_L(x_A \approx 1/2, \epsilon_a=0,T)L/2k_BT$ for the binary Lennard-Jones mixture at $T=1.0$ (b) versus $2/D$. Linear dimension $L=40$ lattice spacings in case (a) and L = 30 Lennard Jones units in case (b). Part (a) is taken from \cite{99}, part (b) from \cite{50}. Note that the ordinate intercept in case (a) was fixed at the literature value \cite{139} $\gamma=0.434$, and in case (b) at the estimate $\gamma_{AB}=0.72$ (Fig.~\ref{fig11}). The slope of the straight lines yields line tension estimates $\tau(\theta = \pi/2)\approx -0.26 \pm 0.01$ (a) and $-0.52 \pm 0.01$ (b).}
\end{figure}

While for the systems shown in Fig.~\ref{fig14} the existence of a Tolman length \cite{51} is precluded simply due to the symmetry between the coexisting phases, for the simple Lennard-Jones fluid there is no such symmetry between vapor and liquid, and a Tolman length $\delta$ should exist. Since the radius of curvature is negative for a bubble, the sign of the correction in Eq.~\ref{eq1} for droplets and bubbles must differ. Fig.~\ref{fig15} shows plots of $\gamma_{v \ell}(R)/\gamma_{v \ell} (\infty)-1$  vs.~$1/R$ for two temperatures. Indeed one sees that this curvature correction for droplets differs from its counterpart for bubbles, ruling out that Eq.~\ref{eq2} holds. On the other hand, in the available region the data are not compatible with a simple linear relation in $1/R$ \{Eq.~\ref{eq1}\} either, while a fit to Eq.~\ref{eq3} which involves two characteristic lengths $\delta$ and $\ell$ is possible. Taking these data together with results for cylinders, for which one expects instead of Eq.~\ref{eq3} a relation
\begin{equation}\label{eq62}
\gamma(R)= \gamma(\infty)/1[1+\delta/R+2 (\ell_c/R)^2],
\end{equation}
where the expectation is that the same length $\delta$ in the leading $1/R$ correction must occur, while the lengths for spheres $(\ell)$ and cylinders $(\ell_c)$ in the quadratic corrections may differ. Estimates for $\delta \approx - 0.1 \sigma$ were extracted \cite{50} at both temperatures that were studied, while the lengths $\ell, \ell_c$ were both found to be of order $\sigma$, as for the binary symmetric LJ mixture. The fact that $\delta$ is negative, and its absolute value is only of the order $0.1 \sigma$, is compatible both with density functional calculations \cite{50} and with estimates extracted from different approaches \cite{57,58}. However, Anisimov \cite{153} has derived that near the critical point one expects $\delta$ to diverge as
\begin{equation}\label{eq63}
\delta =  -  \hat{\delta} (1-T/T_c)^{\beta-\nu} \;,
\end{equation}
where $\beta \approx 0.325$ and $\nu \approx 0.63$ are the critical exponents of order parameter and correlation length in the (three-dimensional) Ising universality class, and he related the amplitude prefactor $\hat{\delta}$ to an amplitude characterizing the asymmetry of the liquid-vapor phase diagram near criticality. However, with the methods described in the present paper a test of Eq.~\ref{eq63} seems prohibitively difficult.

\begin{figure}\centering
\includegraphics[width=0.6\linewidth]{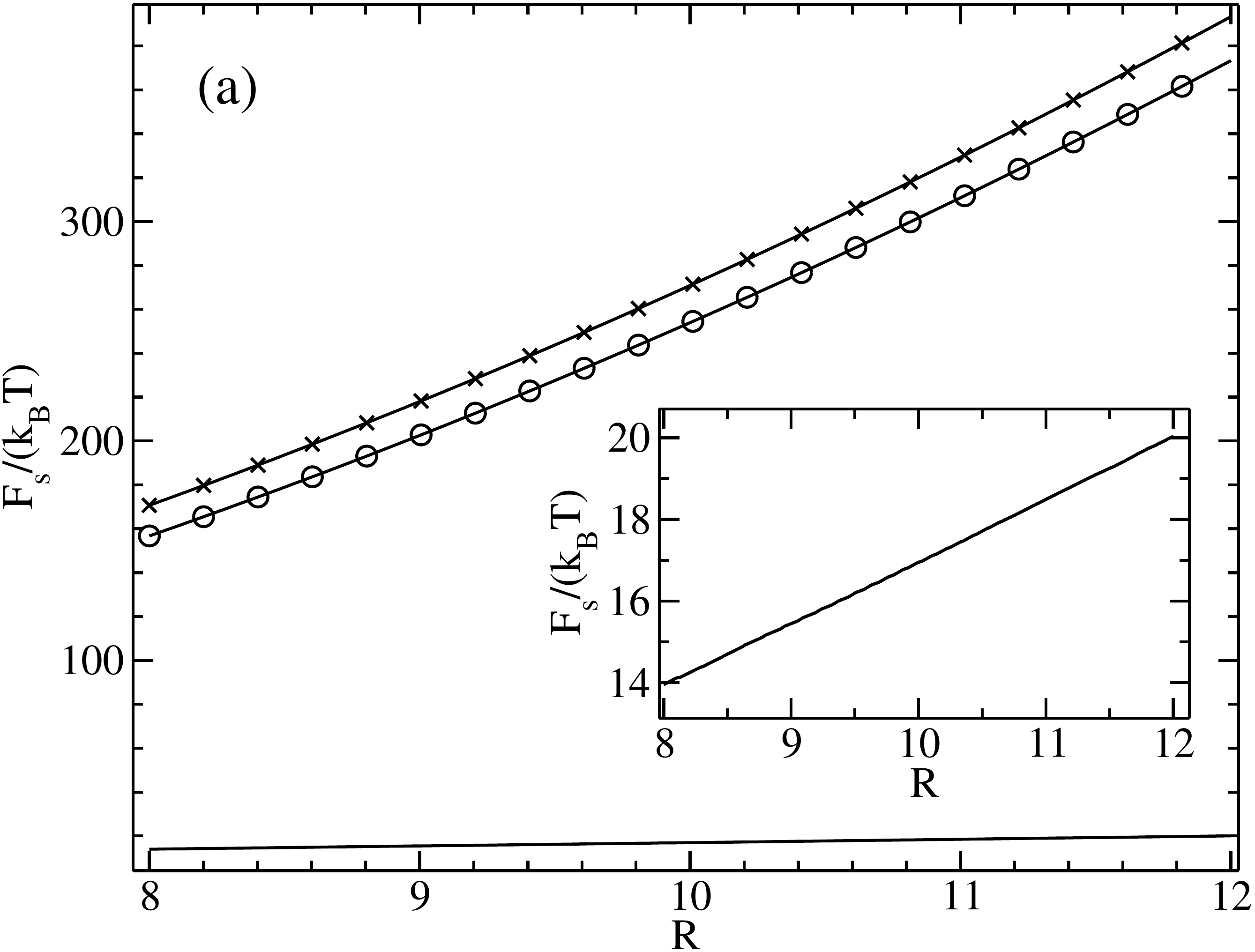}
\includegraphics[width=0.6\linewidth]{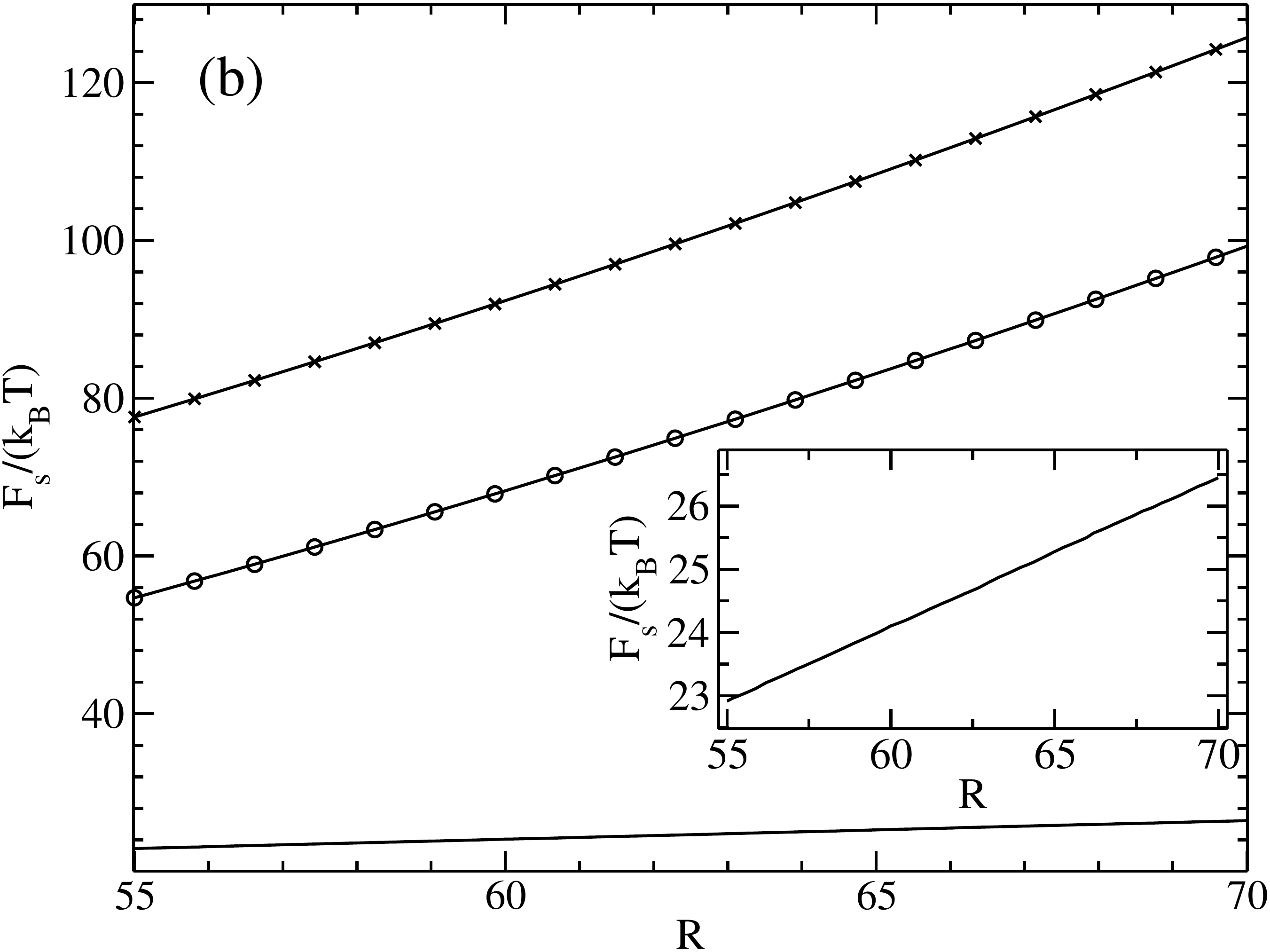}
\caption{\label{fig18} Surface free energy $F(R,\theta)/k_BT$ of wall attached droplets in the Ising model at $k_BT/J=3.0$ plotted vs. the droplet radius $R$, for $H_1/J=0$ ($\theta =90^\circ$), case (a), and $H_1/J=0.73$ $(\theta \approx 23^\circ)$, case (b). Upper curve in each case shows $4 \pi R^2 \gamma(R)f(\theta)$ with $\gamma (R)$ taken from the corresponding calculation in the bulk. Lower curve show data obtained as $L^3\Delta f$ \{Eq.~\ref{eq71}\}, using $R$ from Eq.~\ref{eq70}. Insert shows that the difference between both curves varies linearly with $R$, hence allowing to extract a meaningful estimate of $\tau(\theta)$, using Eq.~\ref{eq71}.}
\end{figure}

\section{Wall-attached droplets and methods for estimating line tensions}

We now study the order parameter distribution of systems such as considered in the previous section (see e.g. Fig.~\ref{fig9}) but we remove the periodic boundary condition in z- direction such that we simulate a $L \times L \times D$ geometry with periodic boundaries in x- and y-directions, while in the z-direction the system is confined by walls under incomplete wetting conditions. If we restrict attention to strictly symmetric systems (Ising model, or symmetric binary LJ model), it is straightforward to choose a strictly ``antisymmetric'' boundary condition (as used already in Sec.~3 to obtain information on contact angles), which has the advantage that phase coexistence in this thin film geometry is not shifted relative to the bulk. Then the analysis outlined in Fig.~12 can be straightforwardly extended to study the free energy excess due to wall-attached droplets or slab configurations confined by planar interfaces (Fig.~\ref{fig16}). For the Ising system shown here, the free energy excess is defined in analogy to Eq.~\ref{eq48} as (note that no bulk field $H$ is applied here)
\begin{equation}\label{eq64}
f_{L,D}(m,H_1,T)=-[k_BT/(L^{d-1}D)] \ln[P_{LH_1T})(m)/P_{LH_1T}(m_\textrm{coex})],
\end{equation}
and the field $H_L(m,H_1,T)$ simply is the derivative of this function, in full analogy to Eq.~\ref{eq54},
\begin{equation}\label{eq65}
H_{L,D}(m,H_1,T)=[\partial f_L(m,H_1,T)/\partial m]_{T,H_1} .
\end{equation}
Due to the choice of the antisymmetric boundary fields, we still have a strict symmetry of the effective free energy and an antisymmetry of its derivative,
\begin{equation}\label{eq66}
f_{L,D}(m,H_1,T)=f_{L,D}(-m,H_1,T) ,
\end{equation}
\begin{equation}\label{eq67}
H_{L,D}(-m,H_1,T)=-H_{L,D}(m,H_1,T).
\end{equation}

\begin{figure}\centering
\includegraphics[width=0.8\linewidth]{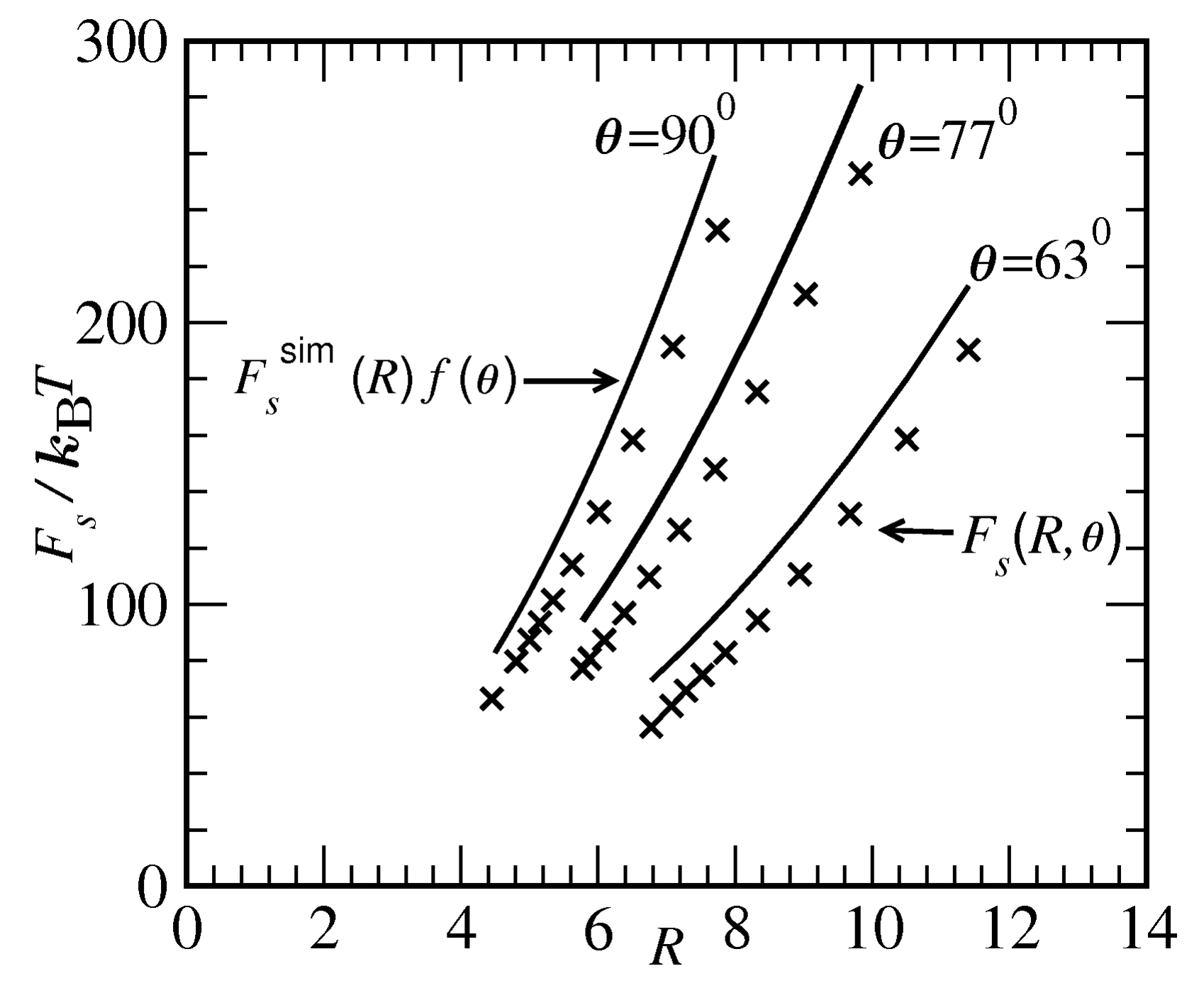}
\caption{\label{fig19} Surface free energy $F_s(R,\theta)/k_BT$ of wall-attached droplets in the symmetric binary LJ mixture at $T=1$, $\rho^* =1$ (crosses) plotted vs. the droplet radius $R$, for three choices of the contact angle (obtained as described in Sec.~3). The full curves are the corresponding predictions $F_s(R,\theta)=4 \pi R^2 \gamma_{AB}(R)f(\theta)$. From \cite{101}.}
\end{figure}

Again, the simplest case is the slab configuration in Fig.~\ref{fig16}, in particular for the case $H_1=0$, where the contact angle is $\theta = \pi/2$. Due to the symmetry of our model (see Sec.~3) we have two domain walls of area $LD$ (in $d=3$), each oriented perpendicular to the walls. Thus we expect that in this case the height of the free energy hump in Fig.~\ref{fig16} can be written as
\begin{equation}\label{eq68}
f_{L,D}(m\approx 0, H_1=0,T)= \frac 2 L (\gamma + 2 \tau (\theta = \frac \pi 2)/D), \quad L \rightarrow \infty, D \rightarrow \infty .
\end{equation}
Taking data at fixed (large) $L$ and varying $D$, this relation was tested for both the Ising model and the binary Lennard-Jones mixture. It was found that in both cases Eq.~\ref{eq68} is compatible with the simulation, implying that the line tension $\tau(\theta = \pi/2)$ is negative for both models. As a caveat, we mention that Eq.~\ref{eq68} disregards the problem that for the equivalent $L^2D$ geometry with periodic boundary conditions instead of walls we also expect a finite size correction, similarly to what is seen in the $L^3$ geometry (Fig.~\ref{fig11}). However, it is reasonable to assume that for a system with $L \ll D$ this ``intrinsic'' size effect is even smaller than for the $L^3$  geometry, and since the slope in Fig.~\ref{fig11} is an order of magnitude smaller than the slope in Fig.~\ref{fig17}b), this problem is ignored henceforth. Another issue (see also the discussion of Schimmele et al. \cite{45}) is the fact that in the separation of what is attributed to be a ``surface effect'' and what is a ``line effect'', certain conventions about the meaning of lengths characterizing the systems are inevitable. We have taken the Ising model surface area to be identical to the number of spins it contains, $LD$. However, if one would measure the distance from the row $n=1$ to the row at $n=D$ as the relevant length, one could say the surface area is only $L(D-1)$.

\begin{figure}\centering
\includegraphics[width=0.6\linewidth]{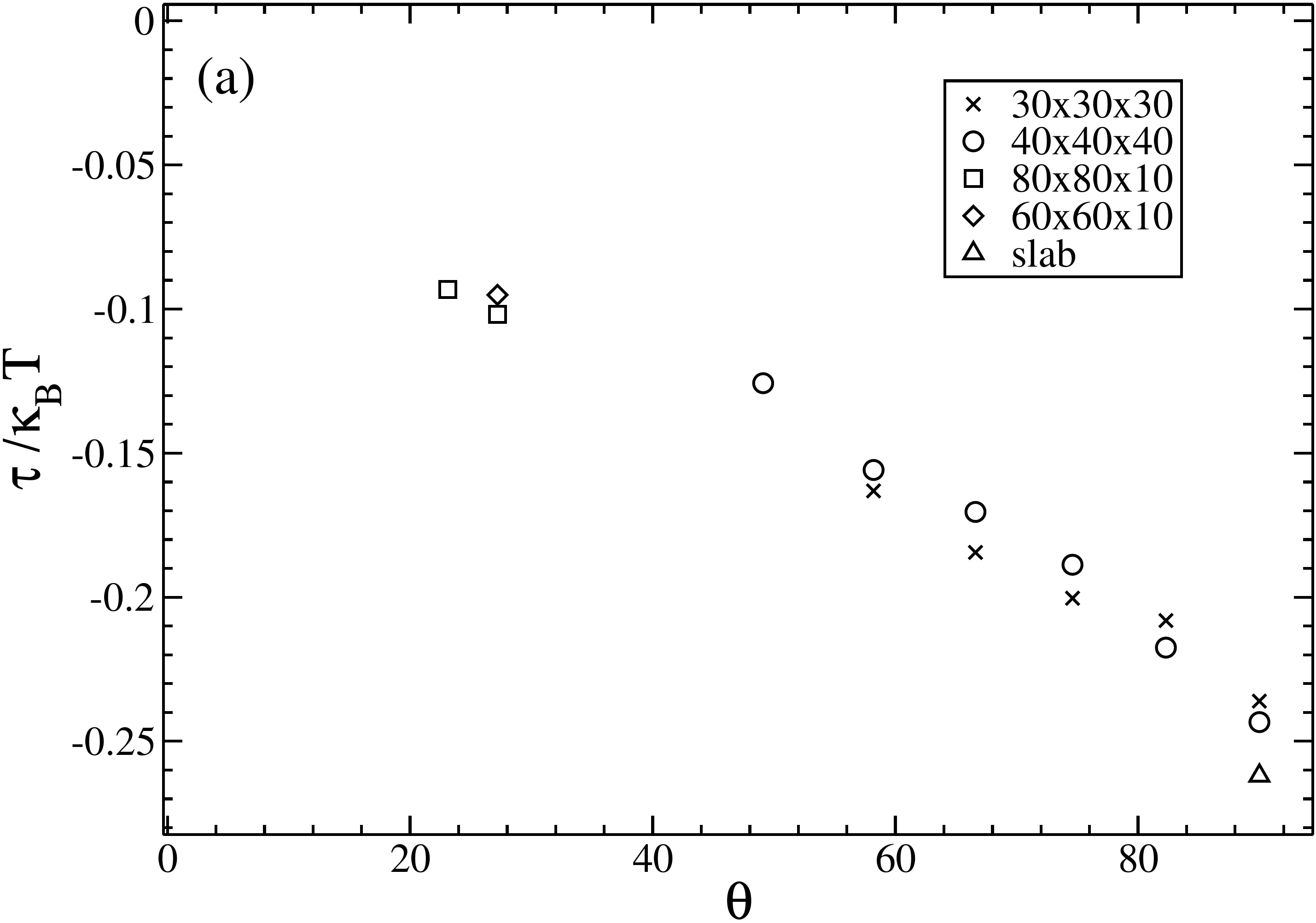}
\includegraphics[width=0.8\linewidth]{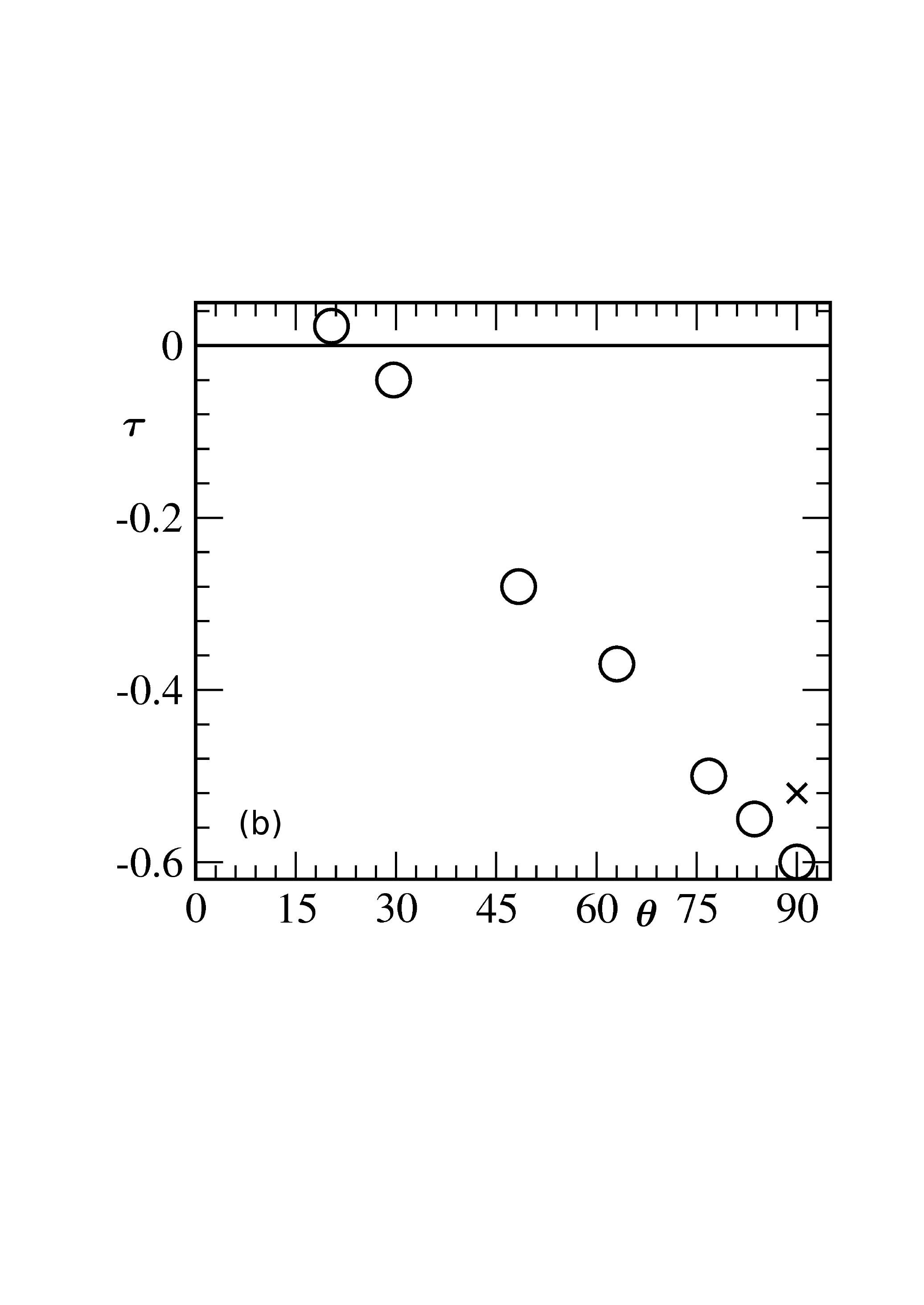}
\caption{\label{fig20} Line tension $\tau (\theta)$ plotted vs. $\theta$, for the Ising model with $J_s/J=1$ at $k_BT/J=3.0$ (a) and the binary symmetric LJ mixture at $T=1, \rho^*=1$. From \cite{94,101}. $\Delta$ in (a) and $\times$ in (b) are derived from a slab geometry with $\theta=90^\circ$.}
\end{figure}

If one assumes that the droplets shown schematically in Fig.~\ref{fig16} have a sphere-cap shape, the analysis based on Eqs.~\ref{eq57}-\ref{eq59} \{cf.~also Figs.~\ref{fig12}, \ref{fig13}\} can easily be generalized: The volume of the droplet is no longer $4 \pi R^3/3$, as used in Eq.~\ref{eq58}, but reduced by a factor $f(\theta)$
\begin{equation}\label{69}
V_{\textrm{sphere-cap}}= (4 \pi R^3/3)f(\theta), \quad f(\theta)=(2+\cos \theta)(1-\cos \theta)^2 /4,
\end{equation}
and hence instead of Eq.~\ref{eq58} we now have (for the Ising model)
\begin{equation}\label{eq70}
\Delta m = m'-m=(m''+m')4 \pi R^3f(\theta)/(3 L^2D)
\end{equation}

The same reduction factor $f(\theta)$ applies to the relation replacing Eq.~\ref{eq59},
\begin{equation}\label{eq71}
L^2 D\Delta f \equiv F_s(R,\theta)=4 \pi R^2\gamma (R) + 2 \pi R \tau(\theta) \sin \theta \;;
\end{equation}
interpreting the (absolute) excess free energy $L^2 D \Delta f$ as a total surface excess free energy of the sphere cap. We also allow for a correction proportional to the line tension $\tau (\theta)$. Note that the factor $2 \pi R \sin \theta$ is just the length of the 3-phase contact line of the sphere-cap shaped droplet.

Eq.~\ref{eq71} makes the explicit assumption that the same function $\gamma(R)$ or $\gamma _{AB}(R)$ for spherical droplets in the bulk (see previous section) also applies for sphere-cap shaped droplets attached to a flat wall. The simulation results for both the Ising model and the symmetrical binary Lennard-Jones mixture in fact are compatible with this assumption (Figs.~\ref{fig18}, \ref{fig19}). The resulting estimates $\tau(\theta)$ for both models are shown in Fig.~\ref{fig20}. Note that in the Ising case (with $J_s=J$) one has a second order wetting transition, and hence one expects $\tau(\theta \rightarrow 0) \rightarrow 0$ \cite{42}. In contrast, the LJ mixture exhibits a first order wetting transition (see Sec.~3), and in this case one expects $\tau (\theta \rightarrow 0)\rightarrow + \infty$. However, since $\tau$ is negative for large $\theta$, it is implied that $\tau$ must change sign before the wetting transition is reached \cite{42,43}. Indeed the numerical data (Fig.~20b) are compatible with this expectation.

\section{Outlook}
In the present article, we have reviewed a selection of methods which have been used to simulate phase coexistence between fluid phases, for simple models of statistical mechanics, and we have discussed the information on interfacial free energies and related properties (contact angles at walls, line tension) that one can extract from these studies. Neither the detailed interfacial structure nor the study of wetting transitions has been discussed. However, even with this restricted scope we have by no means attempted to present an exhaustive coverage of the subject, but rather we have described the techniques using a few examples, taken from the research groups of the authors for the sake of simplicity. It needs to be stated, that very valuable research on related topics can be found in the literature from other groups, sometimes also for other models than the models (lattice gas, Lennard-Jones fluid, symmetric binary LJ mixture) that were discussed here. E.g., a particularly well-studied model system is the square-well fluid. Van Swol and Henderson \cite{154,155} have investigated wetting and drying at fluid-wall interfaces for this model, and compared their simulations to density functional calculations. Due to the lack of any particularly symmetries for this model (or the Lennard-Jones fluid considered in the pioneering study by Sikkenk et al. \cite{156}), it was difficult to obtain very accurate results on interfacial free energies, contact angles etc. at this time, considering also the fact that now orders of magnitude more computer power can be invested than available for this early work \cite{154,155,156}. Significant progress has also been made via a better understanding of the statistical mechanics of the fluid-wall interactions \cite{157}, and this work is also basic for a very interesting recent simulation study \cite{158} where a slab geometry of a liquid bridge between two walls with contact angles $\theta_1, \theta_2$ such that $\theta_1 + \theta_2=\pi$ has been analyzed. In this work, it became possible to derive estimates for the line tension $\tau(\theta)$ for several values of the contact angle $\theta$.

We also add that we have only addressed the problem of phase equilibria of two fluid phases in the presence of a solid flat wall. We have neither considered other geometries (e.g. wedge geometries, where ``filling transitions'' can occur \cite{24}) nor have we considered interfacial phenomena where the solid wall is replaced by another fluid phase; in this case three fluid-fluid interfaces also can meet at a contact line, but the corresponding line tension \cite{15} was out of consideration here. Also the problem of contact angles and wall-attached droplets for non-volatile fluids was out of our scope. E.g., a typical example is a droplet formed by a polymer melt at a substrate \cite{160}: the vapor surrounding the polymer droplet typically does not contain any polymers at all, so strictly speaking the system is not in full equilibrium (which would imply that the polymers completely evaporate), but in practice this does not occur at all at the time scales of interest. Some of the methods described here could be carried over to a study of such more complex systems as well.

\underline{Acknowledgements}: This research was supported in part by the Deutsche Forschungsgemeinschaft (grants N$^o$ SFB TR6/A5 and SPP 1296-BI 314/19). We are greatly indebted to Martin Oettel for fruitful discussions. We would also like to thank JSC J\"ulich for grant of computing time.

\end{document}